\documentclass[aip]{revtex4-1}

\usepackage[utf8]{inputenc}
\usepackage[T1]{fontenc}

\usepackage[caption = false]{subfig}
\usepackage{pgfplots}
\usepackage{pgfplotstable}
\usepackage{import} 
\usepackage{siunitx}
\usepackage{catchfile}
\usepackage{latexcolors} 
\usepgfplotslibrary{groupplots}
\usepackage{standalone} 
\usepackage{derivative}
\usepackage{amsfonts}

\pgfplotsset{compat=newest, 
	table/search path={figs} 
}

\usepgfplotslibrary{external}
\usetikzlibrary{math, shapes.geometric, shapes.misc, calc, arrows.meta, fit, backgrounds} 

\tikzsetexternalprefix{figcache/}
\tikzset{external/mode=list and make}

\usepackage[abbreviations]{glossaries-extra}
\setabbreviationstyle{long-short} 
\usepackage[colorlinks,linkcolor=blue,citecolor=blue,urlcolor=red]{hyperref}

\usepackage{cleveref}

\usepackage{xcolor}
\usepackage{xargs}

\newcommandx{\unsure}[2][1=]{\todo[linecolor=red,backgroundcolor=red!25,bordercolor=red,#1]{#2}}
\newcommandx{\change}[2][1=]{\todo[linecolor=blue,backgroundcolor=blue!25,bordercolor=blue,#1]{#2}}
\newcommandx{\info}[2][1=]{\todo[linecolor=OliveGreen,backgroundcolor=OliveGreen!25,bordercolor=OliveGreen,#1]{#2}}

\newcommand{\addmarkandpin}[4]{\node[font=\footnotesize, fill, inner sep=0pt,outer sep=0pt,circle,minimum size=3,pin={[pin edge={#3}]#4:{$#2$}}] at (#1) {};}

\newcommand{\PreserveBackslash}[1]{\let\temp=\\#1\let\\=\temp}

\newcolumntype{C}[1]{>{\PreserveBackslash\centering}p{#1}}
\newcolumntype{R}[1]{>{\PreserveBackslash\raggedleft}p{#1}}
\newcolumntype{L}[1]{>{\PreserveBackslash\raggedright}p{#1}}
\pgfmathsetseed{11}

\tikzset{
	fluid node/.style={
			fill=bluegray,
			shape=circle,
			scale=0.9
		}
}
\tikzset{
	ghost node/.style={
			fill=pastelorange,
		}
}

\tikzset{
	single layer node/.style={
			fill=wildwatermelon,
			shape=star,
			star points=6,
			scale=0.7
		}
}

\tikzset{
	inert node/.style={
			fill=tropicalrainforest,
			shape=diamond,
			scale=0.4
		}
}

\tikzset{
	fictitious node/.style={
			fill=slategray,
			shape=regular polygon,
			regular polygon sides=6,
			scale=0.8
		}
}

\tikzset{
	ghost mirror node/.style={
			shape=cross out,
			line width=2,
			draw=red,
			scale=0.8
		}
}

\pgfplotsset{
	yticklabel style={
			/pgf/number format/fixed,
		},
	xticklabel style={
			/pgf/number format/fixed,
		},
	xtick align=center,
	ytick align=center,
	every axis/.append style={
			font=\footnotesize,
		}
}

\makeatletter
\tikzset{
	@arc through/.style 2 args={
		to path={
			\pgfextra
			\pgfextract@process\pgf@tostart{\tikz@scan@one@point\pgfutil@firstofone(\tikztostart)\relax}%
			\pgfextract@process\pgf@tothrough{\tikz@scan@one@point\pgfutil@firstofone#1}%
			\pgfextract@process\pgf@totarget{\tikz@scan@one@point\pgfutil@firstofone(\tikztotarget)\relax}%
			\pgfextract@process\pgf@topointMidA{\pgfpointlineattime{.5}{\pgf@tostart}{\pgf@tothrough}}%
			\pgfextract@process\pgf@topointMidB{\pgfpointlineattime{.5}{\pgf@totarget}{\pgf@tothrough}}%
			\pgfextract@process\pgf@tocenter{%
				\pgfpointintersectionoflines{\pgf@topointMidA}
				{\pgfmathrotatepointaround{\pgf@tothrough}{\pgf@topointMidA}{90}}
				{\pgf@topointMidB}{\pgfmathrotatepointaround{\pgf@tothrough}{\pgf@topointMidB}{90}}}%
			\pgfcoordinate{arc through center}{\pgf@tocenter}%
			\pgfpointdiff{\pgf@tocenter}{\pgf@tostart}%
			\pgfmathveclen@{\pgfmath@tonumber\pgf@x}{\pgfmath@tonumber\pgf@y}%
			\edef\pgf@toradius{\pgfmathresult pt}
			\pgfmathanglebetweenpoints{\pgf@tocenter}{\pgf@tostart}%
			\let\pgf@tostartangle\pgfmathresult
			\pgfmathanglebetweenpoints{\pgf@tocenter}{\pgf@totarget}%
			\let\pgf@toendangle\pgfmathresult
			\ifdim\pgf@tostartangle pt>\pgf@toendangle pt\relax
			\pgfmathsetmacro\pgf@tostartangle{\pgf@tostartangle-360}%
			\fi
			#2%
			\pgfmathsetmacro\pgf@toendangle{\pgf@toendangle-360}%
			\fi
			\endpgfextra
			arc [radius=+\pgf@toradius, start angle=\pgf@tostartangle, end angle=\pgf@toendangle] \tikztonodes
	}},
	arc through ccw/.style={@arc through={#1}{\iffalse}},
	arc through cw/.style={@arc through={#1}{\iftrue}},
}
\makeatother
\newabbreviation
{sph}
{SPH}
{Smoothed Particle Hydrodynamics}

\newabbreviation
{wcsph}
{WCSPH}
{Weakly Compressible \gls{sph}}

\newabbreviation
{isph}
{ISPH}
{Incompressible \gls{sph}}

\newabbreviation
{mms}
{MMS}
{Method of Manufactured Solutions}

\newabbreviation
{sisph}
{SISPH}
{Simple Iterative \gls{sph}}

\newabbreviation
{dtsph}
{DTSPH}
{Dual-Time \gls{sph}}

\newabbreviation
{edac}
{EDAC}
{Entropically Damped Artificial Compressibility}

\newabbreviation
{dbc}
{DBC}
{Dynamic Boundary Condition}

\newabbreviation
{mdbc}
{mDBC}
{modified \gls{dbc}}

\newabbreviation
{lust}
{LUST}
{Local Uniform STencil}

\newabbreviation
{tvdrk2}
{TVD-RK2}
{Total Variaton Diminishing - Runge Kutta 2}

\newabbreviation
{fvm}
{FVM}
{Finite Volume Method}

\newabbreviation
{fou}
{FOU}
{First Order Upwind}

\newabbreviation
{mls}
{MLS}
{Moving Least Squares}

\newabbreviation
{tvf}%
{TVF}%
{Transport Velocity Formulation}

\newabbreviation
{epec}%
{EPEC}%
{Evaluate Predict Evaluate Correct}
\begin{document}


\title{Robust Solid Boundary Treatment for Compressible Smoothed Particle Hydrodynamics} 



\author{Navaneet Villodi} 
\email[Corresponding Author: ]{navaneet@iitb.ac.in}
\affiliation{Department of Aerospace Engineering, Indian Institute of
Technology Bombay, Powai, Mumbai 400076}

\author{Prabhu Ramachandran} 
\email[]{prabhu@aero.iitb.ac.in}
\homepage[\href{http://www.aero.iitb.ac.in/~prabhu}{http://www.aero.iitb.ac.in/{\fontfamily{ptm}\selectfont\textasciitilde}prabhu}]{}
\affiliation{Department of Aerospace Engineering, Indian Institute of
Technology Bombay, Powai, Mumbai 400076}


\date{\today}

\begin{abstract}
    The unavailability of accurate boundary treatment methods for compressible \gls{sph} severely limits its ability to simulate flows in and around bodies. To this end, challenges specific to compressible flows with \gls{sph} are carefully considered. Based on these, robust and widely applicable boundary treatment methods for compressible \gls{sph} are proposed. These are accompanied by a novel technique to prevent particle penetration at boundaries. The proposed methods are shown to be significantly better than other recent approaches. A wide variety of test problems, many of which are not shown to be simulated with \gls{sph} thus far, are employed to highlight the strengths and weaknesses of the proposed methods. The implementation is open source and the results are automated in the interest of reproducibility. Overall, this research contributes to the advancement of \gls{sph} as a viable alternative to mesh-based methods for compressible flow simulations.
\end{abstract}


\maketitle 

\section{Introduction} \label{sec:intro}

Smoothed Particle Hydrodynamics(\gls{sph}) is quite capable of simulating compressible fluid flows and has been
widely used for simulating various phenomena
involving compressible hydrodynamics \citep{rosswogAstrophysicalSmoothParticle2009,
springelSmoothedParticleHydrodynamics2010a}. Many applications
usually involve flow in and around bodies. To simulate these, proper boundary
treatment methods are required to accurately model the physics at the
boundary. 

The boundaries that need to be dealt with can be broadly classified into two
categories: permeable boundaries and solid wall boundaries. Permeable
boundaries are those that allow fluid to enter or exit the computational
domain. Permeable boundary treatment methods for incompressible flows should
not pose major setbacks for use with compressible flows. Solid wall boundaries
are commonly dealt with using ghost particles~\citep[for
e.g.,][]{takedaNumericalSimulationViscous1994,morrisModelingLowReynolds1997,colagrossiNumericalSimulationInterfacial2003,yildizSPHMultipleBoundary2009,maciaTheoreticalAnalysisNoSlip2011,marroneDSPHModelSimulating2011,
adamiGeneralizedWallBoundary2012, marroneAccurateSPHModeling2013,
antuonoCloneParticlesSimplified2023}. All of these have been proposed for
incompressible and weakly compressible flows. The solid wall boundary
treatment methods proposed in the incompressible and weakly compressible
\gls{sph} literature do not work well for compressible flows. The following
are the major issues that remain unresolved in the context of compressible
flows and are addressed in this work:

\begin{enumerate}  
	\item Particle penetration: Compressible flows result in scenarios that
	encourage particles to creep into the boundary. Popular methods from the
	incompressible \gls{sph} literature like that of
	\citet{adamiGeneralizedWallBoundary2012} suffer from this (see
	\cref{fig:ccor oadami}). A detailed discussion on the reasons behind this
	is presented in \cref{sec:challenges}.
	
	\item Inaccuracies: Particle penetration brings with it inaccuracies and
	instabilities. The use of short-range repulsive forces is effective in
	avoiding particle penetration
	\citep{sillaSPHSimulationOblique2017,englestadInvestigationsNovelBoundary2020}.
	However, these results also leave much to be desired. The methods
	presented in the present work furnish demonstrably better results (see
	\cref{tab:ccor-err}).
	
	\item Lack of simulation data: There is limited simulation data available
	for compressible flows with solid boundaries using \gls{sph} as most of
	the compressible flow test cases and benchmark problems that are used to
	validate \gls{sph} schemes are without solid wall boundaries. The work of
	\citet{sunAccurateSPHVolume2021} showcases some interesting simulations
	involving compressible flows with boundaries. They claim to have used the
	method of \citet{marroneDSPHModelSimulating2011}. As this method was
	introduced to be used with incompressible flows, we found the details
	regarding modifications, if any, to be missing. Moreover, their problems
	do not focus on shocks and shock-boundary interactions. The present work
	includes a wide variety of problems including but not limited to shocks
	and shock-boundary interactions.
\end{enumerate}

The existing boundary treatment methods in \gls{sph} are augmented with following major ideas to address the above issues:

\begin{enumerate}
	\item Judicious extrapolation: Differently from
	\citet{adamiGeneralizedWallBoundary2012,marroneDSPHModelSimulating2011},
	instead of just pressure and velocity, all relevant quantities are
	extrapolated from the fluid domain to the ghost particles in a manner that
	is consistent with the boundary conditions, as explained in
	\cref{sec:bc-strategy}. 
	
	\item Ghost volume constancy: While the density of a ghost particle is
	allowed to vary, its volume should be constant for the 
	ghost particles to be an effective partition of space. This
	improves the quality of results and curtails the particle penetration
	problem to a large extent. The rationale is explained in
	\cref{sec:ghost-volume-constancy}.
	
	\item Penetration shield: For extreme cases where the above is not
	sufficient, a penetration shield is introduced. The particles that are on
	a collision course with the boundary are steered away making use of the
	\gls{tvf}\citep{TVFAdami2013}, as explained in
	\cref{sec:penetration-shield}. 
\end{enumerate}

The rest of this paper is organized as follows. \Cref{sec:lit} presents a
detailed overview of permeable and solid boundary treatment methods in
\gls{sph}. \Cref{sec:sph} lays down the governing equations and the
discretization scheme used in this study. We also outline what is expected at
the boundary and the boundary treatment strategy that is used in this study.
Further, the challenges that are unique to the treatment of boundaries in
compressible \gls{sph} are dissected. Remedial measures to address these
issues are also presented therein. \Cref{sec:results} demonstrates that the
proposed techniques are capable of simulating a diverse collection of
benchmark problems. These include problems involving oblique shocks, normal
shocks, bow shocks, subsonic to supersonic transition, complex wave
interactions, flow over bodies with sharp edges, flow over moving bodies, and
three-dimensional flows. \Cref{sec:summary} summarizes the work. Finally, \cref{sec:conclusions} presents some suggestions for future and concludes the work.

\section{State of Boundary Treatment Methods}\label{sec:lit}

With \gls{sph}, incompressible flows are usually simulated using \gls{wcsph} and \gls{isph} schemes. A considerable amount of progress has been made in the treatment of boundary conditions with \gls{wcsph} and \gls{isph}. The recent work of \citet{negiHowTrainYour2022a}, wherein the popular boundary treatment techniques have been rigorously benchmarked using \gls{mms}, serves as a structured review of these boundary treatment methods in \gls{sph}. We also touch upon some notable boundary treatment methods in the subsequent subsections.

\subsection{Permeable boundaries}\label{sec:permiable}

Implementation of permeable boundaries in \gls{sph} generally involves the use
of buffer zones~\citep{lastiwkaPermeableNonreflectingBoundary2009a,
federicoSimulating2DOpenchannel2012b, tafuniVersatileAlgorithmTreatment2018a,
negiImprovedNonreflectingOutlet2020, holmesNovelPressureInlet2021}. These buffer zones facilitate the entry and exit of fluid
particles. There also
exist methods that do not conform to this framework~\citep{zhangLagrangianFreestreamBoundary2023, ferrandUnsteadyOpenBoundaries2017a}.
\citet{werdelmannApproachPermeableBoundary2021} presents a concise summary of
permeable boundary treatment methods for \gls{sph} along with their own novel
framework for the same.

As far as permeable boundaries are concerned, the implementations from
\gls{isph} and \gls{wcsph} can be applied, mostly without any fundamental
modifications, to compressible flow simulations. In supersonic flows, all the
quantities can be set at the inlet as all characteristics at the inlet are
known to be directed into the domain. Therefore, an extrapolation procedure,
like that employed by \citet{tafuniVersatileAlgorithmTreatment2018a} is not
necessary for supersonic flows.

On the account that existing permeable boundary treatment methods can be
relied upon, we chose to focus on the shortcomings with the treatment of solid
boundaries and make use of \citet{federicoSimulating2DOpenchannel2012b}'s
method for permeable boundaries. This choice is motivated by the ease of
implementation. We are aware of the fact that \citet{negiHowTrainYour2022a}
found this method to be not the most optimal method for convergence. We also
do not discount the potential issues this method may pose, for instance, being
ineffective when the velocity components are normal to the interface~\citep
{werdelmannApproachPermeableBoundary2021}.

\subsection{Solid wall boundaries}\label{ssec: solid wall boundaries} 

In this section, we touch upon the notable solid boundary treatment methods
from the literature.

Earliest attempts to implement solid boundary conditions in \gls{sph} relied 
on using short-range repulsive forces~\citep[for
e.g.,][]{monaghanSimulatingFreeSurface1994,
monaghanSolitaryWavesCretan1999,
monaghanSPHParticleBoundary2009}.
Fixed particles representing the boundary were conferred with the ability to
exert a repulsive force on fluid particles in the vicinity. These repulsive
forces could be exerted by a single layer of particles at the interface as
shown in \cref{fig:boundary single}. Having a single layer of particles at the
interface is appealing as it makes the representation of complex geometries
easier, albeit at the cost of kernel truncation errors. In fact,
\citet{campbellNewAlgorithmsBoundary1989} had emphasized the retention of
boundary integral or wall normalization terms while deriving the discretized
governing equations to remedy the errors due to kernel truncation when a
single layer of particles is used. The discretionary nature of the form of
force that is employed is another notable disadvantage of this approach. The
methods of
\citet{marongiuNumericalSimulationFlow2007,hashemiModifiedSPHMethod2012} are
some other notable methods that employ a single layer of particles at the
interface. As expected, these methods exhibit poor
convergence~\citep{negiHowTrainYour2022a}.

\begin{figure}
	\centering
	\begin{tikzpicture}
		\draw node[matrix, draw, align=center]  {
			\draw[dashed, thick]  (0,0) -- (0.5,0) node[anchor=west]{\scriptsize interface};  &
			\node[fluid node, label=east:\scriptsize fluid] {};  &
			\node[single layer node, label=east:\scriptsize single layer] {}; &
			\node[ghost node, label=east:\scriptsize ghost]{}; &
			\node[ghost mirror node, label=east:\scriptsize ghost-mirror] {};  &
			\node[fictitious node, label=east:\scriptsize ficticious]{}; &
			\node[inert node, label=east:\scriptsize inert] {};\\
		};
	\end{tikzpicture}

	\begin{tabular}{cc}
	\subfloat[]{
		\centering
		\input{macros.tex}
\begin{tikzpicture}
	\tikzmath{
	let \dx = 1;
	let \nx = 3;
	let \ny = 5;
	for \i in {0,..,\nx - 1} {
	for \j in {0,..,\ny - 1} {
	if ((\i==0) && (\j==2)) then {
			\x=(-\i - 1)*\dx;
			\y=(\j + 0.5)*\dx;
		} else {
			\x=(-\i - 1)*\dx + rand*0.15;
			\y=(\j + 0.5)*\dx + rand*0.15;
		};
	{
	\draw (\x,\y) node[fluid node] {};
	};
	};
	};
	{
	\draw[dashed, thick] (0,0) -- (0,\ny*\dx);
	};
	for \y in {0.5*\dx,1.5\dx,...,\ny*\dx} {
			{
					\draw (0,\y) node[single layer node] {};
				};
		};
	}
\end{tikzpicture}
		\label{fig:boundary single}
	}&
	\subfloat[]{
		\centering
		\input{macros.tex}
\begin{tikzpicture}
	\tikzmath{
	let \dx = 1;
	let \nx = 3;
	let \ny = 5;
	for \i in {0,..,\nx - 1} {
	for \j in {0,..,\ny - 1} {
	if ((\i==0) && (\j==2)) then {
			\x=(-\i - 0.5)*\dx;
			\y=(\j + 0.5)*\dx;
		} else {
			\x=(-\i - 0.5)*\dx + rand*0.15;
			\y=(\j + 0.5)*\dx + rand*0.15;
		};
	{
	\draw (\x,\y) node[fluid node] {};
	};
	};
	};
	{
	\draw[dashed, thick] (0,0) -- (0,\ny*\dx);
	};
	for \y in {0,0.5*\dx,...,\ny*\dx} {
			{
					\draw (0,\y) node[inert node] {};
				};
		};
	\i=0;
	\j=2;
	\xo=(-\i - 0.5)*\dx;
	\yo=(\j + 0.5)*\dx;
	{
	\draw (\xo,\yo) node[shape=circle, draw=black, thick, label=north:$i$]{};
	\draw (\xo,\yo) node[dotted, thick, shape=circle, draw=black, line width=1, minimum size=95]{};
	};
	for \i in {1,..,\nx - 2} {
	for \j in {1,..,\ny - 2} {
	\x=(\i - 0.5)*\dx;
	\y=(\j + 0.5)*\dx;
	{
	\draw (\x,\y) node[fictitious node]{};
	\draw[dashdotted, thick] (\xo,\yo) -- (\x,\y);
	};
	};
	};
	}

\end{tikzpicture}
		\label{fig:boundary ficticious}
	}\\

	\subfloat[]{
		\centering
		\input{macros.tex}
\begin{tikzpicture}
	\tikzmath{
	let \dx = 1;
	let \nx = 3;
	let \ny = 4;
	for \i in {0,..,\nx - 1} {
	for \j in {0,..,\ny - 1} {
	if ((\i==0) && (\j==2)) then {
			\x=(-\i - 0.5)*\dx;
			\y=(\j + 0.5)*\dx;
		} else {
			\x=(-\i - 0.5)*\dx + rand*0.15;
			\y=(\j + 0.5)*\dx + rand*0.15;
		};
	{
	\draw (\x,\y) node[fluid node] {};
	\draw (-\x,\y) node[ghost node] {};
	};
	};
	};
	{
	\draw[dashed, thick] (0,0) -- (0,\ny*\dx);
	};
	}
\end{tikzpicture}
		\label{fig:boundary mirrored ghost}
	} &
	%
	\subfloat[]{
		\centering
		\input{macros.tex}
\begin{tikzpicture}
	\tikzmath{
	let \dx = 1;
	let \nx = 3;
	let \ny = 4;
	for \i in {0,..,\nx - 1} {
	for \j in {0,..,\ny - 1} {
	if ((\i==0) && (\j==2)) then {
			\x=(-\i - 0.5)*\dx;
			\y=(\j + 0.5)*\dx;
		} else {
			\x=(-\i - 0.5)*\dx + rand*0.15;
			\y=(\j + 0.5)*\dx + rand*0.15;
		};
	{
	\draw (\x,\y) node[fluid node] {};
	};
	};
	};
	{
	\draw[dashed, thick] (0,0) -- (0,\ny*\dx);
	};
	for \x in {0.5*\dx,1.5\dx,...,\nx*\dx} {
			for \y in {0.5*\dx,1.5\dx,...,\ny*\dx} {
					{
							\draw (\x,\y) node[ghost node] {};
						};
				};
		};
	}
\end{tikzpicture}
		\label{fig:boundary stationary ghost}
	} \\

	\subfloat[]{
		\centering
		\input{macros.tex}
\begin{tikzpicture}
	\tikzmath{
	let \dx = 1;
	let \nx = 3;
	let \ny = 4;
	for \i in {0,..,\nx - 1} {
	for \j in {0,..,\ny - 1} {
	if ((\i==0) && (\j==2)) then {
			\x=(-\i - 0.5)*\dx;
			\y=(\j + 0.5)*\dx;
		} else {
			\x=(-\i - 0.5)*\dx + rand*0.15;
			\y=(\j + 0.5)*\dx + rand*0.15;
		};
	{
	\draw (\x,\y) node[fluid node] {};
	};
	};
	};
	{
	\draw[dashed, thick] (0,0) -- (0,\ny*\dx);
	};
	for \x in {0.5*\dx,1.5\dx,...,\nx*\dx} {
			for \y in {0.5*\dx,1.5\dx,...,\ny*\dx} {
					{
							\draw (\x,\y) node[ghost node] {};
						};
				};
		};
	for \x in {-0.5*\dx,-1.5\dx,...,-\nx*\dx} {
			for \y in {0.5*\dx,1.5\dx,...,\ny*\dx} {
					{
							\draw (\x,\y) node[ghost mirror node] {};
						};
				};
		};
	}
\end{tikzpicture}
		\label{fig:boundary ghost mirror}
	} &
	%
	\subfloat[]{
		\centering
		\input{macros.tex}
\begin{tikzpicture}
	\tikzmath{
	let \dx = 1;
	let \nx = 3;
	let \ny = 4;
	for \i in {0,..,\nx - 1} {
	for \j in {0,..,\ny - 1} {
	if ((\i==0) && (\j==2)) then {
			\x=(-\i - 0.5)*\dx;
			\y=(\j + 0.5)*\dx;
		} else {
			\x=(-\i - 0.5)*\dx + rand*0.15;
			\y=(\j + 0.5)*\dx + rand*0.15;
		};
	{
	\draw (\x,\y) node[fluid node] {};
	};
	};
	};
	{
	\draw[dashed, thick] (0,0) -- (0,\ny*\dx);
	};
	for \x in {0.5*\dx,1.5\dx,...,\nx*\dx} {
			for \y in {0.5*\dx,1.5\dx,...,\ny*\dx} {
					{
							\draw (\x,\y) node[ghost node] {};
						};
				};
		};
	for \y in {0,0.25*\dx,...,\ny*\dx} {
			{
					\draw (0,\y) node[inert node] {};
				};
		};
	}
\end{tikzpicture}
		\label{fig:boundary inert}
	} 
	\end{tabular}
	\caption{Various particle arrangements at the boundary interface. (a) Boundary represented by a single layer of fixed particles at the interface. (b) Boundary treatment using ficticious and inert particles. Fictitious particles generated for a fluid particle $i$ are shown. The region of dependence is marked by the dotted circle. (c) Ghost particles mirroring fluid particles about the interface. (d) Ghost particles which are stationary with respect to the interface. (e) Ghost particles which are stationary with respect to the interface paired with accomplices which mirror them about the interface. (f) Ghost particles which are stationary with respect to the interface and inert particles at the interface.}

	\label{fig: boundary arrngement}
\end{figure}

The semi-analytical boundary treatment method of
\citet{kulasegaramVariationalFormulationBased2004} has its governing equations
derived from a variational formulation with wall renormalization terms
incorporated. Therefore, additional correction factors appear in their
equations. These correction factors were computed using a polynomial
approximation. Later, \citet{feldmanDynamicRefinementBoundary2007} showed that
the correction factors can be computed exactly.
\citet{kulasegaramVariationalFormulationBased2004}'s approximation was
found to be in good agreement with the exact values computed by
\citet{feldmanDynamicRefinementBoundary2007}. Subsequently,
\citet{ferrandUnifiedSemianalyticalWall2013} laid out a better way to compute
and evolve these correction factors in-simulation. Semi-analytical boundary treatment methods
have been under development ever since~\citep[see for
e.g.,][]{mayrhoferInvestigationWallBounded2013,
leroyUnifiedSemianalyticalWall2014,mayrhoferUnifiedSemianalyticalWall2015,chironFastAccurateSPH2019,boregowdaInsightsUsingBoundary2023a}.
They have also been extended to open boundaries\citep{leroyNewOpenBoundary2016, ferrandUnsteadyOpenBoundaries2017a}.
Nonetheless, they have not yet garnered the widespread adoption that ghost
particle based methods enjoy.

The work of \citet{takedaNumericalSimulationViscous1994} is one of the early
endeavors to impose a no-slip condition that explored the use of mirrored
ghost particles, i.e., ghost particles generated by mirroring fluid particles
about the interface as depicted in \cref{fig:boundary mirrored ghost}.
However, their approach is challenging to implement for complex non-planar
interfaces. Later, \citet{morrisModelingLowReynolds1997} introduced a method
that places all the particles on a regular lattice throughout the
computational domain and designates the particles that fall within a solid
object as ghost particles. \citet{morrisModelingLowReynolds1997}'s method
borrowed \citet{takedaNumericalSimulationViscous1994}'s approach for
estimating ghost particles' velocity for no-slip. This method makes
representing interfaces less complicated at the cost of imperfect
representation of curved boundaries. Since the ghost particles of
\citet{takedaNumericalSimulationViscous1994} and
\citet{morrisModelingLowReynolds1997} do not inherit the pressure and
density of their fluid counterparts, the accuracy of the pressure gradient
near the boundary is expected to be inaccurate. Unlike them,
\citet{colagrossiNumericalSimulationInterfacial2003} made use of mirrored
ghost particles that inherit the pressure and density of their fluid
counterparts to impose the free-slip condition. Subsequently,
\citet{yildizSPHMultipleBoundary2009} employed mirrored ghost particles to
impose a no-slip condition, also highlighting some limitations of
\citet{morrisModelingLowReynolds1997}'s method and attempting to improve
those. Nevertheless, their technique also sticks out as rather elaborate and
onerous to implement. Later, \citet{maciaTheoreticalAnalysisNoSlip2011} came
up with a consistent formulation for the Laplacian operator as an extension to
\citet{takedaNumericalSimulationViscous1994}'s work. This corrected
formulation was seminal for the implementation of the no-slip boundary
condition.

\citet{ferrariNew3DParallel2009} introduced a new boundary treatment method
that makes use of virtual particles generated by locally mirroring fluid
particles about points located on the interface, as shown in
\cref{fig:boundary ficticious}. These virtual or fictitious particles are
generated for each particle near the boundary and are not shared, i.e. one
fluid particle cannot access another fluid particle's virtual particles. Since
the points or particles situated on the interface are present only to act as
local points of symmetry for the generation of fictitious particles and do not
interact with the fluid particles, they can be termed inert.
\citet{vacondioSmoothedParticleHydrodynamics2012} improved
\citet{ferrariNew3DParallel2009}'s method by enabling an additional layer
of virtual particles and introducing better treatment of corners.
\citet{fourtakasApproximateZerothFirstorder2015} introduced further
enhancements to the fictitious particle generation algorithm, mainly focussing
on ensuring better support for the fluid particles. Recently,
\citet{fourtakasLocalUniformStencil2019} introduced a new method in which they
discarded the use of inert particles altogether in favor of using triangles to
discretise the boundary interfaces. They also replaced locally mirrored
particles with a \gls{lust} of fictitious particles that surround every
particle. The particles in \gls{lust} that are located within the fluid domain
are turned off while the contributions from the rest are used.

\gls{dbc} consider multiple layers of ghost particles to model the solid
boundaries, as shown in \cref{fig:boundary stationary ghost}. The density at
the ghost particle is updated using summation density. With this density,
\citet{crespoBoundaryConditionsGenerated2007b} shows that the pressure at the
ghost particle can be evaluated using the first term from the Taylor expansion
of the equation of state. This pressure acts naturally through the pressure
gradient term in the momentum equation to influence the acceleration of the
interacting fluid particle. As we understand, computing pressure this way,
compared to using the actual equation lends a marginal reduction of
computational effort.

The ghost particles that are used in this case do not mirror a fluid particle
about the interface. They remain stationary unless they represent a moving
boundary. Utilizing this property,
\citet{renNonlinearSimulationsWaveinduced2015} show that dynamic boundary
conditions can be used for fluid-rigid body coupling problems just like
\citet{akinciVersatileRigidfluidCoupling2012, liuISPHSimulationCoupled2014}.
Later, \citet{liImprovedDynamicBoundary2021} contended that the particle
penetration problem with \citet{crespoBoundaryConditionsGenerated2007b}'s
approach can be addressed by enhancing the forces between fluid particles and
boundary particles. They proposed a procedure to improve the uniformity of the
repulsive forces and recommended employing a higher-order expansion of the
equation of state for this. Recently,
\citet{englishModifiedDynamicBoundary2022} presented a method which they named
as \gls{mdbc}. Their setup is akin to that used by
\citet{marroneDSPHModelSimulating2011,marroneAccurateSPHModeling2013}, as
depicted in \cref{fig:boundary ghost mirror} and explained later in
\cref{sec:bc-strategy}, with the difference that they extrapolate only the
density from ghost-mirror to ghost. The actual equation of state is used to
compute the pressure. They demonstrate that their method results in a
hydrostatic pressure that is less noisy compared to the former \gls{dbc}.
However, we note that their method targets the Neumann boundary condition for
density about the interface instead of the Neumann boundary condition for
pressure.

Non-homogenous Neumann boundary conditions are being actively explored, especially for heat transfer simulations.
For example, \citet{sikarudiNeumannRobinBoundary2016} explored two methods of implementing non-homogenous Neumann boundary conditions without ghost particles.
The work of \citet{wangModelingHeatTransfer2019} is another example where they demonstrated three ways to treat non-homogenous Neumann boundary conditions making use of ghost particles.

It is clear that most methods make use of ghost particles in some form or
other. Of these, the technique of \citet{adamiGeneralizedWallBoundary2012} has
garnered wide adoption. 
\citet{valizadehStudySolidWall2015} studied and compared variations of
\citet{monaghanSPHParticleBoundary2009}'s and
\citet{adamiGeneralizedWallBoundary2012}'s  methods with a host of test
problems. Their results declare
\citet{adamiGeneralizedWallBoundary2012}'s method as the better one. On
the other hand, the study by \citet{negiHowTrainYour2022a} reveals that the
solid boundary treatment method of \citet{marroneDSPHModelSimulating2011}
yields better convergence. As noted earlier, most of these boundary treatment
methods have been employed in the context of incompressible or weakly
compressible SPH. The present work seeks to identify suitable methods for
compressible fluid flow problems.  The approach employed in this paper is
based on these two methods. We present the details in \cref{sec:bc-strategy},
but before that we need to introduce the basic discretization that is used in
this study.

\section{Formulation} \label{sec:sph}
\subsection{Governing equations}
Inviscid compressible flow is governed by the Euler equations, which are given as,
\begin{equation}
	\label{eq:continuity} \odv{\rho}{t} =-\rho \ \nabla \cdot \boldsymbol{u},
\end{equation}
\begin{equation}
	\label{eq:momentum} \odv{\boldsymbol{u}}{t} =-\frac{1}{\rho}\ \nabla p,
\end{equation}
\begin{equation}
	\label{eq:energy} \odv{e}{t} =-\frac{p}{\rho}\ \nabla \cdot \boldsymbol{u}.
\end{equation}
Here, \(\mathrm{d} / \mathrm{d} t\) represents the material derivative, \(\rho\) is the density, \(p\) is the pressure, \(\boldsymbol{u}\) is the velocity, and \(e\) is the thermal energy per unit mass.

With the ideal gas assumption, this system is closed with an equation of state,
\begin{equation}
	\label{eq:igeos} p=(\gamma-1) \rho e ,
\end{equation}
where \(\gamma\) is the ratio of the specific heat of the gas at constant pressure to its specific heat at constant volume.
\(\gamma\) is constant for a calorifically perfect gas.

\subsection{Semi-discretised governing equations}

We assume readers' familiarity with the basics of \gls{sph} discretization and
proceed to list out the discretized form of the above equations as per the
compressible \(\delta\)-\gls{sph} scheme of \citet{sunAccurateSPHVolume2021}. These discretized equations also contain additional stabilizing terms.
It may be noted that there was no particular motivation behind the selection
of the compressible \(\delta\)-\gls{sph} scheme as the base for this study. We
also see no reason for any other compressible scheme to not work with the
proposed boundary treatment explained in further sections.

The \(\delta\)-\gls{sph} makes use of renormalized kernel
gradients, introduced by \citet{randlesSmoothedParticleHydrodynamics1996}, and
employs an anti-symmetric, conservative discretization for the gradient of
pressure and a symmetric, non-conservative discretization for the divergence
of velocity. If \(\mathcal{N}_i\) be the set of particles in the neighborhood of
a particle, indexed \(i\), the divergence velocity, \(\nabla \cdot
\boldsymbol{u}\), at \(i\) is expressed as a summation over its neighbor
particles, \(\{j: j \in \mathcal{N}_i\}\), as
\begin{equation}
	\langle\nabla \cdot \boldsymbol{u}\rangle_i^L=\sum_j\left(\boldsymbol{u}_j-\boldsymbol{u}_i\right) \cdot \nabla_i W_{i j}^C \frac{m_j}{\rho_j} ,
\end{equation}
where
\begin{equation}
	\nabla_i W_{i j}^C=\mathbb{L}_i \nabla_i W_{i j} ,
\end{equation}
\begin{equation}
	\mathbb{L}_i=\left[\sum_k\left(\boldsymbol{r}_j-\boldsymbol{r}_k\right) \otimes \nabla_i W_{i k} \frac{m_k}{\rho_k}\right]^{-1}.
\end{equation}
Here, \(W_{i j}\) is a shorthand for the \gls{sph} kernel, \(W(\left|\boldsymbol{r}_{i}-\boldsymbol{r}_{j}\right|, h_{ij})\); \(\boldsymbol{r}\) is used to represent position vectors; \(\otimes\) represents the outer product; \(h\) is the smoothing length; and \(h_{ij} = (h_i + h_j)/2\).

Similarly, the gradient of a general scalar variable \(f\) may be expressed as
\begin{equation}
	\label{eq:grad-scalar-or} \langle\nabla f\rangle_i^{L}=\sum_j\left(f_j - f_i\right)\nabla_i W_{i j}^C \frac{m_j}{\rho_j} .
\end{equation}
However, the following anti-symmetric approach is used for the gradient of pressure,
\begin{equation} 
	\label{eq:grad-scalar} \langle\nabla f\rangle_i^{L 2}=\sum_j\left(f_i \nabla_i W_{i j}^C-f_j \nabla_j W_{i j}^C\right) \frac{m_j}{\rho_j} ,
\end{equation}
\Cref{eq:grad-scalar} is used to compute the gradient of pressure. The superscripts \(L\) and \(L 2\) are merely labels to distinguish between the two gradient operators.

Now, the discretized counterpart of continuity equation~\labelcref{eq:continuity}, along with an additional diffusion term reads
\begin{equation}
	\label{eq:continuity-delta} \odv{\rho_i}{t} =-\rho_{i}\langle\nabla\cdot \boldsymbol{u}\rangle_{i}^{L}+\delta\sum_{j}\phi_{i j}c_{i j}h_{i j}\mathcal{D}_{i j}\cdot\nabla_{i}W_{i j}\frac{m_j}{\rho_j} .
\end{equation}
Here, the second term on the RHS imparts diffusion for density.
\(\delta\) is set as 0.1, as proposed by \citet{antuonoFreesurfaceFlowsSolved2010}.
The parameter \(\phi_{i j}\) is set as 1 if the interacting phases are the same, else 0.
$c$ is the speed of sound, computed as \(c_i=\sqrt{\gamma p_i / \rho_i}\) and symmetrised as \(c_{ij} = (c_i + c_j)/2\). \(\mathcal{D}_{i j}\) is given as
\begin{equation}
	\label{eq:dcontinuity}
	\mathcal{D}_{i j}=\frac{2 \boldsymbol{r}_{j i}}{\left\|\boldsymbol{r}_{j i}\right\|^2}\left[\left(\rho_j-\rho_i\right)-\frac{1}{2}\left(\langle\nabla \rho\rangle_i^L+\langle\nabla \rho\rangle_j^L\right) \cdot \boldsymbol{r}_{j i}\right],
\end{equation}
where \(\boldsymbol{r}_{ij} = \boldsymbol{r}_i - \boldsymbol{r}_j=-\boldsymbol{r}_{ji}\).
Instead of \cref{eq:continuity-delta}, one may also use summation density with an iterative solution for smoothing lengths just like \citet{priceSmoothedParticleHydrodynamics2012, puriComparisonSPHSchemes2014}.

Similar to the continuity equation, the discretized counterpart of momentum equation~\labelcref{eq:momentum}, with an additional artificial viscosity term reads
\begin{equation}
	\label{eq:dmomentum} \odv{\boldsymbol{u}_i}{t}=-\frac{1}{\rho_i}\langle\nabla p\rangle_i^{L 2}+\sum_j \frac{\rho_j}{\rho_{i j}} \Pi_{i j} \nabla_i W_{i j} \frac{m_j}{\rho_j},
\end{equation}
where
\begin{equation}
	\Pi_{i j}=\alpha c_{i j} \frac{h_{i j} \boldsymbol{u}_{i j} \cdot \boldsymbol{r}_{i j}}{\left\|\boldsymbol{r}_{i j}\right\|^2}-\beta\left(\frac{h_{i j} \boldsymbol{u}_{i j} \cdot \boldsymbol{r}_{i j}}{\left\|\boldsymbol{r}_{i j}\right\|^2}\right)^2 \text{ if } \boldsymbol{u}_{i j} \cdot \boldsymbol{r}_{i j} < 0 \text{ else } 0.
\end{equation}
Here, \(\rho_{ij} = (\rho_i + \rho_j)/2\) and \(\boldsymbol{u}_{ij} = \boldsymbol{u}_i - \boldsymbol{u}_j\).
The parameters $\alpha$ and $\beta$ are set as 1 and 2 respectively.

Finally, the discretized counterpart of energy equation~\labelcref{eq:energy}, with the additional artificial viscosity term and an additional artificial conduction term may be expressed as
\begin{equation}
	\begin{split}
		\label{eq:denergy} \odv{e_i}{t}=&-\frac{p_i}{\rho_i}\langle\nabla \cdot \boldsymbol{u}\rangle_i^L-\frac{1}{2} \sum_j \frac{\rho_j}{\rho_{i j}} \Pi_{i j} \boldsymbol{u}_{i j} \cdot \nabla_i W_{i j} \frac{m_j}{\rho_j}\\ &+\kappa \sum_j \phi_{i j} c_{i j} h_{i j} \mathcal{E}_{i j} \cdot \nabla_i W_{i j} \frac{m_j}{\rho_j}.
	\end{split}
\end{equation}
Here, the second term on the RHS encapsulates the contribution of artificial viscosity.
The third term is the artificial conduction term.
\(\kappa\) is a constant set as 0.1. \(\mathcal{E}_{i j}\) is given as
\begin{equation}
	\mathcal{E}_{i j}=2\left(e_j-e_i\right) \boldsymbol{r}_{ji} /\left\|\boldsymbol{r}_{ji}\right\|^2.
\end{equation}

The equation of state is straightforwardly discretized as
\begin{equation}
	p_i=(\gamma-1) \rho_i e_i.
\end{equation}

When the continuity equation\labelcref{eq:continuity-delta} is used to update density, smoothing length is updated using,
\begin{equation}
	\label{eq:h-updt} \odv{h}{t} = - \frac{h}{d \cdot \rho} \odv{\rho}{t},
\end{equation}
Here, \(d\) can be 1, 2, or 3 depending upon whether the problem is one, two or three-dimensional, respectively.

\subsection{\gls{tvf}} \label{sec:tvf}

The particles can be moved with a transport velocity,
\(\tilde{\boldsymbol{u}}\) which is different from the Lagrangian velocity,
\(\boldsymbol{u}\) by making use of the \gls{tvf} formulation. We refer the
readers to the work of \citet{sunConsistentApproachParticle2019} for more
details and to the work of
\citet{adepuCorrectedTransportvelocityFormulation2023} for a detailed
derivation. The gist is that if we define
\begin{equation}
	\delta \boldsymbol{u} = \boldsymbol{u} - \tilde{\boldsymbol{u}},
\end{equation}
then the accelerations in \cref{eq:continuity-delta,eq:dmomentum,eq:denergy} can practically be expressed incorporating transport velocity as
\begin{equation}
	\begin{split}
		\label{eq:continuity-delta-tvf} \odv[style-inf-num=\mathrm{\tilde{d}}]{\rho_i}{t} =&-\rho_{i}\langle\nabla\cdot \boldsymbol{u}\rangle_{i}^{L}-\rho_{i}\langle\nabla\cdot \delta\boldsymbol{u}\rangle_{i} +\langle\nabla\cdot \left(\rho \delta\boldsymbol{u}\right)\rangle_{i}\\ &+\delta\sum_{j}\phi_{i j}c_{i j}h_{i j}\mathcal{D}_{i j}\cdot\nabla_{i}W_{i j}\frac{m_j}{\rho_j},
	\end{split}
\end{equation}
\begin{equation}
	\begin{aligned}
		\label{eq:dmomentum-tvf} \odv[style-inf-num=\mathrm{\tilde{d}}]{\boldsymbol{u}_i}{t}= & -\frac{1}{\rho_i}\langle\nabla p\rangle_i^{L 2}+\rho_i\langle \nabla \cdot \left(\boldsymbol{u}\otimes \delta \boldsymbol{u} \right) \rangle_{i} - \rho_i \boldsymbol{u}_i \langle \nabla \cdot \delta \boldsymbol{u} \rangle_i \\  & +\sum_j \frac{\rho_j}{\rho_{i j}} \Pi_{i j} \nabla_i W_{i j} \frac{m_j}{\rho_j},
	\end{aligned}
\end{equation}
and
\begin{equation}
	\begin{split}
		\label{eq:denergy-tvf} \odv[style-inf-num=\mathrm{\tilde{d}}]{e_i}{t}=&-\frac{p_i}{\rho_i}\langle\nabla \cdot \boldsymbol{u}\rangle_i^L - e_i \langle\nabla\cdot \delta\boldsymbol{u}\rangle_{i} +\langle\nabla\cdot \left(e \delta\boldsymbol{u}\right)\rangle_{i} \\ &-\frac{1}{2} \sum_j \frac{\rho_j}{\rho_{i j}} \Pi_{i j} \boldsymbol{u}_{i j} \cdot \nabla_i W_{i j} \frac{m_j}{\rho_j}\\ &+\kappa \sum_j \phi_{i j} c_{i j} h_{i j} \mathcal{E}_{i j} \cdot \nabla_i W_{i j} \frac{m_j}{\rho_j},
	\end{split}
\end{equation}
where
\begin{equation}
	\langle\nabla \cdot \left(f\delta\boldsymbol{u}\right)\rangle_{i}=\sum_j\left(f_j\delta\boldsymbol{u}_j+f_i\delta\boldsymbol{u}_i\right) \cdot \nabla_i W_{i j} \frac{m_j}{\rho_j} ,
\end{equation}
and
\begin{equation}
	\langle\nabla \cdot \left(\boldsymbol{u}\otimes\delta\boldsymbol{u}\right)\rangle_{i}=\sum_j\left(\boldsymbol{u}_j\otimes\delta\boldsymbol{u}_j+\boldsymbol{u}_i\otimes\delta\boldsymbol{u}_i\right) \cdot \nabla_i W_{i j} \frac{m_j}{\rho_j} .
\end{equation}

Here, \(\mathrm{\tilde{d}}/\mathrm{d} t\) represents the material derivative with respect to the transport velocity. 
It can be easily noted that when \(\delta \boldsymbol{u} = 0\), then \cref{eq:continuity-delta-tvf,eq:dmomentum-tvf,eq:denergy-tvf} reduce to \cref{eq:continuity-delta,eq:dmomentum,eq:denergy}, respectively.

We make use of \gls{tvf} in the penetration shield that is proposed ahead. The
\gls{tvf} also finds use in particle shifting. Essentially, the shifting
velocity is embodied as a transport velocity. While shifting techniques
regularise particle distributions, it is also found that shifting does not play well with the shocks. One could use a shock detector
\citep[like,][]{morrisSwitchReduceSPH1997,cullenInviscidSmoothedParticle2010,
readSPHSSmoothedParticle2012,rosswogSimpleEntropybasedDissipation2020} and
avoid shifting near the shocks or try more sophisticated shifting
algorithms~\citep[like,][]{khayyerProjectionbasedParticleMethod2019,rastelliImplicitIterativeParticle2022},
however, we mark this as a subject for future work and stick with boundary treatment methods for now.

\subsection{Boundary Treatment Strategy}\label{sec:bc-strategy}

In the case of inviscid flows, we
are aiming for a free-slip and no-penetration boundary condition. This entails
that the fluid particles in the immediate vicinity of the interface should
have a velocity that is tangential to the interface and the pressure Neumann
condition is to be satisfied at the interface. These can be achieved by
setting the velocities and pressures of the ghost particles such that they
mirror the component of velocity normal to the interface and the pressure of
the fluid particles, about the interface. This is done in two steps: 

\subsubsection{Extrapolation}\label{sec:extrapolation} 

The properties \(\boldsymbol{u}, e, p, \text{and } h\) are
extrapolated from fluid to ghost. If \(\alpha\) is variable, that is also extrapolated. We consider two approaches for this:
\begin{enumerate}
	\item \textit{Without ghost-mirror particles}: This is based on method of
	      \citet{adamiGeneralizedWallBoundary2012}. This approach relies on
	      interpolation to extrapolate properties from fluid to ghost
	      particles. We will let \cref{eq:adami-shepard} clarify this
	      seemingly paradoxical statement. The ghost particles are placed
	      across the prescribed boundary as shown in \cref{fig:boundary
	      stationary ghost}. To extrapolate a property \(f\) from fluid to
	      ghost particles, the following expression is evaluated
	      \begin{equation}
		      f_i=\frac{\sum_j f_j W_{i j}}{\sum_j W_{i j}}.
		      \label{eq:adami-shepard}
	      \end{equation}
	      In this expression, \(i\) represents ghost particles and \(j\)
	      represents fluid particles in the neighborhood of the corresponding
	      ghost particle. So, the summation is over the neighboring fluid
	      particles instead of all the neighboring particles.
	      \citet{adamiGeneralizedWallBoundary2012} recommends using the
	      equation of state to obtain density with the extrapolated
	      properties. Others are able to get good results by obtaining density
	      using summation
	      density\citep[e.g.][]{ramachandranEntropicallyDampedArtificial2019,
	      mutaEfficientAccurateAdaptive2022,
	      haftuParallelAdaptiveWeaklycompressible2022} for the solid
	      particles. In the present study, we stick with the former approach.
	      In plots hereafter, this variant is abbreviated as WOM.
	\item \textit{With ghost-mirror particles}: This is based on the method of
	      \citet{marroneDSPHModelSimulating2011}. In this approach, the ghost
	      particles are placed across the prescribed boundary. Another set of
	      particles is placed in the fluid region mirroring the location of
	      the ghost particles about the boundary, as shown in
	      \cref{fig:boundary ghost mirror}. These are the ghost-mirror
	      particles. Fluid particles interact with ghost particles but not
	      with ghost-mirror particles. The ghost-mirror particles exist for
	      the sole purpose of interpolation of properties. The properties are
	      interpolated using a \gls{mls} interpolator. Practically, this boils
	      down to SPH interpolation by using kernel correction of
	      \citet{liuRestoringParticleConsistency2006}. Thus, to extrapolate a
	      scalar property $f$ from fluid to ghost particles, the following
	      expression is evaluated
	      \begin{equation}
		      \label{eq:marrone} f_i=\sum_j f_j W_{i j}^{LC} V_j.
	      \end{equation}
	      where $ W_{i j}^{LC}$ is the kernel with the correction of
	      \citet{liuRestoringParticleConsistency2006} applied. Then, these
	      properties are copied over from ghost-mirror to the corresponding
	      ghost particles. This variant is abbreviated as WM in plots
	      hereafter.
\end{enumerate}

\subsubsection{Post extrapolation}\label{sec:post-extrapolation}

The extrapolated pressure and velocity need to be modified further to ensure
that the boundary conditions are enforced. Let us denote the velocity obtained
by extrapolation as \(\boldsymbol{{u}_{\text{extrapolated}}}\). Then, the
velocity of the ghost particles is set as
\begin{equation}
	\boldsymbol{u}_i = 2 \boldsymbol{u}_{i,\text{prescribed}} - \boldsymbol{{u}}_{i,\text{extrapolated}},
\end{equation}
where $\boldsymbol{u}_{\text{prescribed}}$ is the prescribed velocity of the boundary.
For free-slip, this prescribed velocity equals the velocity of the fluid tangential to the boundary.
So, given the normal to the boundary $\hat{\boldsymbol{n}}$, the prescribed velocity is
\begin{equation}
	\label{eq:flip velocity normal} \boldsymbol{u}_{i,\text{prescribed}} = \boldsymbol{u}_{i,\text{extrapolated}} - \boldsymbol{u}_{i,\text{extrapolated}} \cdot \hat{\boldsymbol{n}}_i,
\end{equation}
In case the interface is moving with a velocity, \(\boldsymbol{{u}_{\text{interface}}}\), the prescribed velocity would be
\begin{equation}
	\label{eq:flip velocity normal interface} \boldsymbol{u}_{i,\text{prescribed}} = \boldsymbol{u}_{i,\text{interface}} + \boldsymbol{u}_{i,\text{extrapolated}} - \boldsymbol{u}_{i,\text{extrapolated}} \cdot \hat{\boldsymbol{n}}_i,
\end{equation}

The extrapolation of pressure onto ghosts aims at ensuring that the pressure gradient normal to the interface is zero, i.e., \(\pdv{p}/{n}=0\).
However, this condition is valid only when there are no body forces and the interface is not accelerating.
We consider the body forces to be zero.
For accelerating interfaces, this needs to be adjusted to ensure that the pressure conferred by the ghost particles on fluid particles is consistent with the acceleration of the interface.
This is done by setting the pressure of the ghost particles as
\begin{equation}
	p_i = p_{i,\text{extrapolated}} + 2 \Delta s_{i,\text{g2i}} \left. \pdv{p}{n}\right|_i,
\end{equation}
where \(\Delta s_\text{g2i}\) is the distance to the interface.
Exploiting \cref{eq:momentum}, \(\pdv{p}/{n}\) may be estimated as,
\begin{equation}
	\left.\pdv{p}{n}\right|_i = -\rho_i \boldsymbol{a}_i \cdot \hat{\boldsymbol{n}}_i.
\end{equation}
where \(\boldsymbol{a} = \odv{\boldsymbol{u}}/{t}\) is the acceleration of the ghost particle representing the accelerating interface.
While \citet{marroneDSPHModelSimulating2011} does not consider accelerating interfaces, \citet{adamiGeneralizedWallBoundary2012, antuonoCloneParticlesSimplified2023} do include similar terms to account for the acceleration of the interface.
This concludes our description of the general strategy for the treatment of solid boundaries.

\subsection{The challenges and remedies}\label{sec:challenges}

The fluid and ghost particles are initialized with some spacing following the
prescribed mass and density. The particle density and the fluid density are
related. In fact, if a constant mass discretization is assumed, the particle
density is solely responsible for representing fluid density. Given that the
density of a fluid parcel is expected to not change significantly as it moves
with the flow velocity in incompressible flows, the average spacing between
the fluid particles is expected to be fairly constant. Therefore, in
incompressible flows, if the ghost particles and fluid particles that interact
are set up well, they do not end up in a situation where the fluid particles
are spaced much closer than the ghost particles or vice-versa. Even if
adaptive resolution is employed, the particle spacing is varied smoothly
\citep[see for
e.g.,][]{vacondioVariableResolutionSPH2013,
yangAdaptiveResolutionMultiphase2019,
mutaEfficientAccurateAdaptive2022,haftuParallelAdaptiveWeaklycompressible2022}.
However, the density can vary significantly in compressible flows and so can
the spacing. As a result, scenarios that may result in a substantial
difference in spacing at the interface may evolve. This promotes interpolation
errors.

Furthermore, the local average particle spacing, \(\Delta s_{\text{avg}}\), is
expected to be consistent with the effective volume associated with the
particle,
\begin{equation}
	m / \rho \propto (\Delta s_\text{avg})^d.
\end{equation}
The density of the ghost particles may be computed using summation density, 
or using the equation of state after its pressure and energy are set.
Either way, a change in density invariably changes the effective volume 
associated with the ghost particle. This volume may turn out to be inconsistent 
with the actual particle spacing. In other words, the ghost particles end up 
being an ineffective partition of space altogether.

The equation of state plays an interesting role in the repulsion mechanism 
that is described by many authors 
\citep[for e.g.,][]{crespoBoundaryConditionsGenerated2007b, 
englishModifiedDynamicBoundary2022}.
In \gls{wcsph}, when particles come closer and density increases, the pressure 
increase is manyfold.
This transpires by virtue of the stiff equation of the state that is used in 
\gls{wcsph}.
Consequently, a repulsion is generated via the pressure term in the momentum 
equation.
This repulsion curtails the tendency of fluid particles to leak through the 
ghost particles.
For instance, leakage can be observed in the simulation of 3D dam break using 
\gls{sisph} \citep{mutaEfficientOpenSource2020}, a derivative of \gls{isph} 
but not with \gls{edac}-\gls{sph} or \gls{dtsph} 
\citep{ramachandranDualtimeSmoothedParticle2021}, both of which are 
\gls{wcsph} derivatives.
The pressure and density are linearly related in ideal gas equation of state 
that is commonly used with compressible \gls{sph}.
Therefore, this repulsion is rather inconsequential.
This repulsion deficiency may be compensated by increasing artificial 
viscosity and using particle shifting techniques. We also need to be mindful of 
the fact that neither of these is free of consequences.

Careful consideration of the above-described idiosyncrasies regarding
compressible flows with boundaries in \gls{sph} equipped us to direct our
efforts into mitigating these. Based on our investigations,
we propose the following remedies, which are simple yet effective in
addressing the issues described above. 

\subsubsection{Ghost volume constancy}\label{sec:ghost-volume-constancy} 

The effective volume \(V=m/\rho\) of the ghost particles must be
maintained constant. This is in line with the idea that the ghost particles
represent a partition of space and if they do not move, there is no reason for
their volume to change. Upon density change, one may ensure volume constancy
by resetting the mass of the ghost particle accordingly. It may also be noted
from the equations in \cref{sec:sph} that \(m\) and \(\rho\) always occur as
\(m/\rho\) inside the summation, except for the \(\rho\) in artificial
viscosity and density diffusion terms. Therefore, it would suffice if the
ghost particles do not have an explicit mass property \(m\) as long as they
have a constant volume property \(V\) and a density property \(\rho\).

\subsubsection{Penetretion shield}\label{sec:penetration-shield}

While maintaining the effective volume of the ghost particles constant is
enough for most cases, it is not foolproof as there may be cases where one
cannot predict how the flow field will evolve. There is still a risk of
fluid particles ending up interacting with ghost particles with wildly
different associated volumes and leaking through. For such situations, we
propose a penetration shield. In this procedure, a fluid particle that is on a
course to penetrate the interface is steered away using transport velocity as
\begin{equation}
	\label{eq:shield}
	\delta \boldsymbol{u}_i = \delta{u}_i \hat{\boldsymbol{n}}_\text{i,in}
\end{equation}
where
\begin{equation}
	\label{eq:shield-1}
	\delta u_i =
	\begin{cases}
		2.0 \frac{ \Delta s_{i,\text{nom}} - \Delta s_{i,\text{f2g}} }{\Delta s_{i,\text{nom}} } \hat{\boldsymbol{n}}_{i,\text{in}} \cdot \boldsymbol{u}_i & \text{if } \hat{\boldsymbol{n}}_{i,\text{in}} \cdot \boldsymbol{u}_i < 0 \text{ and } \Delta s_{i,\text{f2g}} < \Delta s_{i,\text{nom}} \\ 0 & \text{otherwise}
	\end{cases}
\end{equation}
Here, \(\Delta s_\text{nom}\) is the nominal spacing set as
\((m/\rho)^d\), \(\Delta s_{\text{f2g}}\) is the distance to the nearest
ghost, and \(\hat{\boldsymbol{n}}_\text{in}\) represents a unit vector that is
normal to the interface. \(\hat{\boldsymbol{n}}_\text{in}\) for a fluid
particle is updated by \gls{sph} interpolation of the normals carried by
ghost particles. \(-\hat{\boldsymbol{n}}_\text{in}\) would be pointing
towards interface from the fluid side.

It is worth mentioning that this
penetration shield is an elegant way to address the problem of penetration
without resorting to the usage of short-range repulsive forces near
the interface. It may also be noted the penetration shield is like a fallback
for cases where the volume constancy is not sufficient, like in the case of
extreme compressions or rarefactions. This shield actively prevents penetration 
in the biconvex aerofoil case presented in \cref{sec:supaf}.

\subsection{Time stepping and other parameters}

\Cref{eq:continuity-delta-tvf,eq:dmomentum-tvf,eq:denergy-tvf} are integrated
in time using the \gls{epec} integrator \citep{ramachandranEntropicallyDampedArtificial2019}. The time step is
computed as
\begin{equation}
	\Delta t = C_{\text{CFL}}\min \left( \Delta t_{\text{vel}}, \Delta t_{\text{force}} \right),
\end{equation}
where
\begin{equation}
	\Delta t_{\text{vel}} = \frac{h_{\text{min}}}{\max \left( c \right)},
\end{equation}
\begin{equation}
	\Delta t_{\text{force}} = C_{\text{force}} \sqrt{\frac{h_{\text{min}}}{\max \left( \left| \odv{\boldsymbol{u}}{t} \right| \right)}}.
\end{equation}
$C_{\text{CFL}}$ and $C_{\text{force}}$ are constants, both set as 0.5.

Respecting the findings of \citet{negiHowTrainYour2022a}, the Quintic Spline
kernel is used for all the simulations in this study. The smoothing length is
set as 1.5 times the particle spacing for all problems. We found that reducing
this value to 1.2 makes some of the results noisy. A fluid with \(\gamma =
1.4\) is used for all problems, unless explicitly mentioned otherwise. The SPH
results shown in \cref{sec:results} were simulated using
\texttt{PySPH}~\citep{ramachandranPySPHPythonbasedFramework2021a}. The ghost
particles representing the bodies in the hypersonic
cylinder~(\cref{sec:hcyl}), biconvex aerofoil~(\cref{sec:supaf}), rotating
square projectile~(\cref{sec:rotsq}), and Apollo reentry capsule
3D~(\cref{sec:a3d}) problems were created using the particle packing algorithm
of \citet{negiAlgorithmsUniformParticle2021}. The simulations were
orchestrated using
\texttt{automan}~\citep{ramachandranAutomanPythonBasedAutomation2018}. In the
interest of reproducibility, all the code for the present study is available
at \url{https://gitlab.com/pypr/compressible-sph-bc}.

\section{Results}
\label{sec:results}
\subsection{Compression Corner}\label{sec:ccor}

This verification case involves the computation of the supersonic flow field
past a wedge. \Cref{fig:ccor-rho} depicts the flow-fields over a wedge of
half-angle $20^\circ$ at Mach $M=2.5$ as simulated using different approaches.

In the case of \gls{sph}, the particles at the inlet are spaced \qty{0.0625}{\meter}
apart. The meshes for the \gls{fvm} cases were also created with comparable
cell sizes. The simulation is run until $t=\qty{10}{\second}$. \Cref{fig:ccor oadami} is the
result of using our implementation of the original unmodified method of
\citet{adamiGeneralizedWallBoundary2012}. In this case, it can be seen that
particles leak through the wall. In \cref{fig:ccor roe} the result of a simple
\gls{fvm} implementation using \gls{fou} scheme with approximate Riemann
solver of \citet{roeApproximateRiemannSolvers1981}. It can be seen that the
smearing of the shock wave increases with the distance from the corner, as
expected from a first-order scheme. \Cref{fig:ccor eilmer} is a result of
simulation with Eilmer \citep{gibbonsEilmerOpensourceMultiphysics2023}. The
mesh was created with Eilmer's geometry package. This result appears
comparable to the presented \gls{sph} results. \Cref{fig:ccor su2} is a result
of simulation with SU2~\citep{economonSU2OpenSourceSuite2016}. The mesh used
by SU2 was created using Gmsh~\citep{geuzaineGmsh3DFinite2009}. SU2 is
\gls{fvm} based but it uses a cell-vertex scheme and outputs results as point
data. So, a scatter plot is used for SU2 results just like the \gls{sph}
cases. It can be observed that the shock is less smeared in the case of SU2
but overshoot and dispersion wiggles can also be seen.

\Cref{fig:ccor adami} and \cref{fig:ccor marrone} are the result of
simulations using the methods proposed in this paper. There is no leakage of
particles through the wall in either of these cases. The shock wave is more
smeared than \cref{fig:ccor su2} but the smearing does not visibly increase
with distance from the corner as seen in \cref{fig:ccor roe}. The overshoot is
also more subdued than that of the other cases. The same is better quantified
in \cref{fig:ccor-line}. Note that the cell data is plotted as a piecewise
constant over the cell, in this figure.

\begin{figure}
	\centering
	\pgfplotsset{width=0.5\textwidth, height=0.015\textwidth}

	\begin{tikzpicture}
	\begin{axis}[enlargelimits=false,
			ylabel=$\rho$,
			ylabel style={rotate=-90},
			axis on top,
			xtick pos=bottom,
			every major tick/.append style={major tick length=2pt},
			ymajorticks=false,
			scale only axis
		]
		\addplot graphics[
				xmin=1.4,
				xmax=3.4,
				ymin=0.0,
				ymax=1.0
			]{cbar1};
	\end{axis}
\end{tikzpicture}
	\pgfplotsset{width=\textwidth, height=\textwidth}

	\begin{tabular}{cc}
	\subfloat[]{
		\begin{minipage}[c]{.42\textwidth}
			\input{../../macros}
\begin{tikzpicture}

	\begin{axis}[enlargelimits=false,
			name=main,
			axis on top,
			xlabel=$x$,
			ylabel=$y$,
			axis equal image
		]
		\addplot graphics
			[xmin=-1.0,xmax=3.0,ymin=0.0,ymax=3.0]
			{rho_eosnavm_bare};
	\end{axis}
\end{tikzpicture}
		\label{fig:ccor oadami}
	\end{minipage}
	} &
	\subfloat[]{
		\begin{minipage}[c]{.42\textwidth}
		\input{../../macros}
\begin{tikzpicture}
	\begin{axis}[enlargelimits=false,
			name=main,
			axis on top,
			xlabel=$x$,
			ylabel=$y$,
			axis equal image
		]
		\addplot graphics
			[xmin=-1.0,xmax=3.0,ymin=0.0,ymax=3.0]
			{rho_fvm_roe_wedge_bare};
	\end{axis}
\end{tikzpicture}
		\label{fig:ccor roe}
	\end{minipage}
	} \\

	\subfloat[]{
		\begin{minipage}[c]{.42\textwidth}
		\input{../../macros}
\begin{tikzpicture}
	\begin{axis}[enlargelimits=false,
			name=main,
			axis on top,
			xlabel=$x$,
			ylabel=$y$,
			axis equal image
		]
		\addplot graphics
			[xmin=-1.0,xmax=3.0,ymin=0.0,ymax=3.0]
			{rho_eilmer_wedge_bare};
	\end{axis}
\end{tikzpicture}
		\label{fig:ccor eilmer}
	\end{minipage}
	}&

	\subfloat[]{
		\begin{minipage}[c]{.42\textwidth}
		\input{../../macros}
\begin{tikzpicture}
	\begin{axis}[enlargelimits=false,
			name=main,
			axis on top,
			xlabel=$x$,
			ylabel=$y$,
			axis equal image
		]
		\addplot graphics
			[xmin=-1.0,xmax=3.0,ymin=0.0,ymax=3.0]
			{rho_su2_wedge_bare};
	\end{axis}
\end{tikzpicture}
		\label{fig:ccor su2}
	\end{minipage}
	} \\

	\subfloat[]{
		\begin{minipage}[c]{.42\textwidth}
		\input{../../macros}
\begin{tikzpicture}
	\begin{axis}[enlargelimits=false,
			name=main,
			axis on top,
			xlabel=$x$,
			ylabel=$y$,
			axis equal image
		]
		\addplot graphics
			[xmin=-1.0,xmax=3.0,ymin=0.0,ymax=3.0]
			{rho_eosampv_bare};
	\end{axis}
\end{tikzpicture}
		\label{fig:ccor adami}
	\end{minipage}
	} & 

	\subfloat[]{
		\begin{minipage}[c]{.42\textwidth}
		\input{../../macros}
\begin{tikzpicture}
	\begin{axis}[enlargelimits=false,
			name=main,
			axis on top,
			xlabel=$x$,
			ylabel=$y$,
			axis equal image
		]
		\addplot graphics
			[xmin=-1.0,xmax=3.0,ymin=0.0,ymax=3.0]
			{rho_marrone_bare};
	\end{axis}
\end{tikzpicture}
		\label{fig:ccor marrone}
	\end{minipage}
	}
	\end{tabular}

	\caption{Points/cells colored by density for the compression corner problem simulated using various methods/tools. (a) \gls{sph} with original \citet{adamiGeneralizedWallBoundary2012}'s method. (b) First Order Upwind Roe. (c) Eilmer. (d) SU2. (e) \gls{sph} with present method, without ghost-mirror. (f) \gls{sph} with present method, with ghost-mirror.}
	\label{fig:ccor-rho}
	
\end{figure}

\begin{figure}
	\centering
	\pgfplotsset{width=0.6\textwidth, height=0.4\textwidth}
	\centering
	\begin{tikzpicture}
	\pgfplotstableread[col sep=comma]{auto/Wedge/wallline.csv}\datatable
	\begin{axis}[
			xmin=-0.2, xmax=0.6,
			ymin=1.2, ymax=4.1,
			legend pos=south east,
			xlabel=$x$,
			ylabel=$\rho$,
			every axis plot/.append style={thick},
			cycle list name=exotic,
			grid=major
		]

		\addplot+[
			const plot mark mid,
			no marks,
			densely dotted
		] table [
				x=roex,
				y=roerho
			] from \datatable;
		\addlegendentry{FOU-Roe}

		\addplot+[
			const plot mark mid,
			no marks, densely dashed
		] table [x=eilmerx, y=eilmerrho] from \datatable;
		\addlegendentry{Eilmer}

		\addplot+[
			only marks,
			mark=+
		] table [
				x=su2x,
				y=su2rho
			] from \datatable;
		\addlegendentry{SU2}

		\addplot+[
			only marks
		] table [
				x=eosampvx,
				y=eosampvrho
			] from \datatable;
		\addlegendentry{SPH-WOM}

		\addplot+[
			only marks,
			mark=x
		] table [
				x=marronex,
				y=marronerho
			] from \datatable;
		\addlegendentry{SPH-WM}

	\end{axis}
\end{tikzpicture}
	\caption{Density along the wall about the corner.}
	\label{fig:ccor-line}
\end{figure}

For the purpose of comparison with
\citet{englestadInvestigationsNovelBoundary2020}, the shock angle, \(\beta\)
needs to be extracted. This is carried out by exploiting the variation of
density across the shock. The gradient of density is estimated using
\cref{eq:grad-scalar}. The particles near the shock can be identified if their
density gradient magnitude is close to the maximum density gradient within the
domain. The condition
\begin{equation}
	\left| \langle\nabla \rho\rangle_i \right|>0.6 \max_i\left(\left| \langle\nabla \rho\rangle_i \right|\right)
\end{equation}
is used to mark the particles near the shock. We find that the factor of 0.6
works well.  The wave angle is computed from the slope of the least squares
1\textsuperscript{st} order polynomial fit to these points. Comparison of
error in wave angle for different configurations of compression corner problem
is presented in \cref{tab:ccor-err}. For these results, the particle spacing
is matched with \citet{englestadInvestigationsNovelBoundary2020} to keep the
comparison fair. It can be observed that errors are consistently below 1\% and
considerably lower than obtained reported by
\citet{englestadInvestigationsNovelBoundary2020}.

\begin{table}
	\centering
	\setlength{\tabcolsep}{12pt}
	\begin{tabular}{ccccccc}
		\hline$M$ & $\gamma$ & $\theta$ & $\beta_{\text {th }}$ & $\beta_{\mathrm{EC}}(\%$ error $)$ & $\beta_{\mathrm{WOM}}(\%$ error $)$ & $\beta_{\mathrm{WM}}(\%$ error $)$ \\
		\hline \hline 5 & 1.4 & $10^{\circ}$ & $19.38^{\circ}$ & $18.0^{\circ}(-7.10)$ & $19.20^{\circ}(-0.90)$ & $19.24^{\circ}(-0.69)$ \\
		\hline 5 & 1.4 & $15^{\circ}$ & $24.32^{\circ}$ & $21.8^{\circ}(-10.37)$ & $24.10^{\circ}(-0.91)$ & $24.12^{\circ}(-0.81)$ \\
		\hline 5 & 1.4 & $20^{\circ}$ & $29.80^{\circ}$ & $27.0^{\circ}(-9.40)$ & $29.54^{\circ}(-0.86)$ & $29.55^{\circ}(-0.83)$ \\
		\hline 5 & 1.4 & $25^{\circ}$ & $35.78^{\circ}$ & $32.6^{\circ}(-8.89)$ & $35.48^{\circ}(-0.84)$ & $35.49^{\circ}(-0.82)$ \\
		\hline 2 & 1.4 & $10^{\circ}$ & $39.31^{\circ}$ & $39.0^{\circ}(-0.80)$ & $39.16^{\circ}(-0.38)$ & $39.19^{\circ}(-0.32)$ \\
		\hline 5 & 1.3 & $20^{\circ}$ & $28.76^{\circ}$ & $27.0^{\circ}(-6.11)$ & $28.55^{\circ}(-0.73)$ & $28.54^{\circ}(-0.74)$ \\
		\hline
	\end{tabular}
	\caption{Comparison of error in wave angle $\beta$ for different
	configurations of compression corner problem. $\beta_\text{th}$ is the
	wave angle obtained using oblique shock theory
	\cite[]
	{andersonModernCompressibleFlow2021}. $\beta_\text{EC}$
	represents the values obtained by
	\citet{englestadInvestigationsNovelBoundary2020} with their boundary
	treatment method. $\beta_{\text{WOM}}$ and $\beta_{\text{WM}}$ represent
	the values obtained using the methods from the present study, without
	ghost-mirrors and with ghost-mirrors, respectively.}
	\label{tab:ccor-err}
\end{table}

\subsection{Reflecting Shocktube} \label{sec:rsod}

\begin{figure}
	\centering
	\pgfplotsset{width=0.5\textwidth, height=0.2\textwidth}
	\input{../../macros}
\begin{tikzpicture}
	\tikzset{
		text node/.style={
				font=\footnotesize
			}
	}
	\begin{axis}[enlargelimits=false,
			name=main,
			axis on top,
			xlabel=$x$,
			ylabel=$t$,
			scale only axis,
		]
		\addplot graphics
			[xmin=-0.49875,xmax=0.499366,ymin=0.0,ymax=0.375085]
			{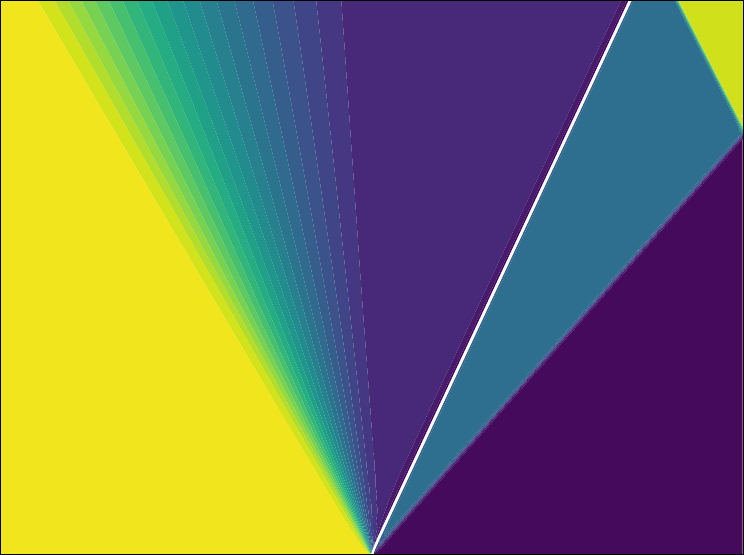};
	\end{axis}
	\path[<->, draw] (rel axis cs: 0.64, 0.41) to[out = 135, in = 0]
	(rel axis cs: 0.75, 1.1) node[text node, left] {Contact Wave};
	\path[<->, draw] (rel axis cs: 0.77, 0.40) to[out = 135, in = 0]
	(rel axis cs: 0.85, 1.2) node[text node, left] {Shockwave};
	\path[<->, draw] (rel axis cs: 0.96, 0.86) to[out = 45, in = 0]
	(rel axis cs: 0.9, 1.3) node[text node, left] {Reflected Shockwave};
	\path[<->, draw] (rel axis cs: 0.50, 0.0) to[out = 90, in = 0]
	(rel axis cs: 0.3, 1.2) node[text node, left] {Initial Discontinuity};
	\path[dashed,<->, draw] (rel axis cs: 0.27, 0.5) to[out = 0, in = 180]
	(rel axis cs: 0.49, 0.5);
	\path[<->, draw] (rel axis cs: 0.35, 0.5) to[out = 90, in = 0]
	(rel axis cs: 0.2, 1.1) node[text node, left] {Expansion Fan};
	\coordinate (A0) at (rel axis cs: 0,0);
	\coordinate (A1) at (rel axis cs: 1,1);
	\path ($(A1)-(A0)$);
	\pgfgetlastxy{\Awidth}{\Aheight}
	\coordinate (A0) at (rel axis cs: 0,0);
	\coordinate (A1) at (rel axis cs: 1,1);
	\path ($(A1)-(A0)$);
	\pgfgetlastxy{\Awidth}{\Aheight}

	\begin{axis}[enlargelimits=false,
			title=$\rho$,
			axis on top,
			at={(main.south east)},
			anchor=south west,
			ytick pos=right,
			yticklabel pos=right,
			xshift=0.05*\Awidth,
			every major tick/.append style={major tick length=2pt},
			xmajorticks=false,
			width=0.05*\Awidth,
			height=\Aheight ,
			scale only axis
		]
		\addplot graphics
			[xmin=0.0,xmax=0.1,ymin=0.12,ymax=0.54]
			{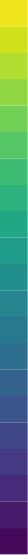};
	\end{axis}
\end{tikzpicture}
	\caption{$xt$-plot for reflecting shocktube.}
	\label{fig:rsod-xt}
\end{figure}

This case involves the reflection of a moving normal shock from the wall. The
considered domain for the shock tube is \qty{1}{\meter} long and \qty{0.02}{\meter} wide. The
initial particle spacing is \qty{0.002}{\meter}. The particles are initialized with the
classic Sod shocktube~\citep{sodSurveySeveralFinite1978} conditions. The
velocity is set as \qty{0}{\meter\per\second}. Pressure and density are initialized as
\begin{equation}
	\begin{bmatrix}
		\rho_L \\ p_L \\
	\end{bmatrix}
	=
	\begin{bmatrix}
		\qty{1}{\kilo\gram\per\cubic\meter} \\ \qty{1}{\kilo\gram\per\cubic\meter} \\
	\end{bmatrix}
	,
	\begin{bmatrix}
		\rho_R \\ p_R \\
	\end{bmatrix}
	=
	\begin{bmatrix}
		\qty{0.125}{\pascal} \\ \qty{0.1}{\pascal} \\
	\end{bmatrix}
	,
\end{equation}
where the subscripts $L$ and $R$ denote the left and right states. To avoid a
spurious pressure blip, the quantities are smoothed about the initial
discontinuity. This smoothed pressure of a particle $p_i$ can be expressed as
\begin{equation}
	p_i = \frac{p_L - p_R} {1 + \exp\left({\frac{2(x_i - x_0)}{3\Delta x}}\right)} + p_R,
\end{equation}
where $x_i$ is the $x$-component of is position vector, $x_0$ is the location
of initial discontinuity and $\Delta x$ is the particle spacing. Density is
also smoothed, likewise. The simulation is run until the shockwave hits the
right wall and reflects as shown in \cref{fig:rsod-xt}. The walls at either
end of the shock tube are dealt with using boundary treatment methods
described in the present paper. The boundaries in the $y$-direction are set to
be periodic. The periodicity is implemented internally in
PySPH~\citep{ramachandranPySPHPythonbasedFramework2021a}. The results in this
section illustrate that the presented boundary treatment method works well
with periodic boundaries.

\begin{figure}
	\centering
	\pgfplotsset{width=0.8\linewidth, height=0.075\textwidth}
	\input{../../macros}
\begin{tikzpicture}
	\begin{axis}[enlargelimits=false,
		name=eosampv,
		axis on top,
		xlabel=$x$,
		ylabel=$y$,
		scale only axis,
		xtick scale label code/.code={$\times10^{#1}$},
		ytick scale label code/.code={$\times10^{#1}$},
		title= Without ghost-mirror
	]
	\addplot graphics[xmin=-0.5,xmax=0.5,ymin=0.0,ymax=0.08]
	{rho_eosampv_bare};
	\coordinate (A1) at (rel axis cs: 1,1);
\end{axis}

\begin{axis}[enlargelimits=false,
	at={(eosampv.south)},
	anchor=north,
	yshift=-5em,
	name=marrone,
	axis on top,
	xlabel=$x$,
	ylabel=$y$,
	scale only axis,
	xtick scale label code/.code={$\times10^{#1}$},
	ytick scale label code/.code={$\times10^{#1}$},
	title= With ghost-mirror
]
		\addplot graphics
			[xmin=-0.5,xmax=0.5,ymin=0.0,ymax=0.08]
			{rho_marrone_bare};
			\coordinate (A0) at (rel axis cs: 0,0);
\end{axis}

	\newdimen\xmax
	\pgfextractx{\xmax}{\pgfpointanchor{A1}{center}}
	\newdimen\ymax
	\pgfextracty{\ymax}{\pgfpointanchor{A1}{center}}
	\newdimen\xmin
	\pgfextractx{\xmin}{\pgfpointanchor{A0}{center}}
	\newdimen\ymin
	\pgfextracty{\ymin}{\pgfpointanchor{A0}{center}}

		\begin{axis}[enlargelimits=false,
			title=$\rho$,
			axis on top,
			at={(marrone.south east)},
			anchor=south west,
			ytick pos=right,
			yticklabel pos=right,
			xshift=2em,
			every major tick/.append style={major tick length=2pt},
			xmajorticks=false,
			width=1em,
			height=\ymax-\ymin,
			scale only axis
		]
		\addplot graphics
			[xmin=0.0,xmax=0.1,ymin=0.26838957772982214,ymax=0.9999992393797743]
			{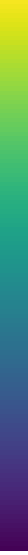};
	\end{axis}

\end{tikzpicture}
	\caption{Reflecting shocktube with particles colored by density.}
	\label{fig:rsod-rho}
\end{figure}

The particle positions at $t=\qty{0.375}{\second}$ colored by density are shown in
\cref{fig:rsod-rho}. The density profile is compared with the exact solution in
\cref{fig:rsod-rho-exact}. The movement of particles results in the formation
of a low particle density region around the contact discontinuity. It may also
be noted that the contact discontinuity in \cref{fig:rsod-rho-exact} appears
smeared because of the smoothed initial condition and reduced particle
density, largely due to the latter.

\IfFileExists{figcache/rsod_rho_line.pdf}{
	\begin{figure}
		\centering
		\includegraphics[width=0.7\textwidth]{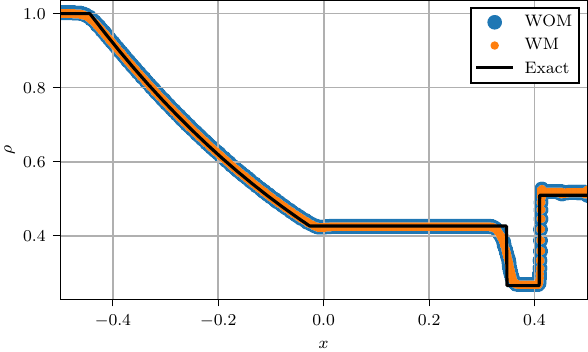}
		\caption{Density variation with respect to exact for reflecting shocktube case}
		\label{fig:rsod-rho-exact}
	\end{figure}
}{Figure not found}

\subsection{Hypersonic Cylinder} \label{sec:hcyl}

The bow shock formed upstream of a blunt cylinder-shaped body is usually
examined as a standard test case for carbuncle instability. Mach 10 flow over
the unit radius cylinder, ahead of a $14.04^\circ$ cylinder cone is considered
as shown in \cref{fig:hcyl-setup}. Particle spacing at the inlet is 
\qty{0.025}{\meter}. The simulation is run till $t = \qty{3.5}{\second}$.

\begin{figure}
	\centering
	\pgfplotsset{width=0.6\textwidth, height=0.6\textwidth}
	\begin{tikzpicture}
	\begin{axis}[
			axis equal,
			thick,
			axis lines=none,
			xmin = -2.75,
			xmax = 1,
			ymin = -3.5,
			ymax = 3.5
		]
		\addplot[
		domain = rad(90+14.04) : rad(270-14.04),
		smooth,
		data cs=polar,
		] (deg(x), 1);
		\addplot [
			domain = -sin(14.04) : 3]
		{1 + 1 - cos(14.04) + 0.25*x}
		coordinate [pos=0.15] (A)
		coordinate [pos=0.65] (B);
		\draw (A) -| (B)
		node [pos=0.25, anchor=north]{$4$}
		node [pos=.75,anchor=west]{$1$};
		\addplot [
			domain = -sin(14.04) : 3]
		{-1 - 1 + cos(14.04) - 0.25*x};
		\draw [->](0.0,0.0) -- (0.5,0.0);
		\node[above] at (0.0,0.5){$y$};
		\draw [->](0.0,0.0) -- (0.0,0.5);
		\node[right] at (0.5,0.0){$x$};
		\draw [fill=blue,draw=none,fill opacity=0.5] (-1.75,-3.5) rectangle (-0.24253562503633297,3.5);
		\draw [->] (-2.3,3.0) -- (-1.75,3.0)node[midway,above]{$M_\infty$};
	\end{axis}
\end{tikzpicture}
	\caption{Setup for hypersonic cylinder case. The flow domain is shaded blue.}
		\label{fig:hcyl-setup}
\end{figure}

The shape of the shock front is modeled as a hyperbola, expressed as
\begin{equation} 
	\label{eq:billing} x=R+\delta-R_{C}\cot^{2}\beta\left[\left(1+\frac{y^{2}\tan^{2}\beta}{R_{c}^{2}}\right)^{1/2}-1\right].
\end{equation}
Here, $\beta$ is the wave angle for a the turn angle $\theta$. The turn angle
is $14.04^\circ$ in this case. $\delta$ is the standoff distance, i.e.\ the
shortest distance from the tip of the nose to the shockfront. $R$ is the
radius of the nose and $R_C$ is the radius of curvature of shockwave at the
vertex of the hyperbola. $\delta$ and $R_C$ are correlated as
\begin{equation}
	\frac{\theta}{R}=0.386\exp\left(\frac{4.67}{M_{\infty}^{2}}\right),
\end{equation}
and
\begin{equation}
	\frac{\theta}{R}=1.386\exp\left[\frac{1.8}{(M_{\infty} - 1)^{0.75}}\right],
\end{equation}
where $M_{\infty}$ is the freestream Mach number. $M_{\infty}$ is 10 in this
case. The readers may also refer to the work of
\citet{billigShockwaveShapesSphericaland1967} or \citet[chap.
5]{andersonHypersonicHighTemperatureGas2019} for a more detailed explanation.

The shock front agrees fairly well with the profile
given by \cref{eq:billing}. Upon closer inspection, it may be observed that 
the curvature of the profile given by \cref{eq:billing} tends to exceed 
the simulation result, towards the outlet. However, plots presented by 
\citet{billigShockwaveShapesSphericaland1967} are indicative of the fact 
that such minor variations can be expected. From \cref{fig:hcyl-rho} it 
is quite clear that there is no evidence of carbuncle instability. 

\begin{figure}
	\centering
	\pgfplotsset{width=0.6\textwidth, height=0.6\textwidth}
	\begin{tikzpicture}
	\edef\ang{5.889550610181072}
	\edef\M{10.0}
	\edef\R{1.0}
	\pgfmathsetmacro{\Rc}{1.386*exp(1.8/(\M-1)^0.75*\R)}
	\pgfmathsetmacro{\dlt}{0.386*exp(4.67/\M^2)*\R}
	\begin{axis}[
		name = eosampv,
		xlabel=$x$,
		ylabel=$y$,
		enlargelimits=false,
		scale only axis,
		axis equal image,
					xmin=-1.75,
			xmax=-0.24253562503633297,
			ymin=-3.0,
			ymax=3.0,
			title={Without ghost-mirror}]
	\addplot graphics[
		xmin=-1.75,
		xmax=-0.24253562503633297,
		ymin=-3.0,
		ymax=3.0] {rho_eosampv_bare};
		\coordinate (A0) at (rel axis cs: 0,0);
		\addplot[
		red,
		line width=1pt,
		smooth,
		] (-(\R + \dlt - \Rc * cot(\ang)^2 * ((1 + (x * tan(\ang) / \Rc) ^ 2) ^ 0.5 - 1)),x);
	\end{axis}

		\begin{axis}[
			at = {(eosampv.south east)},
			anchor = south west,
			xshift = 6em,
			xlabel=$x$,
			ylabel=$y$,
			enlargelimits=false,
			scale only axis,
			axis equal image,
						xmin=-1.75,
				xmax=-0.24253562503633297,
				ymin=-3.0,
				ymax=3.0,
				title=With ghost-mirror]
		\addplot graphics[
			xmin=-1.75,
			xmax=-0.24253562503633297,
			ymin=-3.0,
			ymax=3.0] {rho_marrone_bare};
			\coordinate (A1) at (rel axis cs: 1,1);
				\addplot[
			red,
			line width=1pt,
			smooth,
			] (-(\R + \dlt - \Rc * cot(\ang)^2 * ((1 + (x * tan(\ang) / \Rc) ^ 2) ^ 0.5 - 1)),x);
			
		\end{axis}




	\newdimen\xmax
	\pgfextractx{\xmax}{\pgfpointanchor{A1}{center}}
	\newdimen\ymax
	\pgfextracty{\ymax}{\pgfpointanchor{A1}{center}}
	\newdimen\xmin
	\pgfextractx{\xmin}{\pgfpointanchor{A0}{center}}
	\newdimen\ymin
	\pgfextracty{\ymin}{\pgfpointanchor{A0}{center}}

	\begin{axis}[
			name=cbar,
			enlargelimits=false,
			ylabel=$\rho$,
			ylabel style={rotate=-90},
			axis on top,
			at={(eosampv.north west)},
			anchor=south west,
			ytick pos=left,
			xtick pos=bottom,
			every major tick/.append style={major tick length=2pt},
			ymajorticks=false,
			yshift=5em,
			width=\xmax-\xmin,
			height=1em,
			scale only axis
		]
		\addplot graphics[
				xmin=1.0815842480446578,
				xmax=15.792344034377303,
				ymin=0.0,
				ymax=1.0
			]{cbar1};
	\end{axis}
\end{tikzpicture}
	\caption{Hypersonic cylinder with particles colored by density.
		The red line is the location of the shock front given by \cref{eq:billing}.
	}
	\label{fig:hcyl-rho}
\end{figure}

\subsection{Convergent Divergent Nozzle} \label{sec:condi}

This is a problem in which the transition from subsonic flow to supersonic
flow is demonstrated. The nozzle profile can be obtained using
\begin{equation}
	y =
	\begin{cases}
		y_0 & \text{if } x_0 \leq x \leq x_1 \\ \sqrt{R_{tu}^2 - (x - x_1)^2} & \text{if } x_1 < x \leq x_2 \\ R_{th} + R_{cu} -\sqrt{R_{cu}^2 - x ^2} & \text{if } x_3 < x \leq x_4 \\ y_4 + (x - x_4) \tan(\theta) & \text{if } x_4 < x \leq x_5
	\end{cases}
	.
\end{equation}
The symbols in the above equation are to be read along with the markings in
\cref{fig:nozzle-profile}. This profile is obtained from examples in the
Eilmer repository, which in turn is a simplified adaption of the profile from
the work of \citet{backComparisonMeasuredPredicted1965b}.
We simulate a rectangular, non-axisymmetric nozzle with this profile.
\begin{figure}
	\centering
	\pgfplotsset{
		table/search path={figs/}
	}
	\pgfplotsset{width=0.8\textwidth, height=0.6\textwidth}
	\input{macros}
\begin{tikzpicture}
	\pgfplotstableread[col sep=comma]{auto/ConDiNozzle/profile_mach.csv}\profiletable
	\begin{axis}[name=condi,
			xlabel=$x$,
			ylabel=$y$,
			scale only axis,
			axis equal image,
			ymin = 0.00,
			ymax = 0.08,
			axis lines=none
		]
		\addplot[
			smooth,
			no marks,
			fill=blue,
			opacity=0.5
		] table [
				x=x,
				y=y,
			] from \profiletable |- (-0.0762,0) -- cycle;
		\addplot[
			only marks,
			mark size=1.5,
		] coordinates {
				(-0.0762,0)
				(-0.0538129,0)
				(-0.02651725,0)
				(-0.02651725,0)
				(0,0)
				(0.0101897,0)
				(0.0762,0)
			};
		\path[dashed, draw](-0.0762,0) -- (-0.0762,0.0405257);
		\addmarkandpin{-0.0762,0.0405257}{(x_0,y_0)}{red}{90}
		\path[dashed, draw](-0.0538129,0) -- (-0.0538129,0.0405257);
		\addmarkandpin{-0.0538129,0.0405257}{(x_1,y_1)}{red}{90}
		\path[dashed, draw](-0.02651725,0.0299546) -- (-0.02651725,0.0299546);
		\addmarkandpin{-0.02651725,0.0299546}{(x_2,y_2)}{red}{90}
		\path[dashed, draw](-0.02651725,0) -- (-0.02651725,0.0299546);
		\addmarkandpin{-0.02651725,0.0299546}{(x_2,y_2)}{red}{90}
		\path[dashed, draw](0,0) -- (0,0.019685);
		\addmarkandpin{0,0.019685}{(x_3,y_3)}{red}{90}
		\path[dashed, draw](0.0101897,0.02100265) -- (0.0101897,0);
		\addmarkandpin{0.0101897,0.02100265}{(x_4,y_4)}{red}{75}
		\path[dashed, draw](0.0762,0.0387140) -- (0.0762,0);
		\addmarkandpin{0.0762,0.0387140}{(x_5,y_5)}{red}{90}
		\path [->, draw] (-0.0538129,0) -- node [midway, fill=white, rounded corners, font=\footnotesize, inner sep=0.1em] {$R_{tu}$} (-0.0407939393939394,0.0383775797406432);
		\path [->, draw] (0,0) -- node [midway, fill=white, rounded corners, font=\footnotesize, inner sep=0.1em] {$R_{tr}$} (0,0.019685);
		\path [{Circle}->, draw] (0,0.059055) -- node [midway, fill=white, rounded corners, font=\footnotesize, inner sep=0.1em] {$R_{cu}$} (-0.0207818181818182, 0.0256168042792699);
		\draw (0.0346363636363636,	0.0275769624423694) arc (15:0:0.0346363636363636-0.0101897);
		\draw (0.0101897,0.02100265) -- node [font=\footnotesize, right, anchor=south, xshift=2.7em] {$\theta$} (0.040,	0.02100265);
	\end{axis}

\end{tikzpicture}
	\caption{Profile of convergent-divergent nozzle.
		Here, $R_{tu}=1.5955$, $R_{tr}=0.755$, $R_{cu}=1.55$, $x_0=-3$, $x_0=0.0$, $x_5=3$.
		These dimensions are in inches.
		$\theta=15^\circ$.
	}
	\label{fig:nozzle-profile}
\end{figure}

The particle spacing at the inlet is $R_{tu}/30$ \si{inch}.
The simulation is run for \qty{0.004}{\second}.
Air at temperature \qty{300}{\kelvin} enters the inlet with a mass flux of \qty{275.16}{\kilo\gram\per\meter}.
For an isentropic quasi 1D convergent-divergent nozzle, the area ratio and Mach number are related as
\begin{equation}
	\frac{A}{A_{tr}}=\left({\frac{\gamma+1}{2}}\right)^{-{\frac{\gamma+1}{2(\gamma-1)}}}\;{\frac{(1+{\frac{\gamma-1}{2}}\;M^2)^{\frac{\gamma+1}{2(\gamma-1)}}}{M}}
\end{equation}
where \(A\) denotes area and \(A_{tr}\) is the area at the throat.
The Mach number at the inlet can be obtained using the area ratio of the inlet to the throat.
The velocity, density and pressure at the inlet are determined using this information.

\begin{figure}
	\centering
	\pgfplotsset{
		table/search path={figs/auto/ConDiNozzle/}
	}
	\pgfplotsset{width=0.6\textwidth, height=0.4\textwidth}
	\input{../../macros}
\begin{tikzpicture}
	\pgfplotstableread[col sep=comma]{profile_mach.csv}\profiletable
	\begin{axis}[name=condi,
			xlabel=$x$,
			ylabel=$y$,
			xtick scale label code/.code={$\times10^{#1}$},
			ytick scale label code/.code={$\times10^{#1}$},
			enlargelimits=false,
			scale only axis,
			axis equal image]
		\addplot graphics
			[xmin=-0.07619999999999999,xmax=0.07619999999999999,ymin=0,ymax=0.0405257]
			{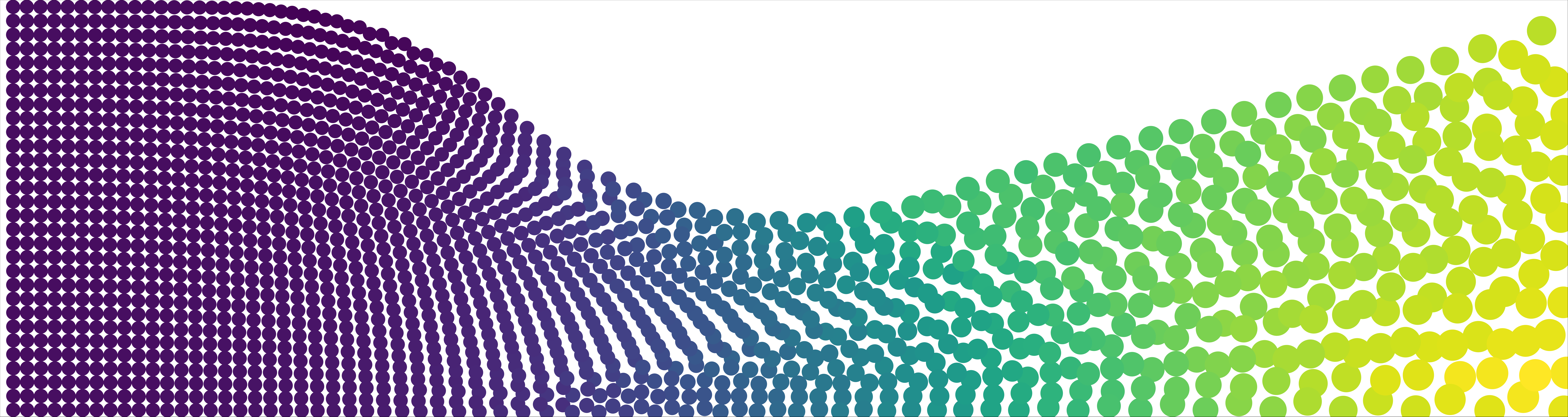};
		\addplot graphics
			[xmin=-0.07619999999999999,xmax=0.07619999999999999,ymin=-0.0405257,ymax=0]
			{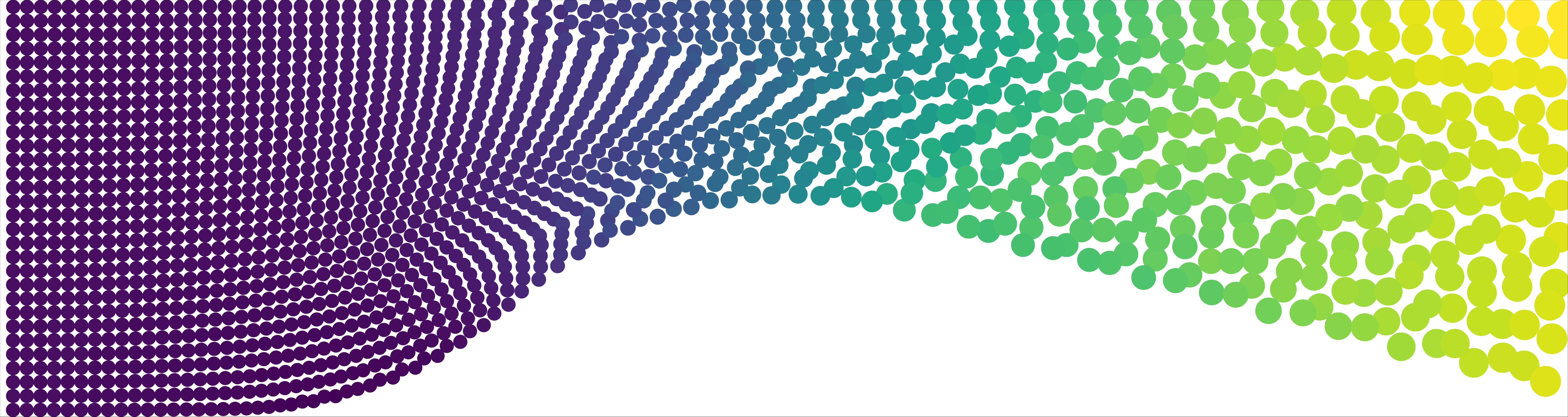};
		\addplot[
			smooth,
			no marks,
		] table [
				x=x,
				y=y,
			] from \profiletable;
		\addplot[
			smooth,
			no marks,
		] table [
				x=x,
				y expr=-\thisrow{y},
			] from \profiletable;
		\node[below, font=\footnotesize] at (rel axis cs: 0.5,.95) {WOM};
		\node[above, font=\footnotesize] at (rel axis cs: 0.5,.05) {WM};
	\end{axis}
	\coordinate (A1) at (rel axis cs: 1,1);
	\coordinate (A0) at (rel axis cs: 0,0);
	\newdimen\xmax
	\pgfextractx{\xmax}{\pgfpointanchor{A1}{center}}
	\newdimen\ymax
	\pgfextracty{\ymax}{\pgfpointanchor{A1}{center}}
	\newdimen\xmin
	\pgfextractx{\xmin}{\pgfpointanchor{A0}{center}}
	\newdimen\ymin
	\pgfextracty{\ymin}{\pgfpointanchor{A0}{center}}

	\begin{axis}[
			name=cbar,
			enlargelimits=false,
			title=$M$,
			axis on top,
			at={(condi.south east)},
			anchor=south west,
			ytick pos=right,
			xtick pos=bottom,
			every major tick/.append style={major tick length=2pt},
			xmajorticks=false,
			xshift=1em,
			width=1em,
			height=\ymax - \ymin,
			scale only axis]
		\addplot graphics[
				xmin=0.0,
				xmax=1.0,
				ymin=0.22566568851470947,
				ymax=2.242220557493974
			]{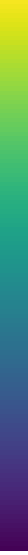};
	\end{axis}
\end{tikzpicture}
	\caption{Convergent-divergent nozzle with particles colored by local Mach number.
		The top and the bottom halves represent the result with and without ghost-mirror respectively.
	}
	\label{fig:condi-M}
\end{figure}

From \cref{fig:condi-M}, it can be observed that there are no qualitatively discernable differences between the two variants, just like the previous cases.
From \cref{fig:condi-mline} it can also be seen that the results are in agreement with a simulation performed using Eilmer.
There is some difference very near the exit.
This could be due to the exit boundary treatment method.
It should be noted that the particle density at the exit for the \gls{sph} simulation is significantly lesser than the cell density near the exit for the mesh used in the Eilmer simulation.

\begin{figure}
	\centering
	\pgfplotsset{
		table/search path={figs/auto/ConDiNozzle/}
	}
	\pgfplotsset{width=0.7\textwidth, height=0.5\textwidth}
	\begin{tikzpicture}
	\begin{axis}[
			enlarge x limits=false,
			legend pos=south east,
			xlabel=$x$,
			xtick scale label code/.code={$\times10^{#1}$},
			ylabel=$M$,
			every axis plot/.append style={thick},
			grid=major,
			cycle list name=exotic,
		]

		\addplot+[
			smooth,
			black,
			no marks,
			dashdotted
		] table [
				col sep=comma,
				x=x,
				y=M
			] {profile_mach.csv};
		\addlegendentry{Quasi 1D}

		\addplot+[
			const plot mark mid,
			no marks
		] table [
				col sep=comma,
				x=x,
				y=M
			] {eilmer_nozzle.csv};
		\addlegendentry{Eilmer}

		\addplot+[
			const plot mark mid,
			only marks,
			mark=+
		] table [
				col sep=comma,
				x=x,
				y=M
			] {eosampv.csv};
		\addlegendentry{SPH-WOM}
		\addplot+[
			const plot mark mid,
			only marks,
			mark=x
		] table [
				col sep=comma,
				x=x,
				y=M
			] {marrone.csv};
		\addlegendentry{SPH-WM}

	\end{axis}
\end{tikzpicture}
	\caption{Convergent divergent nozzle center line local Mach number.}
	\label{fig:condi-mline}
\end{figure}

\subsection{Forward Facing Step} \label{sec:ffs}

This problem deals with the flow of gas at Mach 3 over a forward-facing step
in a duct
\citep{emeryEvaluationSeveralDifferencing1968,woodwardNumericalSimulationTwodimensional1984a}.
This problem is considered challenging due to the complex flow physics
involved, including the formation interaction and reflection of waves. The
computational domain is \qty{3}{\meter} long and \qty{1}{\meter} wide. A step of height \qty{0.2}{\meter} is present at \qty{0.6}{\meter} away from the inlet. The initial particle
spacing is \qty{0.0125}{\meter}. Fluid enters the domain with density
\(\rho=\)\qty{1.4}{\kilo\gram\per\cubic\meter}, velocities
\(u=\)\qty{3.0}{\meter\per\sec}, \(v=\)\qty{0.0}{\meter\per\sec}, and pressure
\(p=\)\qty{1.0}{\pascal}. The simulation is run till $t=\qty{2}{\second}$. As
the flow evolves from the initial condition, we observe that a bow shock
develops ahead of the step. The curvature of the bow reduces and it strikes
the top wall. The curvature continues to reduce and the location of incidence
on the top wall keeps moving upstream. Eventually, a triple point is formed and
the Mach stem keeps traveling upstream. Meanwhile, the reflected shock wave
interacts with the expansion wave from the step-corner, strikes the bottom
wall and reflects further.

The resulting shock pattern at $t=\qty{2}{\second}$ can be read from \cref{fig:ffs-rho}. The 
Mach stem is observed to be about \qty{0.04}{\meter} long and located at about 
\(x=\qty{0.78}{\meter}\). The reflected wavefront strikes the bottom wall at about 
\(x=\qty{1.71}{\meter}\). These are in good agreement with the results that are observed in 
the literature. This indicates that the wall boundaries treated using the 
methods proposed in this paper are able to reflect the shock waves well.

\begin{figure}
	\centering
	\pgfplotsset{width=0.7\textwidth, height=0.5\textwidth}
	\input{../../macros}
\begin{tikzpicture}
	\begin{axis}[enlargelimits=false,
		name=eosampv,
		axis on top,
		xlabel=$x$,
		ylabel=$y$,
		scale only axis,
		axis equal image,
		title=Without ghost-mirror]
		\addplot graphics
			[xmin=0.0,xmax=3.0,ymin=0.0,ymax=1.0]
			{rho_eosampv_bare};
		\addplot[black, no markers] coordinates{
				(0.6,0.0)
				(0.6,0.2)
				(3.0,0.2)
			};
		\coordinate (A1) at (rel axis cs: 1,1);

		\end{axis}
		\begin{axis}[enlargelimits=false,
			at={(eosampv.south east)},
			anchor=north east,
			yshift=-5em,
			name=marrone,
			axis on top,
			xlabel=$x$,
			ylabel=$y$,
			scale only axis,
			axis equal image,
			title=With ghost-mirror]
		\addplot graphics
			[xmin=0.0,xmax=3.0,ymin=0.0,ymax=1.0]
			{rho_marrone_bare};
		\addplot[black, no markers] coordinates{
				(0.6,0.0)
				(0.6,0.2)
				(3.0,0.2)
			};
		\coordinate (A0) at (rel axis cs: 0,0);
	\end{axis}
	\newdimen\xmax
	\pgfextractx{\xmax}{\pgfpointanchor{A1}{center}}
	\newdimen\ymax
	\pgfextracty{\ymax}{\pgfpointanchor{A1}{center}}
	\newdimen\xmin
	\pgfextractx{\xmin}{\pgfpointanchor{A0}{center}}
	\newdimen\ymin
	\pgfextracty{\ymin}{\pgfpointanchor{A0}{center}}

	\begin{axis}[enlargelimits=false,
			title=$\rho$,
			axis on top,
			at={(marrone.south east)},
			anchor=south west,
			ytick pos=right,
			yticklabel pos=right,
			xshift=1em,
			every major tick/.append style={major tick length=2pt},
			xmajorticks=false,
			width=1em,
			height=\ymax-\ymin,
			scale only axis
		]
		\addplot graphics
			[xmin=0.0,xmax=0.1,ymin=0.3686176410854501,ymax=7.432152792728506]
			{rho_cbar_vertical};
	\end{axis}
\end{tikzpicture}
	\caption{Forward facing step with particles colored by density.}
	\label{fig:ffs-rho}
\end{figure}

\subsection{Double Mach Reflection} \label{sec:dmr}

Double Mach Reflection problem was proposed by
\citet{woodwardNumericalSimulationTwodimensional1984a} inspired by
experimental and numerical studies of reflections of planar shocks in the air
from a wedge. This problem involves a Mach 10 shock impinging a rigid wall at
an angle of \(60^\circ\). The impingement results in the formation of a
complex shock reflection structure. It is a self-similar structure that grows
in size as the shock propagates. This problem is considered a difficult case
for most numerical methods. The work \citet{gaoNewSmoothedParticle2023a} is
the only study that we know of, which simulates this problem with \gls{sph}.
However, they do this within the Eulerian framework. As a result, they did not
have to face many challenges that this problem brings when moving particles
are considered.

This problem is set up with an initial particle spacing is \qty{0.0125}{\meter}.
The reflecting ramp lies along the bottom of the problem domain, beginning at $x_0 = \qty[parse-numbers=false]{1/6}{\meter}$.
The angle between the shock and the reflecting boundary of $60^\circ$.
As discussed in \cref{sec:permiable}, the choice of permeable boundary treatment method restricts us from having an exact moving shock wave of Mach 10 prescribed on the top boundary.
We resort to the alternate setup described by \citet{tanInverseLaxWendroffProcedure2010, vevekAlternativeSetupsDouble2019b}.
The undisturbed fluid ahead of the shock has a density of \qty{1.4}{\kilo\gram\per\cubic\meter} and a pressure of \qty{1}{\pascal}.
The simulation is run till $t=\qty{0.2}{\second}$.

\begin{figure}
	\centering
	\pgfplotsset{width=0.7\textwidth, height=0.5\textwidth}
	\input{../../macros}
\begin{tikzpicture}
	\begin{axis}[enlargelimits=false,
			name=main,
			axis on top,
			xlabel=$x$,
			ylabel=$y$,
			scale only axis,
			axis equal image
		]
		\addplot graphics
			[xmin=-0.16666666666666666,xmax=3.8333333333333335,ymin=0.0,ymax=1.0]
			{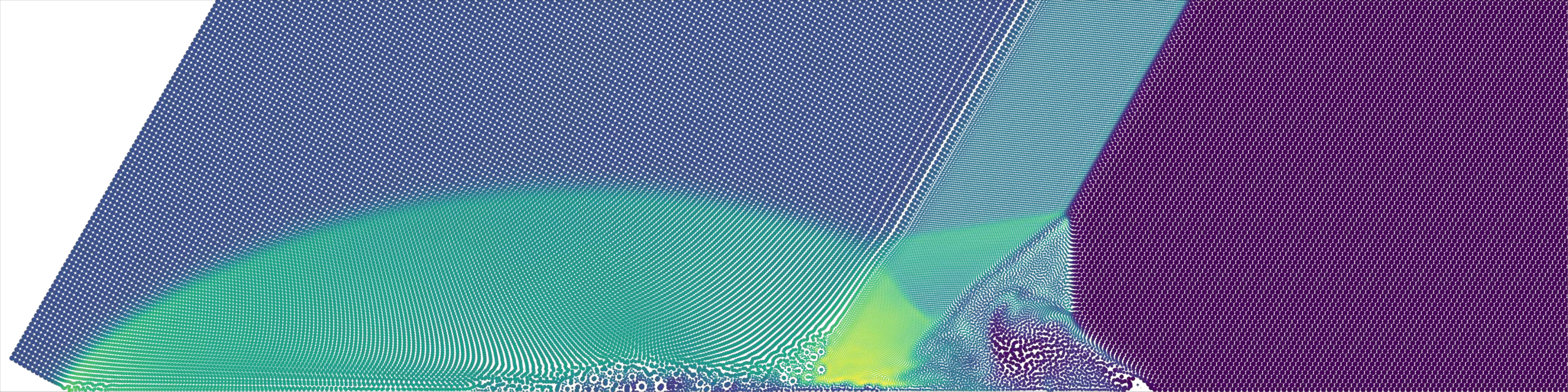};
		\node[fit={(axis cs: 1.8,0.0) (axis cs: 2.0,0.2)}, inner sep=0pt, draw=red, thick] (zoom1rect) {};
		\node[fit={(axis cs: 2.1,0.0) (axis cs: 3.0,0.5)}, inner sep=0pt, draw=red, thick] (zoom2rect) {};
	\end{axis}
	\coordinate (A0) at (rel axis cs: 0,0);
	\coordinate (A1) at (rel axis cs: 1,1);
	\newdimen\xmax
	\pgfextractx{\xmax}{\pgfpointanchor{A1}{center}}
	\newdimen\ymax
	\pgfextracty{\ymax}{\pgfpointanchor{A1}{center}}
	\newdimen\xmin
	\pgfextractx{\xmin}{\pgfpointanchor{A0}{center}}
	\newdimen\ymin
	\pgfextracty{\ymin}{\pgfpointanchor{A0}{center}}
	\begin{axis}[enlargelimits=false,
			title=$\rho$,
			axis on top,
			at={(main.south east)},
			anchor=south west,
			ytick pos=right,
			yticklabel pos=right,
			xshift=1em,
			every major tick/.append style={major tick length=2pt},
			xmajorticks=false,
			width=1em,
			height=\ymax-\ymin,
			scale only axis
		]
		\addplot graphics
			[xmin=0.0,xmax=0.1,ymin=28.16699340870742,ymax=1.3267766517516544]
			{rho_cbar_vertical};
	\end{axis}
	\begin{axis}[
			at = { ($ (main.south west)!0.2!(main.south east) $ ) },
			enlargelimits=false,
			name=zoom1,
			scale only axis,
			axis equal image,
			xmin=1.8,
			xmax=2.0,
			ymin=0.0,
			ymax=0.2,
			height=\ymax-\ymin,
			width=\xmax-\xmin,
			anchor=north,
			yshift = -3em,
			xmajorticks=false,
			ymajorticks=false,
			xlabel=Spot Like Structures]
		\addplot graphics
			[xmin=-0.16666666666666666,xmax=3.8333333333333335,ymin=0.0,ymax=1.0]
			{bad_rho_eosampv_bare};
	\end{axis}
	\begin{axis}[
			at = { ($ (main.south west)!0.75!(main.south east) $ ) },
			enlargelimits=false,
			name=zoom2,
			scale only axis,
			axis equal image,
			xmin=2.1,
			xmax=3.0,
			ymin=0.0,
			ymax=0.5,
			anchor=north,
			height=\ymax-\ymin,
			width=\xmax-\xmin,
			yshift = -3em,
			xmajorticks=false,
			ymajorticks=false,
			xlabel=Kinked Mach Stem]
		\addplot graphics
			[xmin=-0.16666666666666666,xmax=3.8333333333333335,ymin=0.0,ymax=1.0]
			{bad_rho_eosampv_bare};
	\end{axis}

	\draw (zoom1rect) -- (zoom1);
	\draw (zoom2rect) -- (zoom2);
\end{tikzpicture}
	\caption{Double Mach reflection simulated without ghost-mirror.
		The particles are colored by density.
	}
	\label{fig:dmr-rho-bad}
\end{figure}

\Cref{fig:dmr-rho-bad} shows the result of simulation without the use of
ghost-mirrors. It can be seen that there is considerable noise around the
primary slip line and some near-wall disturbance between \(x = \qty{1.5}{\meter}\) and \(x =
\qty{2.0}{\meter}\). The primary match stem appears to be severely kinked. The formation of
spot-like structures can also be noted. These structures are generally
observed at an interface where particles having different masses interact.
This is a known problem with \gls{sph}~\citep{puriComparisonSPHSchemes2014}.
From \cref{fig:dmr-rho}, it can be observed that when this problem is
simulated with the use of ghost-mirrors, the issues are less severe but they
are present nonetheless. It has been demonstrated that spot-like structures
can be mitigated by applying mass diffusion-based fixes proposed by
\citet{readSPHSSmoothedParticle2012,
prasannakumarMultimassCorrectionMulticomponent2018}. By using an adaptive
particle splitting and merging procedure which borrows from
\citet{sunAccurateSPHVolume2021} and
\citet{haftuParallelAdaptiveWeaklycompressible2022} we have results that
demonstrate that the kinked Mach stem, the near-wall disturbance and the
spot-like structures can be very effectively mitigated. To avoid digression,
we reserve the details of this procedure and the results for another article.
The point of showing this example in the present work is to demonstrate that
the boundary condition implementation is effective even in such cases with
complex shock wall interactions and successfully prevents particle
penetration. 

\begin{figure}
	\centering
	\pgfplotsset{width=0.7\textwidth, height=0.5\textwidth}
	\input{../../macros}
\begin{tikzpicture}
	\begin{axis}[name=single_marrone,
			enlargelimits=false,
			xlabel=$x$,
			ylabel=$y$,
			scale only axis,
			axis equal image]

		\addplot graphics
			[xmin=-0.16666666666666666,xmax=3.8333333333333335,ymin=0.0,ymax=1.0]
			{rho_marrone_bare};

	\end{axis}
	\coordinate (A0) at (rel axis cs: 0,0);
	\coordinate (A1) at (rel axis cs: 1,1);
	\newdimen\xmax
	\pgfextractx{\xmax}{\pgfpointanchor{A1}{center}}
	\newdimen\ymax
	\pgfextracty{\ymax}{\pgfpointanchor{A1}{center}}
	\newdimen\xmin
	\pgfextractx{\xmin}{\pgfpointanchor{A0}{center}}
	\newdimen\ymin
	\pgfextracty{\ymin}{\pgfpointanchor{A0}{center}}
	\begin{axis}[enlargelimits=false,
			title=$\rho$,
			axis on top,
			at={(single_marrone.south east)},
			anchor=south west,
			ytick pos=right,
			yticklabel pos=right,
			xshift=1em,
			every major tick/.append style={major tick length=2pt},
			xmajorticks=false,
			width=1em,
			height=\ymax-\ymin,
			scale only axis
		]
		\addplot graphics
			[xmin=0,xmax=1,ymin=1.3267766517516544,ymax=28.16699340870742]
			{rho_cbar_vertical};
	\end{axis}
\end{tikzpicture}
	\caption{Double Mach reflection simulated with ghost-mirror.
		The particles are colored by density.
	}
	\label{fig:dmr-rho}
\end{figure}

\subsection{Biconvex Aerofoil} \label{sec:supaf}

This problem involves a biconvex aerofoil at \(0^\circ\) angle of attack in
Mach 4.04 flow. A simple biconvex aerofoil with a chord, \(c_o\) of \qty{1}{\metre} and
a thickness-to-chord ratio is 0.1 is considered. The particle spacing at the
inlet is \qty{0.0125}{\metre}. The angle of attack is \(0^\circ\).
Freestream pressure, \(p_\infty\), and density, \(\rho_\infty\), are taken as
\qty{1}{\pascal} and \qty{1.4}{\kilo\gram\per\cubic\meter} respectively. The simulation is run till \(t=\qty{1}{\second}\).

Unlike most of the previously presented problems that demonstrate the
strengths of \gls{sph} with the proposed boundary treatment method, this
problem brings forward some of the shortcomings. Meshes for geometries like
these, that have sharp tips can be made to have high resolution at tips.
However, we cannot achieve the same in \gls{sph} without an adaptive
resolution procedure. Another issue is that the particles generated by the
packing algorithm are not perfectly symmetric. The effect of lack of symmetry
reduces with increasing resolution. So, we discretize the airfoil using ghost
particles with a spacing that is half that of the fluid particles. We do not
decrease the spacing further to keep the errors due to lack of consistency in
the particle spacing about the interface, as discussed in
\cref{sec:challenges}, in check.

The particle positions resulting from the simulation, colored by density, are
shown in \cref{fig:supaf-rho}. It was observed that the streak lines near the
body are not static. The fluid particles close to the body become disordered
as they pass by, differently based on the boundary treatment employed. The
pressure over the body is shown in \cref{fig:supaf-p}. Interpolation without
the use of ghost-mirrors yields better results in this problem. It can be
observed that the pressure over the aerofoil does not match perfectly with the
inviscid theory. Along with the issues mentioned above, this could also be
attributed to the usage of artificial viscosity. Despite all these issues, we
can observe from \cref{fig:supaf-p-conv} that the results are convergent.

\begin{figure}
	\centering

	\pgfplotsset{width=0.35\textwidth, height=0.4\textwidth}
	\input{../../macros}
\begin{tikzpicture}
	\begin{axis}[enlargelimits=false,
		name=eosampv,
		axis on top,
		xlabel=$x$,
		ylabel=$y$,
		scale only axis,
		axis equal image,
		title=Without ghost-mirror]
		\addplot graphics
			[xmin=-0.75,xmax=0.75,ymin=-0.5,ymax=0.5]
			{rho_eosampv_0.0075_bare};
		\coordinate(A) at ( axis cs:-0.5, 0);
		\coordinate(B) at (axis cs:0,0.05);
		\coordinate(C) at (axis cs:0.5,0);
		\draw[thick, draw=black]
		(A) to[arc through cw=(B)] (C);
		\coordinate(A) at ( axis cs:0.5, 0);
		\coordinate(B) at (axis cs:0,-0.05);
		\coordinate(C) at (axis cs:-0.5,0);
		\draw[thick, draw=black]
		(A) to[arc through cw=(B)] (C);
		\coordinate (A1) at (rel axis cs: 1,1);

	\end{axis}

		\begin{axis}[enlargelimits=false,
			at={(eosampv.south east)},
			anchor=south west,
			xshift=6em,
			name=marrone,
			axis on top,
			xlabel=$x$,
			ylabel=$y$,
			scale only axis,
			axis equal image,
			title=Without ghost-mirror]
		\addplot graphics
			[xmin=-0.75,xmax=0.75,ymin=-0.5,ymax=0.5]
			{rho_marrone_0.0075_bare};
		\coordinate(A) at ( axis cs:-0.5, 0);
		\coordinate(B) at (axis cs:0,0.05);
		\coordinate(C) at (axis cs:0.5,0);
		\draw[thick, draw=black]
		(A) to[arc through cw=(B)] (C);
		\coordinate(A) at ( axis cs:0.5, 0);
		\coordinate(B) at (axis cs:0,-0.05);
		\coordinate(C) at (axis cs:-0.5,0);
		\draw[thick, draw=black]
		(A) to[arc through cw=(B)] (C);
		\coordinate (A0) at (rel axis cs: 0,0);
	\end{axis}
	\newdimen\xmax
	\pgfextractx{\xmax}{\pgfpointanchor{A1}{center}}
	\newdimen\ymax
	\pgfextracty{\ymax}{\pgfpointanchor{A1}{center}}
	\newdimen\xmin
	\pgfextractx{\xmin}{\pgfpointanchor{A0}{center}}
	\newdimen\ymin
	\pgfextracty{\ymin}{\pgfpointanchor{A0}{center}}
	\begin{axis}[enlargelimits=false,
			title=$\rho$,
			axis on top,
			at={(marrone.south east)},
			anchor=south west,
			ytick pos=right,
			yticklabel pos=right,
			xshift=1em,
			every major tick/.append style={major tick length=2pt},
			xmajorticks=false,
			width=1em,
			height=\ymax-\ymin,
			scale only axis,
			ytick scale label code/.code={\pgfmathparse{int(-#1)}$ \rho \cdot 10^{\pgfmathresult}$},
			every y tick scale label/.style={at={(0.5,1.0)}, anchor = south},
		]
		\addplot graphics
			[xmin=0.0,xmax=0.1,ymin=0.18188718531581113,ymax=2.863324893650888]
			{rho_cbar_vertical};
	\end{axis}
\end{tikzpicture}
	\caption{Mach 4.04 flow over biconvex aerofoil at \(0^\circ\) angle of attack.
		The particle spacing at the inlet is \qty{7.e-3}{\meter}.
		The particles are colored by density.
	}
	\label{fig:supaf-rho}
\end{figure}

\begin{figure}
	\centering
	\pgfplotsset{table/search path={figs}}
	\subfloat[]{
		\centering
		\pgfplotsset{width=0.49\textwidth, height=0.4\textwidth}
		\begin{tikzpicture}
	\pgfplotstableread[col sep=comma]{auto/BiconvexAerofoil/pressure_theo.csv}\theodata
	\pgfplotstableread[col sep=comma]{auto/BiconvexAerofoil/pressure_sim_eosampv.csv}\simedata
	\pgfplotstableread[col sep=comma]{auto/BiconvexAerofoil/pressure_sim_marrone.csv}\simmdata
	\begin{axis}[
			legend style={
					font=\scriptsize,
				},
			xmin=0.03,
			xmax=0.97,
			ymin=0.0,
			ymax=3.0,
			xlabel=$x/c_o$,
			ylabel=$p/p_\infty$,
			grid=major,
			cycle list name=exotic,
		]
		\addplot+[
			no marks,
			smooth,
			thick,
			color=black,
		] table [
				x=xbyc,
				y=pbypinf
			] from \theodata;
		\addlegendentry{Inviscid Theory}

		\addplot+[
			only marks,
		] table [
				x=xbyc,
				y=0.00125,
			] from \simedata;
		\addlegendentry{WOM}

		\addplot+[
			only marks,
		] table [
				x=xbyc,
				y=0.00125,
			] from \simmdata;
		\addlegendentry{WM}

	\end{axis}
\end{tikzpicture}
		\label{fig:supaf-p}
	}
	\subfloat[]{
		\centering
		\pgfplotsset{width=0.49\textwidth, height=0.4\textwidth}
		\begin{tikzpicture}
	\pgfplotstableread[col sep=comma]{auto/BiconvexAerofoil/pressure_theo.csv}\theodata
	\pgfplotstableread[col sep=comma]{auto/BiconvexAerofoil/pressure_sim_eosampv.csv}\simedata
	\begin{axis}[
			legend style={
					font=\scriptsize,
				},
			xmin=0.03,
			xmax=0.97,
			ymin=0.0,
			ymax=3.0,
			xlabel=$x/c_o$,
			ylabel=$p/p_\infty$,
			grid=major,
			cycle list name=exotic,
		]
		\addplot+[
			no marks,
			smooth,
			thick,
			color=black,
		] table [
				x=xbyc,
				y=pbypinf
			] from \theodata;
		\addlegendentry{Inviscid Theory}

		\addplot+[
			only marks,
		] table [
				x=xbyc,
				y=0.005,
			] from \simedata;
		\addlegendentry{$5.00\times10^{-2}$}

		\addplot+[
			only marks,
		] table [
				x=xbyc,
				y=0.0025,
			] from \simedata;
		\addlegendentry{$2.50\times10^{-2}$}

		\addplot+[
			only marks,
		] table [
				x=xbyc,
				y=0.00125,
			] from \simedata;
		\addlegendentry{$1.25\times10^{-2}$}
	\end{axis}
\end{tikzpicture}
		\label{fig:supaf-p-conv}
	}
	\caption{Pressure on the upper surface of biconvex aerofoil. (a) Comparison of the with and without ghost-mirror variants at spacing  \qty{1.25e-2}{\meter}. (b) Comparison of without ghost-mirror variant at different spacings.}
\end{figure}

\subsection{Rotating Square Projectile} \label{sec:rotsq}

This problem is also borrowed from the examples in the Eilmer repository. This
problem involves a square projectile rotating about its center of mass in a
Mach 6 flow. The side of the square is \qty{0.02}{\meter}. The initial
particle spacing is \qty{0.002}{\meter}. The freestream pressure, temperature
and velocity are \qty{760}{\pascal}, \qty{71}{\kelvin} and
\qty{1005.0}{\metre\per\second} respectively. The projectile is initially at
rest and the angle of the square face relative to incoming flow, \(\theta_0\)
is \qty{0}{\radian}. The rotation of the projectile is prescribed using
angular velocity, \(\omega\) as a function of time as
\begin{equation}
	\omega(t) = A \cos{\left(\frac{2\pi t}{t_f}\right)},
\end{equation}
where, \(A\) is the maximum angular velocity, \qty{2000}{\radian\per\second}
and \(t_f\) is the final time, \qty{2}{\milli\second}.

In \cref{fig:rotsq}, the bow shock formed in front of the projectile can be
seen. From \cref{fig:rotsq-force-comp-eilmer} it can be observed that the
force exerted on the projectile is in agreement with the results obtained
using Eilmer. The mesh used for simulation with Eilmer is shown in
\cref{fig:rotsq-mesh}. \Cref{fig:rotsq-force-comp-shield} shows that the force
exerted on the projectile is not crippled by the penetration shield. It can be
seen that the force is marginally less noisy when the penetration shield is
used.

\begin{figure}
	\centering
	\pgfplotsset{width=0.6\textwidth, height=0.5\textwidth}
	\begin{tikzpicture}
		\begin{axis}[
				xlabel={\(x\)},
				ylabel={\(y\)},
				scale only axis,
				axis equal image,
				xmin=-0.2,
				xmax=0.2,
				ymin=-0.2,
				ymax=0.2,
			]
			\addplot graphics [xmin=-0.538145,xmax=0.538145,ymin=-0.538145,ymax=0.538145] {figs/auto/RotatingSquare/mesh};
		\end{axis}
	\end{tikzpicture}
	\caption{Mesh around the rotating square projectile body for Eilmer.}
	\label{fig:rotsq-mesh}
\end{figure}

\IfFileExists{figcache/eosampv_rotsq.pdf}{
	\begin{figure}
		\centering
		
		\begin{tikzpicture}
		\pgfplotsset{width=0.98\textwidth, height=0.35\textwidth}
		\begin{axis}[
				name=eosampv,
			    hide axis,
				enlargelimits=false,
				title=Without ghost-mirror
		]
			\addplot graphics[xmin=0,xmax=1,ymin=0,ymax=1]{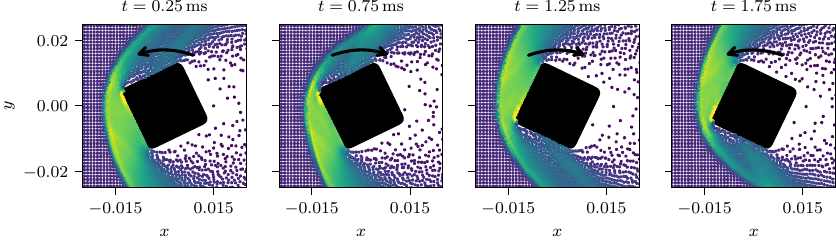};
			\coordinate (A1) at (rel axis cs: 1,1);
		\end{axis}
		\begin{axis}[
			name = marrone
			at=(eosampv.south west),
			anchor=north west,
			hide axis,
			enlargelimits=false,
			title=With ghost-mirror,
			yshift = -3.5em
	]
		\addplot graphics[xmin=0,xmax=1,ymin=0,ymax=1]{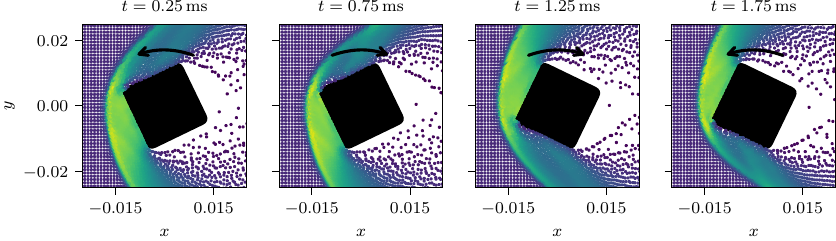};
		\coordinate (A0) at (rel axis cs: 0,0);
	\end{axis}

\newdimen\xmax
\pgfextractx{\xmax}{\pgfpointanchor{A1}{center}}
\newdimen\ymax
\pgfextracty{\ymax}{\pgfpointanchor{A1}{center}}
\newdimen\xmin
\pgfextractx{\xmin}{\pgfpointanchor{A0}{center}}
\newdimen\ymin
\pgfextracty{\ymin}{\pgfpointanchor{A0}{center}}
\begin{axis}[enlargelimits=false,
		title=$\rho$,
		axis on top,
		at={(A1)},
		anchor=north west,
		ytick pos=right,
		yticklabel pos=right,
		xshift=1em,
		every major tick/.append style={major tick length=2pt},
		xmajorticks=false,
		width=1em,
		height=0.95\ymax-0.95\ymin,
		scale only axis
	]
	\addplot graphics[
		ymin=0.001845390763566193,
		ymax=0.31471236714739226,
		xmin=0.0,
		xmax=1.0
		]{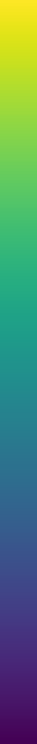};
\end{axis}

	\end{tikzpicture}

		\caption{Time snapshots showing particles colored by density for rotating square projectile problem.}
		\label{fig:rotsq}
	\end{figure}
}{Figure not found}

\begin{figure}
	\centering
	\subfloat[]{
		\centering
		\pgfplotsset{width=0.49\textwidth, height=0.4\textwidth}
		\begin{tikzpicture}
	\pgfplotstableread[col sep=comma]{auto/RotatingSquare/marrone_strict.csv}\mdata
	\pgfplotstableread[col sep=comma]{auto/RotatingSquare/eosampv_strict.csv}\adata
	\pgfplotstableread[col sep=comma]{auto/RotatingSquare/eilmer.csv}\edata
	\begin{axis}[
			xlabel=$t$(ms),
			ylabel=$F$(N/m\textsuperscript{2}),
			legend columns=3,
			legend style={at={(0.5,1.25)},anchor=south},
			legend cell align=center,
			legend style={font=\scriptsize},
			enlarge x limits=false,
			xtick pos=lower, ytick pos=left,
			cycle list name=exotic,
			every axis plot/.append style={semithick},
			no markers,
			grid=both,
			xtick = {0.5e-3, 1.0e-3, 1.5e-3, 2.0e-3},
			xmin = 0.2e-3,
			ymin = -180,
			xmax=2e-3,
			ymax=720,
			xtick scale label code/.code={},
		]
		\addplot+[densely dotted, black] table[x=ts, y=fx]{\edata};
		\addlegendentry{$f_x$(E)}
		\addplot+[dashed, orange] table[x=ts, y=fx]{\adata};
		\addlegendentry{$f_x$(WOM)}
		\addplot+[dashdotdotted, violet] table[x=ts, y=fx]{\mdata};
		\addlegendentry{$f_x$(WM)}
		\addplot+[densely dashed, blue] table[x=ts, y=fy]{\edata};
		\addlegendentry{$f_y$(E)}
		\addplot+[solid, purple] table[x=ts, y=fy]{\adata};
		\addlegendentry{$f_y$(WOM)}
		\addplot+[dashdotted, olive] table[x=ts, y=fy]{\mdata};
		\addlegendentry{$f_y$(WM)}
	\end{axis}
	\begin{axis}[
			xmin = 0.2e-3,
			ymin = -200,
			xmax=2e-3,
			ymax=700,
			xtick = {0.5e-3, 1.0e-3, 1.5e-3, 2.0e-3},
			xticklabels={36.5, 0.0, 36.5, 0.0},
			xtick scale label code/.code={},
			hide y axis,
			x axis line style={opacity=1},
			y axis line style={opacity=1},
			xtick pos=upper, ytick pos=right,
			xlabel=$\theta^\circ$,
			xlabel style={at={(0.5,1.10)},anchor=south},
		]
	\end{axis}
\end{tikzpicture}
		\label{fig:rotsq-force-comp-eilmer}
	}
	\subfloat[]{
		\centering
		\pgfplotsset{width=0.49\textwidth, height=0.4\textwidth}
		\begin{tikzpicture}
	\pgfplotstableread[col sep=comma]{auto/RotatingSquare/eosampv_no_shield.csv}\nsdata
	\pgfplotstableread[col sep=comma]{auto/RotatingSquare/eosampv_strict.csv}\sdata
	\begin{axis}[
			xlabel=$t$(ms),
			ylabel=$F$(N/m\textsuperscript{2}),
			legend columns=2,
			legend style={at={(0.5,1.25)},anchor=south},
			legend cell align=center,
			legend style={font=\scriptsize},
			enlarge x limits=false,
			xtick pos=lower, ytick pos=left,
			cycle list name=exotic,
			every axis plot/.append style={semithick},
			no markers,
			grid=both,
			xtick = {0.5e-3, 1.0e-3, 1.5e-3, 2.0e-3},
			xmin = 0.2e-3,
			ymin = -180,
			xmax=2e-3,
			ymax=720,
			xtick scale label code/.code={},
		]
		\addplot+[solid, black] table[x=ts, y=fx]{\sdata};
		\addlegendentry{$f_x$(WOM-S)}
		\addplot+[densely dashed, teal] table[x=ts, y=fx]{\nsdata};
		\addlegendentry{$f_x$(WOM-NS)}
		\addplot+[densely dashdotted, blue] table[x=ts, y=fy]{\sdata};
		\addlegendentry{$f_y$(WOM-S)}
		\addplot+[densely dotted, purple] table[x=ts, y=fy]{\nsdata};
		\addlegendentry{$f_y$(WOM-NS)}
	\end{axis}
	\begin{axis}[
			xmin = 0.2e-3,
			ymin = -200,
			xmax=2e-3,
			ymax=700,
			xtick = {0.5e-3, 1.0e-3, 1.5e-3, 2.0e-3},
			xticklabels={36.5, 0.0, 36.5, 0.0},
			xtick scale label code/.code={},
			hide y axis,
			x axis line style={opacity=1},
			y axis line style={opacity=1},
			xtick pos=upper, ytick pos=right,
			xlabel=$\theta^\circ$,
			xlabel style={at={(0.5,1.1)},anchor=south},
		]
	\end{axis}
\end{tikzpicture}
		\label{fig:rotsq-force-comp-shield}
	}
	\caption{Comparison of force on the rotating square projectile. (a) Comparison against Eilmer. The E in the legend stands for Eilmer. (b) Comparison of force with and without shield. The S and NS in the legend stand for Shield and No-Shield, respectively.}
	\label{fig:rotsq-force}
\end{figure}

\subsection{Apollo Reentry Capsule 3D} \label{sec:a3d}

For this problem, the geometry of the Apollo reentry capsule is adapted from the work of \citet{mossDSMCSimulationsApollo2006} and is shown in \cref{fig:a3d-geom}.
This problem aims to estimate the coefficient of drag of the body, in order to demonstrate that \gls{sph} can be used for real-world 3D problems with the present boundary treatment method.

\begin{figure}
	\centering
	\pgfplotsset{table/search path={figs/}}
	\pgfplotsset{width=0.7\textwidth, height=0.5\textwidth}
	\input{macros}

\begin{tikzpicture}
	\pgfplotstableread[col sep=comma]{a3d_profile.csv}\profiletable
	\begin{axis}[name=condi,
			xlabel=$x$,
			ylabel=$y$,
			scale only axis,
			axis equal image,
			axis lines=none,
			ymax = 2.3
		]
		\addplot[
			smooth,
			no marks,
			very thick,
		] table [
				x=x,
				y=y,
			] from \profiletable;
		\path[draw] (-0.3,0.0) -- (0.0,0.0);
		\path[draw] (-0.3,1.95576) -- (0.5542701520151482,1.95576);
		\draw[<->] (-0.2, 0.0) --node [midway, fill=white, rounded corners, font=\footnotesize, inner sep=0.1em] {\(l_1\)} (-0.2,1.95576);
		\path[draw] (0.0, 0.0) -- (0.0,2.2);
		\path[draw] (3.4306, 0.0) -- (3.4306,2.2);
		\draw[<->] (0.0, 2.15) --node [midway, fill=white, rounded corners, font=\footnotesize, inner sep=0.1em] {\(l_2\)} (3.4306,2.15);
		\draw[->] (0.34306035214169018,0.9) -- (0.10306035214169018,0.9782067969770973) ;
		\node[right] at (0.34306035214169018,0.85) {$r_0$};
		\draw[->] (0.57,1.7) -- (0.4414069822527363,1.919915009300684) ;
		\node[below] at (0.62,1.7) {$r_1$};
		\draw[->] (3.1995, 0.0) -- (3.4228548174099265,0.059328201892341934) ;
		\node[below] at (3.18, 0.0) {$r_2$};
		\draw[->] (0.0, 0.0) -- (0.2,0.0) ;
		\node[below] at (0.2, 0.0) {$x$};
		\draw[->] (0.0, 0.0) -- (0.0,0.2) ;
		\node[left] at (0.0, 0.2) {$y$};
		\draw[-] (0.0, 0.0) -- (3.4306,0.0) ;
	\end{axis}

\end{tikzpicture}
	\caption{Schematic of the Apollo reentry capsule.
		The dimensions are: \(l_1\)=1.9558~m, \(l_2\)=3.4306~m, \(r_0\)=4.6939~m, \(r_1\)=0.2311~m, \(r_2\)=0.1956~m.
	}
	\label{fig:a3d-geom}
\end{figure}

The streamwise direction is along the \(x\)-axis. The center of the forebody
coincides with the origin. The inlet and outlet are \qty{2}{\meter} upstream
and \qty{6}{\meter} downstream of the origin respectively. Both \(y\) and
\(z\) extents of the domain are \(\pm \qty{8}{\meter}\). The body is at
\(0^\circ\) angle of attack. The flow enters the domain at Mach 2.5 with
stagnation pressure and temperature of \(\qty{1.2e6}{\pascal}\) and
\qty{285}{\kelvin} respectively. The simulation is run until
\(t=\qty{0.15}{\second}\). For SPH simulations, the particle spacing at the
inlet is \(\qty{0.1}{\meter}\).

\begin{figure}
	\centering
	\subfloat[]{
		\centering
		\includegraphics[trim=140 100 140 100, clip, width=0.32\linewidth]{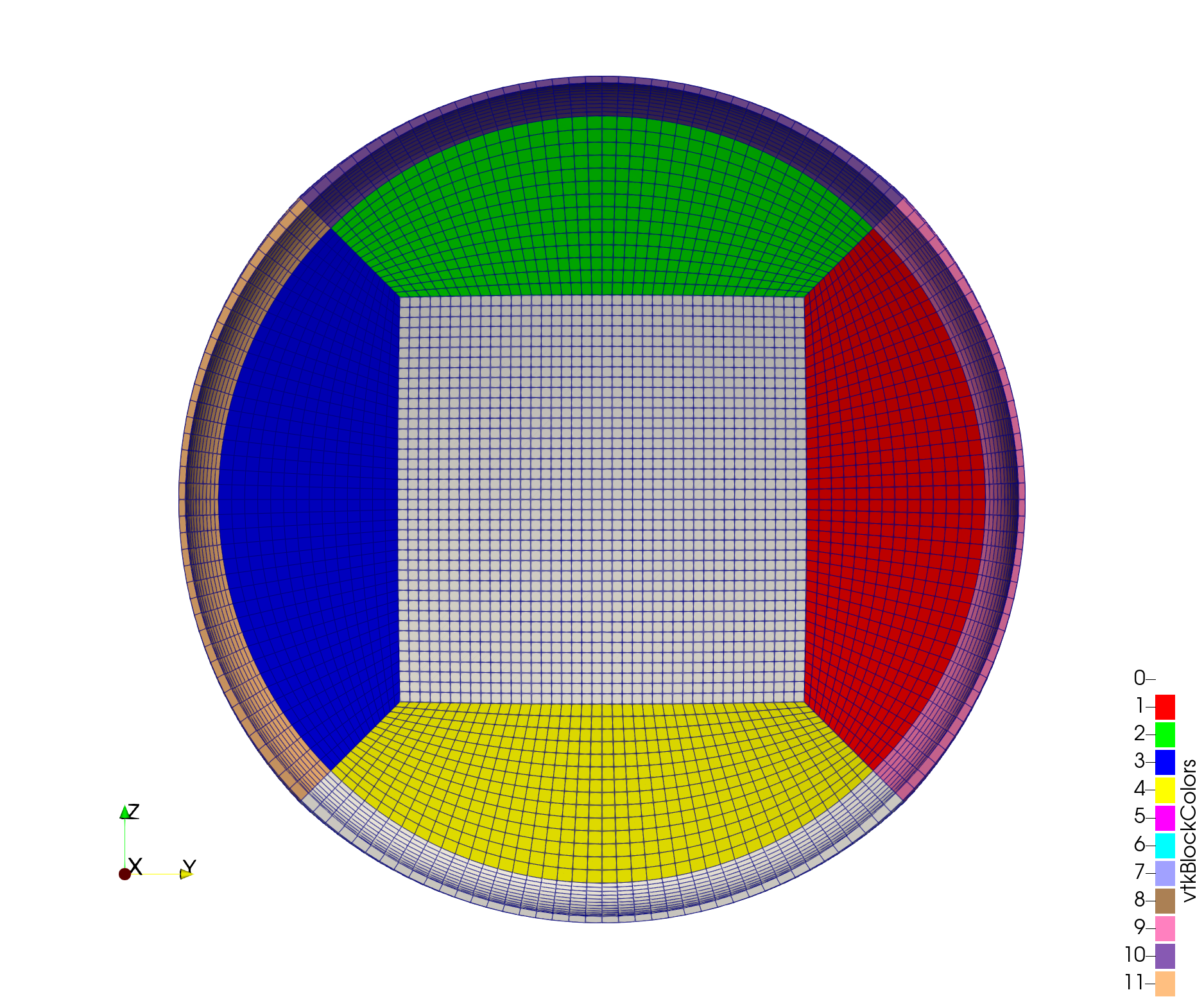}
		\includegraphics[trim=80 100 140 100, clip, width=0.32\linewidth]{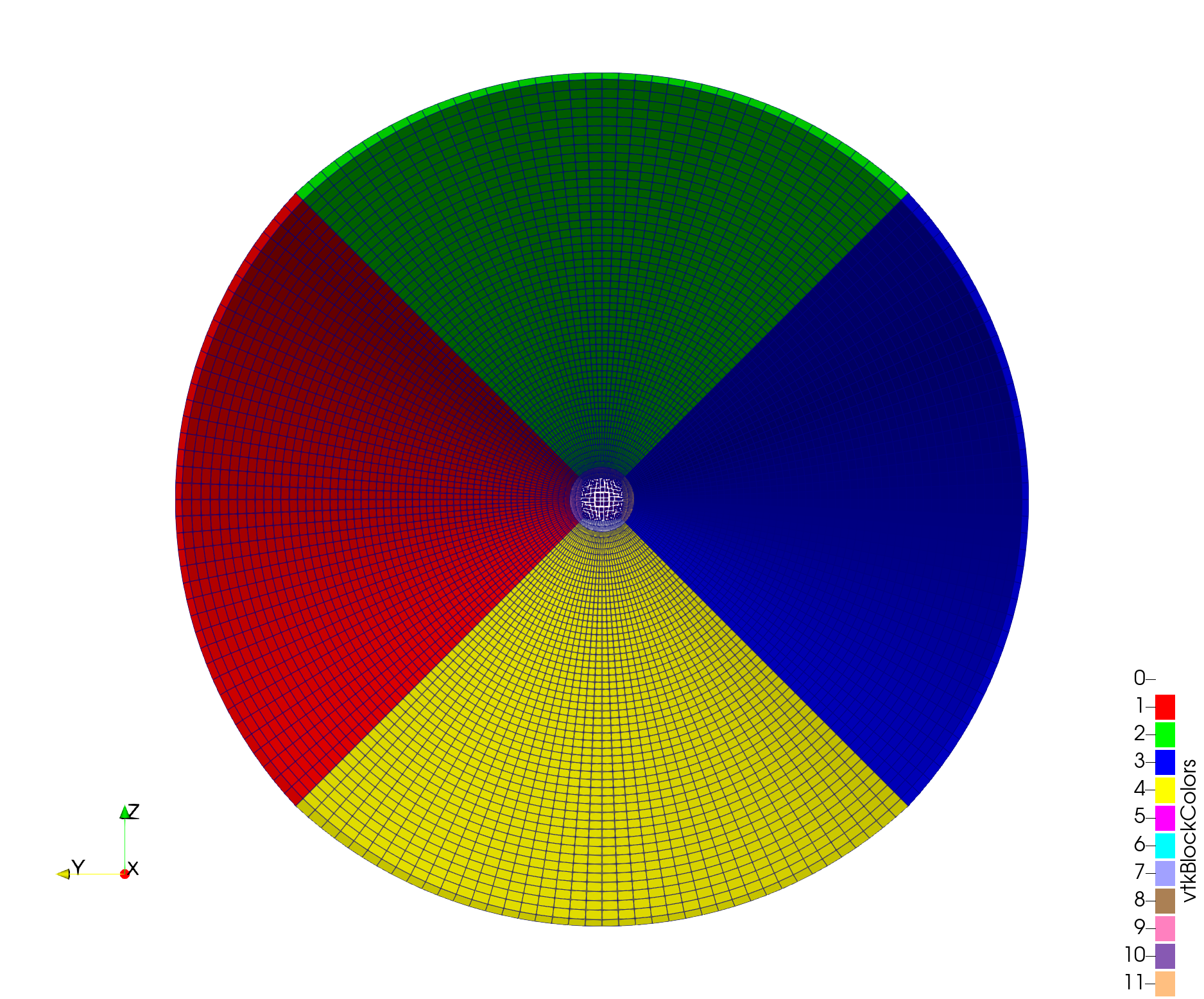}
		\includegraphics[trim=140 50 140 50, clip, width=0.28\linewidth]{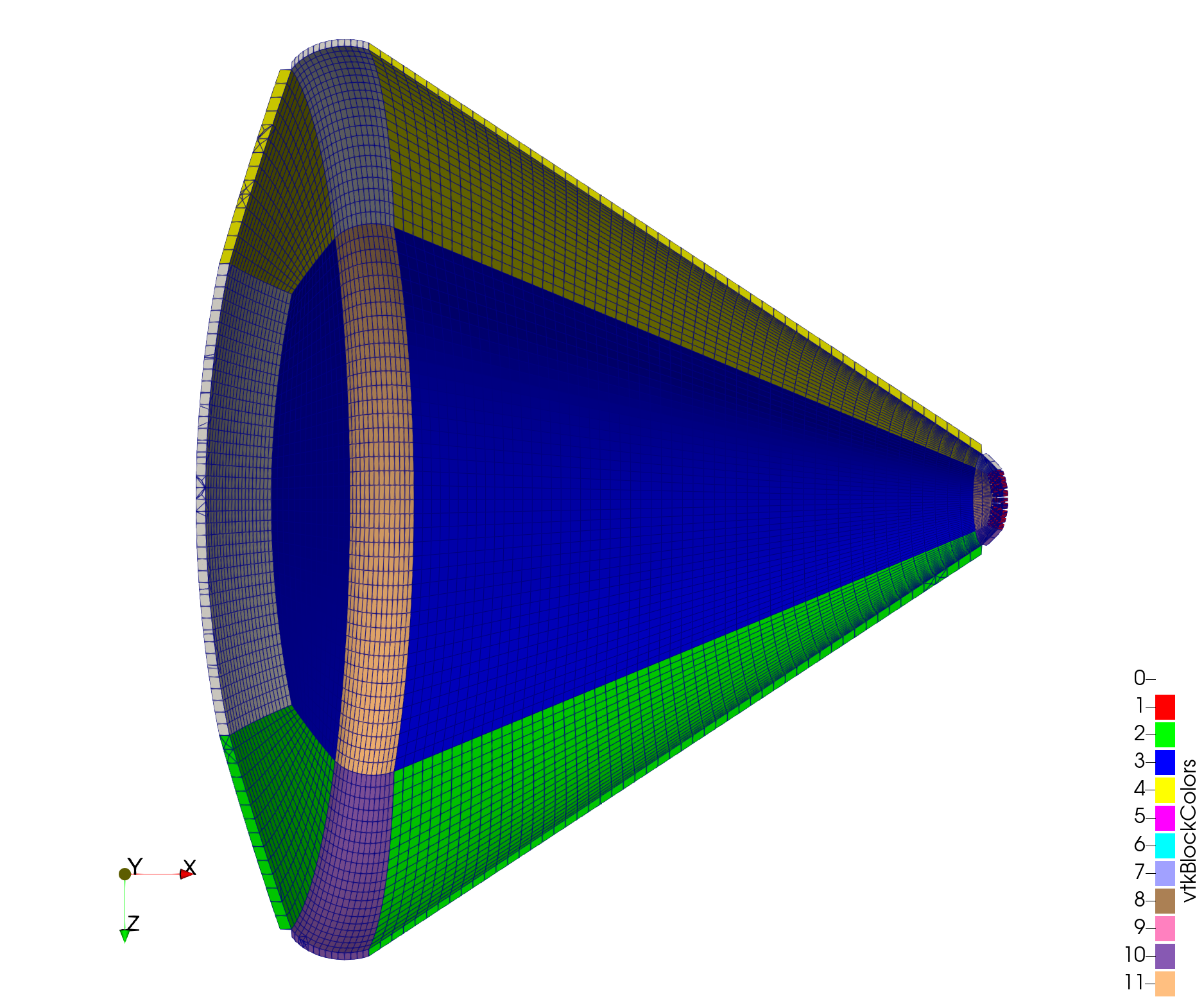}
	}

	\subfloat[]{
		\centering
		\begin{tikzpicture}
			\pgfplotsset{width=0.5\linewidth, height=0.5\linewidth}
			\centering
			\begin{axis}[
					xlabel={\(x\)},
					ylabel={\(y\)},
					scale only axis,
					axis equal image,
					xmin=-1,
					xmax=4.5,
					ymin=0,
					ymax=3.5
				]
				\addplot graphics [xmin=-2,xmax=6,ymin=0,ymax=8]{figs/auto/Apollo3D/eilmer3d_mesh};
			\end{axis}
		\end{tikzpicture}

	}

	\caption{Mesh for Apollo reentry capsule simulation with Eilmer. (a) Fore aft and side views of the body-fitted multiblock mesh, left to right. (b) Slice of multiblock mesh around the body.}
	\label{fig:a3d-mesh}
\end{figure}

For comparison, a similar setup is created in Eilmer. The details of the
multi-block mesh can be read from \cref{fig:a3d-mesh}. The cell sizes vary,
however, the body-fitted cell sizes are roughly \(\qty{0.05}{\meter}\), with
double the resolution at the shoulder. The aftmost end is highly refined due
to the constraint imposed by the corresponding foremost block.

To describe the nature of resolution independence, the expected observation is
for the coefficient of drag to converge to a value as the resolution is
increased. Now, the particle spacing could not have been much greater than
\(\qty{0.1}{\meter}\) as it would obliterate some of the relatively sharp
curves on the body. Also, the particle spacing could not have been much
smaller than the chosen spacing as that would have led to a prohibitively huge
simulation time, given our computational capabilities. This ruled out the
possibility of a resolution independence study. With that in account, the
coefficient of drag is computed for the two variants of the proposed boundary
treatment method and Eilmer is presented in \cref{tab:a3d-cd}. It needs to be
noted that the estimation of forces in this kind of problem may require very
fine resolutions. We do not have a common yardstick to specify the resolution
and compare the results of \gls{sph} and \gls{fvm}. With the result in
\cref{tab:a3d-cd}, we would only like to drive the fact that results are not
too far off, despite the inability to equate the resolution and our present
inability to demonstrate resolution independence.

\begin{figure}
	\centering
	\pgfplotsset{width=0.4\textwidth, height=0.01\textwidth}
	\begin{tikzpicture}
	\begin{axis}[enlargelimits=false,
			ylabel=$\rho$,
			ylabel style={rotate=-90},
			axis on top,
			xtick pos=bottom,
			every major tick/.append style={major tick length=2pt},
			ymajorticks=false,
			scale only axis
		]
		\addplot graphics[
				xmin=-0.05500000000000001,
				xmax=8.0,
				ymin=0.0,
				ymax=1.0
			]{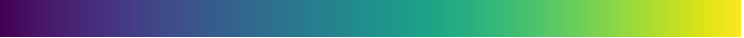};
	\end{axis}
\end{tikzpicture}

	\pgfplotsset{width=0.50\textwidth, height=0.4\textwidth}
	\pgfplotstableread[col sep=comma]{figs/a3d_profile.csv}{\profile}

		\centering
			\begin{tikzpicture}
			\begin{axis}[
					xmin=-2.0,
					xmax=6.0,
					ymin=0.0,
					ymax=4.0,
					xlabel=$x$,
					ylabel=$y$,
					enlargelimits=false,
					scale only axis,
					axis equal image,
					title=Eilmer]
				\addplot graphics
					[xmin=-2.0,xmax=6.0,ymin=-4.0,ymax=4.0,
						includegraphics={trim=266 61 266 61, clip}]
					{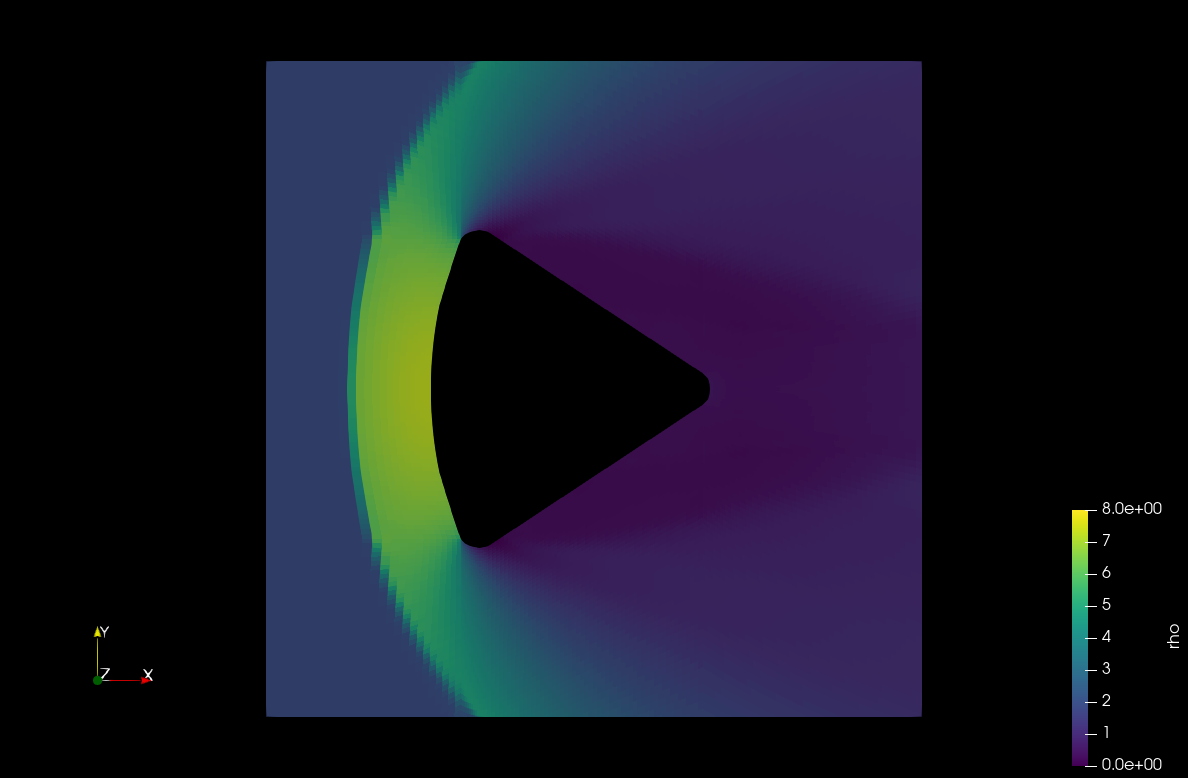};
			\end{axis}
		\end{tikzpicture}
			
		\begin{tikzpicture}
			\begin{axis}[
					xmin=-2.0,
					xmax=6.0,
					ymin=0.0,
					ymax=4.0,
					xlabel=$x$,
					ylabel=$y$,
					enlargelimits=false,
					scale only axis,
					axis equal image,
					title=\gls{sph} Without ghost-mirror]
				\addplot graphics
					[xmin=-2.0,xmax=6.0,ymin=-8.0,ymax=8.0]
					{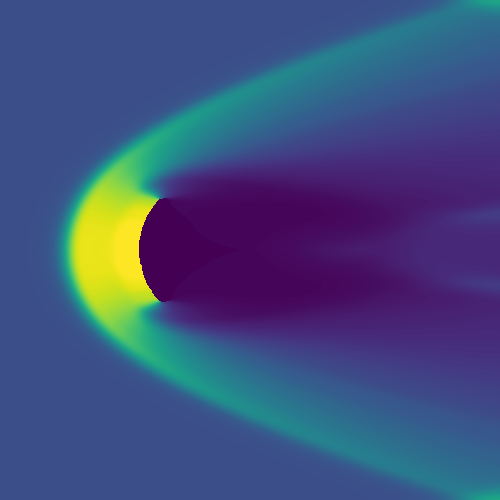};
				\addplot[
					smooth,
					no marks,
					very thick,
					fill=black,
				] table [
						x=x,
						y=y,
					] from \profile -- cycle;
				\addplot[
					smooth,
					no marks,
					very thick,
					fill=black,
				] table [
						x=x,
						y expr = -\thisrow{y},
					] from \profile -- cycle;
			\end{axis}
		\end{tikzpicture}

		\begin{tikzpicture}
			\begin{axis}[
					xmin=-2.0,
					xmax=6.0,
					ymin=0.0,
					ymax=4.0,
					xlabel=$x$,
					ylabel=$y$,
					enlargelimits=false,
					scale only axis,
					axis equal image,
					title=\gls{sph} with ghost-mirror]
				\addplot graphics
					[xmin=-2.0,xmax=6.0,ymin=-8.0,ymax=8.0]
					{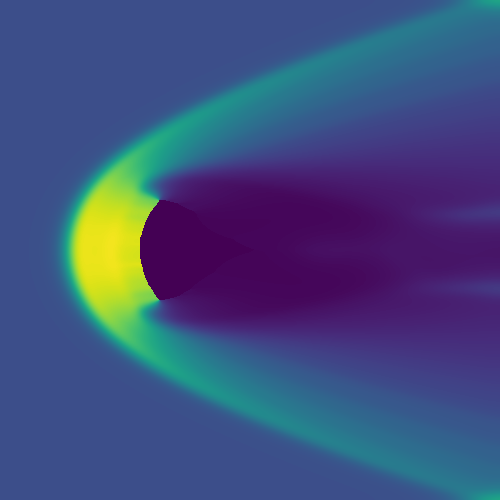};
				\addplot[
					smooth,
					no marks,
					very thick,
					fill=black,
				] table [
						x=x,
						y=y,
					] from \profile -- cycle;
				\addplot[
					smooth,
					no marks,
					very thick,
					fill=black,
				] table [
						x=x,
						y expr = -\thisrow{y},
					] from \profile -- cycle;
			\end{axis}
			\end{tikzpicture}

	\caption{Apollo reentry capsule results with cells colored by density.}
	\label{fig:a3d-rho}
\end{figure}

The density field for the two variants of the proposed boundary treatment
method and Eilmer are shown in \cref{fig:a3d-rho}.  The \gls{sph} results are
shown on a $500 \times 250$ grid. SPLASH
\citep{priceSPLASHInteractiveVisualisation2007} interpolation procedure was used for this. It is observed that a separation bubble is formed behind the shoulder. These low-density regions are very resolution deficient in the \gls{sph} results. Despite the difference in drag, the point of showing this case is to once again demonstrate
that the boundary treatment method can be used on practical three-dimensional geometries.

\begin{table}
	\centering
	\setlength{\tabcolsep}{20pt}
	\begin{tabular}{cc}
		\hline Case & $C_D$ \\
		\hline \hline SPH without ghost-mirror & 1.55 \\
		\hline SPH with ghost-mirror & 1.51 \\
		\hline Eilmer & 1.42 \\
		\hline
	\end{tabular}
	\caption{Coefficient of drag for the Apollo reentry capsule.}
	\label{tab:a3d-cd}
\end{table}

\section{Summary} \label{sec:summary}

We proposed boundary treatment methods for compressible \gls{sph} after identifying
the challenges that are unique to compressible \gls{sph}. The issues highlighted in \cref{sec:challenges} are more pronounced in compressible scenarios. However, we expect that the proposed remedies would be effective in incompressible scenarios as well. We were able to demonstrate significant improvements over the state-of-the-art\citep{englestadInvestigationsNovelBoundary2020}. The proposed boundary treatment
methods do not add much complexity to the existing boundary treatment methods
that are used in \gls{isph} and \gls{wcsph}. The penetration shield which
makes use of the \gls{tvf} elegantly prevents the particles from penetrating
the boundary, without resorting to the usage of short-range forces at the
boundary. 

By and large, the results from both the extrapolation variants are comparable. 
It can be observed from \cref{tab:ccor-err} that extrapolation with ghost-mirror 
generally result in marginally lower errors and does
well in most cases except in the presence of sharp corners. The double mach reflection is a problem where extrapolation with ghost-mirrors clearly produces better results. However, 
the particle sparsity in wakes or separation bubbles can cause the ghost-mirrors to  
end up lacking neighbors. Consequently, the correction matrix for 
calculating \(W_{ij}^{LC}\) may be ill-formed, 
especially when the ratio of smoothing length to spacing is low. 
Extrapolation without ghost-mirrors can be regarded as more robust due to this 
pitfall. 

The proposed methods are
shown to be effective on a diverse set of problems of increasing complexity,
many of which have not been simulated with \gls{sph} before to the best of our
knowledge. The compression corner (\cref{sec:ccor}) showcases a simple
stationary oblique shock, reflecting shocktube (\cref{sec:rsod}) showcases the
reflection of a moving normal shock from a wall, hypersonic cylinder
(\cref{sec:hcyl}) showcases a stationary bow shock in hypersonic flow,
convergent divergent nozzle (\cref{sec:condi}) showcases subsonic to
supersonic transition, and forward facing step (\cref{sec:ffs}) showcases
complex shockwave reflections and interactions. All these problems demonstrate
acceptable results. The double Mach reflection (\cref{sec:dmr}) showcases a
complex shockwave interaction involving complex hypersonic shocks. The
observed artifacts can be resolved by incorporating an adaptive refinement and
derefinement procedure which we will discuss in a forthcoming article. The
biconvex aerofoil (\cref{sec:supaf}) showcases flow over a slender body with
sharp tips. The results are convergent and improvement is likely with an
adaptive refinement and derefinement procedure. The rotating square projectile
(\cref{sec:rotsq}) showcases flow over a moving geometry. The results are in
good agreement Eilmer~\citep{gibbonsEilmerOpensourceMultiphysics2023}. The Apollo reentry capsule
(\cref{sec:a3d}) showcases applicability in 3D. The results are promising
despite the resolution deficiency. This summarizes the problem-wise results.

\section{Future Directions and Concluding Remarks} \label{sec:conclusions}
We note that there is still ample room for improvement and the following are
interesting problems that appear to be worthy of immediate attention:
\begin{itemize}
	\item A better and faster packing algorithm for initializing ghost 
	particles especially with variable resolution.
	\item Incorporation of particle splitting and merging procedure, so that
	more challenging real-world problems can be simulated.
	\item A shock-friendly particle shifting technique.
	\item A possible strategy for the fusion of the two extrapolation variants
	so that extrapolation is done without ghost-mirrors in the presence of
	sharp tips or voids, and with mirrors elsewhere.
\end{itemize}

We do not employ aggressive
problem-specific tuning, chasing the absolute best results possible. So,
individual problems may still have room for improvement. Our objective was
just to show that the methods proposed in this paper have wide applicability,
and perform well even without any problem-specific tuning. We have made an
effort to mention all the intricacies involved. Additionally, we have provided
the source at \url{https://gitlab.com/pypr/compressible-sph-bc} for the
readers to study the actual implementation if they wish to. In the interest of
reproducibility, all \gls{sph} results shown in this manuscript may be 
reproduced using an automation 
framework~\citep{ramachandranAutomanPythonBasedAutomation2018}.

While mesh-based methods are known to be more mature and have better support
for high-order schemes, meshless methods also possess some inherent
indisputable advantages. For instance, with \gls{sph}, one, two or
three-dimensional problems can be simulated with minimal changes to the code.
\gls{fvm} codes do not have this luxury. The Lagrangian nature also accords important benefits\citep{hopkinsNewClassAccurate2015}. For example, in the present work, the rotating square projectile problem illustrated that flows involving moving bodies can be simulated in \gls{sph} by merely updating ghost particles that represent the body. Mesh-based methods warrant more complicated procedures for the same.

If one excludes the time required to set
the problem up and considers only the run-time, the presented \gls{sph}
simulations would register as slower than their \gls{fvm} counterparts.
However, with \gls{fvm} the quality of results is heavily dependent on the
quality of the grid. The grid generation step in mesh-based methods requires a
significant amount of time, operator skill, and effort. Therefore, with an
automated particle packing algorithm for initializing ghost particles
representing the body, we believe that the presented boundary treatment
methods take \gls{sph} a step closer to being an appealing alternative
approach to mesh-based methods for engineering simulations involving
compressible flows, especially when rapid results with minimal effort is a
priority.


%
%

%

\begin{acknowledgments}
	The authors acknowledge the use of the computing resources of the ACE facility, Department of Aerospace Engineering, IIT Bombay.
\end{acknowledgments}

\bibliography{references}

\begin{thebibliography}{93}%
\makeatletter
\providecommand \@ifxundefined [1]{%
 \@ifx{#1\undefined}
}%
\providecommand \@ifnum [1]{%
 \ifnum #1\expandafter \@firstoftwo
 \else \expandafter \@secondoftwo
 \fi
}%
\providecommand \@ifx [1]{%
 \ifx #1\expandafter \@firstoftwo
 \else \expandafter \@secondoftwo
 \fi
}%
\providecommand \natexlab [1]{#1}%
\providecommand \enquote  [1]{``#1''}%
\providecommand \bibnamefont  [1]{#1}%
\providecommand \bibfnamefont [1]{#1}%
\providecommand \citenamefont [1]{#1}%
\providecommand \href@noop [0]{\@secondoftwo}%
\providecommand \href [0]{\begingroup \@sanitize@url \@href}%
\providecommand \@href[1]{\@@startlink{#1}\@@href}%
\providecommand \@@href[1]{\endgroup#1\@@endlink}%
\providecommand \@sanitize@url [0]{\catcode `\\12\catcode `\$12\catcode `\&12\catcode `\#12\catcode `\^12\catcode `\_12\catcode `\%12\relax}%
\providecommand \@@startlink[1]{}%
\providecommand \@@endlink[0]{}%
\providecommand \url  [0]{\begingroup\@sanitize@url \@url }%
\providecommand \@url [1]{\endgroup\@href {#1}{\urlprefix }}%
\providecommand \urlprefix  [0]{URL }%
\providecommand \Eprint [0]{\href }%
\providecommand \doibase [0]{http://dx.doi.org/}%
\providecommand \selectlanguage [0]{\@gobble}%
\providecommand \bibinfo  [0]{\@secondoftwo}%
\providecommand \bibfield  [0]{\@secondoftwo}%
\providecommand \translation [1]{[#1]}%
\providecommand \BibitemOpen [0]{}%
\providecommand \bibitemStop [0]{}%
\providecommand \bibitemNoStop [0]{.\EOS\space}%
\providecommand \EOS [0]{\spacefactor3000\relax}%
\providecommand \BibitemShut  [1]{\csname bibitem#1\endcsname}%
\let\auto@bib@innerbib\@empty
\bibitem [{\citenamefont {Rosswog}(2009)}]{rosswogAstrophysicalSmoothParticle2009}%
  \BibitemOpen
  \bibfield  {author} {\bibinfo {author} {\bibfnamefont {S.}~\bibnamefont {Rosswog}},\ }\bibfield  {title} {\enquote {\bibinfo {title} {Astrophysical {Smooth Particle Hydrodynamics}},}\ }\href {\doibase 10.1016/j.newar.2009.08.007} {\bibfield  {journal} {\bibinfo  {journal} {New Astronomy Reviews}\ }\textbf {\bibinfo {volume} {53}},\ \bibinfo {pages} {78--104} (\bibinfo {year} {2009})}\BibitemShut {NoStop}%
\bibitem [{\citenamefont {Springel}(2010)}]{springelSmoothedParticleHydrodynamics2010a}%
  \BibitemOpen
  \bibfield  {author} {\bibinfo {author} {\bibfnamefont {V.}~\bibnamefont {Springel}},\ }\bibfield  {title} {\enquote {\bibinfo {title} {Smoothed {{Particle Hydrodynamics}} in {{Astrophysics}}},}\ }\href {\doibase 10.1146/annurev-astro-081309-130914} {\bibfield  {journal} {\bibinfo  {journal} {Annual Review of Astronomy and Astrophysics}\ }\textbf {\bibinfo {volume} {48}},\ \bibinfo {pages} {391--430} (\bibinfo {year} {2010})}\BibitemShut {NoStop}%
\bibitem [{\citenamefont {Takeda}\ \emph {et~al.}(1994)\citenamefont {Takeda}, \citenamefont {Miyama},\ and\ \citenamefont {Sekiya}}]{takedaNumericalSimulationViscous1994}%
  \BibitemOpen
  \bibfield  {author} {\bibinfo {author} {\bibfnamefont {H.}~\bibnamefont {Takeda}}, \bibinfo {author} {\bibfnamefont {S.~M.}\ \bibnamefont {Miyama}}, \ and\ \bibinfo {author} {\bibfnamefont {M.}~\bibnamefont {Sekiya}},\ }\bibfield  {title} {\enquote {\bibinfo {title} {Numerical {{Simulation}} of {{Viscous Flow}} by {{Smoothed Particle Hydrodynamics}}},}\ }\href {\doibase 10.1143/ptp/92.5.939} {\bibfield  {journal} {\bibinfo  {journal} {Progress of Theoretical Physics}\ }\textbf {\bibinfo {volume} {92}},\ \bibinfo {pages} {939--960} (\bibinfo {year} {1994})}\BibitemShut {NoStop}%
\bibitem [{\citenamefont {Morris}\ \emph {et~al.}(1997)\citenamefont {Morris}, \citenamefont {Fox},\ and\ \citenamefont {Zhu}}]{morrisModelingLowReynolds1997}%
  \BibitemOpen
  \bibfield  {author} {\bibinfo {author} {\bibfnamefont {J.~P.}\ \bibnamefont {Morris}}, \bibinfo {author} {\bibfnamefont {P.~J.}\ \bibnamefont {Fox}}, \ and\ \bibinfo {author} {\bibfnamefont {Y.}~\bibnamefont {Zhu}},\ }\bibfield  {title} {\enquote {\bibinfo {title} {Modeling {{Low Reynolds Number Incompressible Flows Using SPH}}},}\ }\href {\doibase 10.1006/jcph.1997.5776} {\bibfield  {journal} {\bibinfo  {journal} {Journal of Computational Physics}\ }\textbf {\bibinfo {volume} {136}},\ \bibinfo {pages} {214--226} (\bibinfo {year} {1997})}\BibitemShut {NoStop}%
\bibitem [{\citenamefont {Colagrossi}\ and\ \citenamefont {Landrini}(2003)}]{colagrossiNumericalSimulationInterfacial2003}%
  \BibitemOpen
  \bibfield  {author} {\bibinfo {author} {\bibfnamefont {A.}~\bibnamefont {Colagrossi}}\ and\ \bibinfo {author} {\bibfnamefont {M.}~\bibnamefont {Landrini}},\ }\bibfield  {title} {\enquote {\bibinfo {title} {Numerical simulation of interfacial flows by smoothed particle hydrodynamics},}\ }\href {\doibase 10.1016/S0021-9991(03)00324-3} {\bibfield  {journal} {\bibinfo  {journal} {Journal of Computational Physics}\ }\textbf {\bibinfo {volume} {191}},\ \bibinfo {pages} {448--475} (\bibinfo {year} {2003})}\BibitemShut {NoStop}%
\bibitem [{\citenamefont {Yildiz}\ \emph {et~al.}(2009)\citenamefont {Yildiz}, \citenamefont {Rook},\ and\ \citenamefont {Suleman}}]{yildizSPHMultipleBoundary2009}%
  \BibitemOpen
  \bibfield  {author} {\bibinfo {author} {\bibfnamefont {M.}~\bibnamefont {Yildiz}}, \bibinfo {author} {\bibfnamefont {R.~A.}\ \bibnamefont {Rook}}, \ and\ \bibinfo {author} {\bibfnamefont {A.}~\bibnamefont {Suleman}},\ }\bibfield  {title} {\enquote {\bibinfo {title} {{{SPH}} with the multiple boundary tangent method},}\ }\href {\doibase 10.1002/nme.2458} {\bibfield  {journal} {\bibinfo  {journal} {International Journal for Numerical Methods in Engineering}\ }\textbf {\bibinfo {volume} {77}},\ \bibinfo {pages} {1416--1438} (\bibinfo {year} {2009})}\BibitemShut {NoStop}%
\bibitem [{\citenamefont {Macia}\ \emph {et~al.}(2011)\citenamefont {Macia}, \citenamefont {Antuono}, \citenamefont {Gonzalez},\ and\ \citenamefont {Colagrossi}}]{maciaTheoreticalAnalysisNoSlip2011}%
  \BibitemOpen
  \bibfield  {author} {\bibinfo {author} {\bibfnamefont {F.}~\bibnamefont {Macia}}, \bibinfo {author} {\bibfnamefont {M.}~\bibnamefont {Antuono}}, \bibinfo {author} {\bibfnamefont {L.~M.}\ \bibnamefont {Gonzalez}}, \ and\ \bibinfo {author} {\bibfnamefont {A.}~\bibnamefont {Colagrossi}},\ }\bibfield  {title} {\enquote {\bibinfo {title} {Theoretical {{Analysis}} of the {{No-Slip Boundary Condition Enforcement}} in {{SPH Methods}}},}\ }\href {\doibase 10.1143/PTP.125.1091} {\bibfield  {journal} {\bibinfo  {journal} {Progress of Theoretical Physics}\ }\textbf {\bibinfo {volume} {125}},\ \bibinfo {pages} {1091--1121} (\bibinfo {year} {2011})}\BibitemShut {NoStop}%
\bibitem [{\citenamefont {Marrone}\ \emph {et~al.}(2011)\citenamefont {Marrone}, \citenamefont {Antuono}, \citenamefont {Colagrossi}, \citenamefont {Colicchio}, \citenamefont {Le~Touz{\'e}},\ and\ \citenamefont {Graziani}}]{marroneDSPHModelSimulating2011}%
  \BibitemOpen
  \bibfield  {author} {\bibinfo {author} {\bibfnamefont {S.}~\bibnamefont {Marrone}}, \bibinfo {author} {\bibfnamefont {M.}~\bibnamefont {Antuono}}, \bibinfo {author} {\bibfnamefont {A.}~\bibnamefont {Colagrossi}}, \bibinfo {author} {\bibfnamefont {G.}~\bibnamefont {Colicchio}}, \bibinfo {author} {\bibfnamefont {D.}~\bibnamefont {Le~Touz{\'e}}}, \ and\ \bibinfo {author} {\bibfnamefont {G.}~\bibnamefont {Graziani}},\ }\bibfield  {title} {\enquote {\bibinfo {title} {{$\delta$}-{{SPH}} model for simulating violent impact flows},}\ }\href {\doibase 10.1016/j.cma.2010.12.016} {\bibfield  {journal} {\bibinfo  {journal} {Computer Methods in Applied Mechanics and Engineering}\ }\textbf {\bibinfo {volume} {200}},\ \bibinfo {pages} {1526--1542} (\bibinfo {year} {2011})}\BibitemShut {NoStop}%
\bibitem [{\citenamefont {Adami}\ \emph {et~al.}(2012)\citenamefont {Adami}, \citenamefont {Hu},\ and\ \citenamefont {Adams}}]{adamiGeneralizedWallBoundary2012}%
  \BibitemOpen
  \bibfield  {author} {\bibinfo {author} {\bibfnamefont {S.}~\bibnamefont {Adami}}, \bibinfo {author} {\bibfnamefont {X.}~\bibnamefont {Hu}}, \ and\ \bibinfo {author} {\bibfnamefont {N.}~\bibnamefont {Adams}},\ }\bibfield  {title} {\enquote {\bibinfo {title} {A generalized wall boundary condition for smoothed particle hydrodynamics},}\ }\href {\doibase 10.1016/j.jcp.2012.05.005} {\bibfield  {journal} {\bibinfo  {journal} {Journal of Computational Physics}\ }\textbf {\bibinfo {volume} {231}},\ \bibinfo {pages} {7057--7075} (\bibinfo {year} {2012})}\BibitemShut {NoStop}%
\bibitem [{\citenamefont {Marrone}\ \emph {et~al.}(2013)\citenamefont {Marrone}, \citenamefont {Colagrossi}, \citenamefont {Antuono}, \citenamefont {Colicchio},\ and\ \citenamefont {Graziani}}]{marroneAccurateSPHModeling2013}%
  \BibitemOpen
  \bibfield  {author} {\bibinfo {author} {\bibfnamefont {S.}~\bibnamefont {Marrone}}, \bibinfo {author} {\bibfnamefont {A.}~\bibnamefont {Colagrossi}}, \bibinfo {author} {\bibfnamefont {M.}~\bibnamefont {Antuono}}, \bibinfo {author} {\bibfnamefont {G.}~\bibnamefont {Colicchio}}, \ and\ \bibinfo {author} {\bibfnamefont {G.}~\bibnamefont {Graziani}},\ }\bibfield  {title} {\enquote {\bibinfo {title} {An accurate {{SPH}} modeling of viscous flows around bodies at low and moderate {{Reynolds}} numbers},}\ }\href {\doibase 10.1016/j.jcp.2013.03.011} {\bibfield  {journal} {\bibinfo  {journal} {Journal of Computational Physics}\ }\textbf {\bibinfo {volume} {245}},\ \bibinfo {pages} {456--475} (\bibinfo {year} {2013})}\BibitemShut {NoStop}%
\bibitem [{\citenamefont {Antuono}\ \emph {et~al.}(2023)\citenamefont {Antuono}, \citenamefont {Pilloton}, \citenamefont {Colagrossi},\ and\ \citenamefont {Durante}}]{antuonoCloneParticlesSimplified2023}%
  \BibitemOpen
  \bibfield  {author} {\bibinfo {author} {\bibfnamefont {M.}~\bibnamefont {Antuono}}, \bibinfo {author} {\bibfnamefont {C.}~\bibnamefont {Pilloton}}, \bibinfo {author} {\bibfnamefont {A.}~\bibnamefont {Colagrossi}}, \ and\ \bibinfo {author} {\bibfnamefont {D.}~\bibnamefont {Durante}},\ }\bibfield  {title} {\enquote {\bibinfo {title} {Clone particles: {{A}} simplified technique to enforce solid boundary conditions in {{SPH}}},}\ }\href {\doibase 10.1016/j.cma.2023.115973} {\bibfield  {journal} {\bibinfo  {journal} {Computer Methods in Applied Mechanics and Engineering}\ }\textbf {\bibinfo {volume} {409}},\ \bibinfo {pages} {115973} (\bibinfo {year} {2023})}\BibitemShut {NoStop}%
\bibitem [{\citenamefont {Silla}\ and\ \citenamefont {Bertola}(2017)}]{sillaSPHSimulationOblique2017}%
  \BibitemOpen
  \bibfield  {author} {\bibinfo {author} {\bibfnamefont {M.}~\bibnamefont {Silla}}\ and\ \bibinfo {author} {\bibfnamefont {V.}~\bibnamefont {Bertola}},\ }\bibfield  {title} {\enquote {\bibinfo {title} {{{SPH}} simulation of oblique shocks in compressible flows: {{SPH Simulation}} of {{Oblique Shocks}}},}\ }\href {\doibase 10.1002/fld.4356} {\bibfield  {journal} {\bibinfo  {journal} {International Journal for Numerical Methods in Fluids}\ }\textbf {\bibinfo {volume} {84}},\ \bibinfo {pages} {494--505} (\bibinfo {year} {2017})}\BibitemShut {NoStop}%
\bibitem [{\citenamefont {Englestad}\ and\ \citenamefont {Cassibry}(2020)}]{englestadInvestigationsNovelBoundary2020}%
  \BibitemOpen
  \bibfield  {author} {\bibinfo {author} {\bibfnamefont {T.~J.}\ \bibnamefont {Englestad}}\ and\ \bibinfo {author} {\bibfnamefont {J.~T.}\ \bibnamefont {Cassibry}},\ }\bibfield  {title} {\enquote {\bibinfo {title} {Investigations of a novel boundary condition approach for the accurate prediction of hypersonic oblique shocks in mesh-free {{Lagrangian}} simulations},}\ }\href {\doibase 10.1016/j.ast.2020.106322} {\bibfield  {journal} {\bibinfo  {journal} {Aerospace Science and Technology}\ }\textbf {\bibinfo {volume} {107}},\ \bibinfo {pages} {106322} (\bibinfo {year} {2020})}\BibitemShut {NoStop}%
\bibitem [{\citenamefont {Sun}\ \emph {et~al.}(2021)\citenamefont {Sun}, \citenamefont {Le~Touz{\'e}}, \citenamefont {Oger},\ and\ \citenamefont {Zhang}}]{sunAccurateSPHVolume2021}%
  \BibitemOpen
  \bibfield  {author} {\bibinfo {author} {\bibfnamefont {P.-N.}\ \bibnamefont {Sun}}, \bibinfo {author} {\bibfnamefont {D.}~\bibnamefont {Le~Touz{\'e}}}, \bibinfo {author} {\bibfnamefont {G.}~\bibnamefont {Oger}}, \ and\ \bibinfo {author} {\bibfnamefont {A.-M.}\ \bibnamefont {Zhang}},\ }\bibfield  {title} {\enquote {\bibinfo {title} {An accurate {{SPH Volume Adaptive Scheme}} for modeling strongly-compressible multiphase flows. {{Part}} 1: {{Numerical}} scheme and validations with basic {{1D}} and {{2D}} benchmarks},}\ }\href {\doibase 10.1016/j.jcp.2020.109937} {\bibfield  {journal} {\bibinfo  {journal} {Journal of Computational Physics}\ }\textbf {\bibinfo {volume} {426}},\ \bibinfo {pages} {109937} (\bibinfo {year} {2021})}\BibitemShut {NoStop}%
\bibitem [{\citenamefont {Adami}\ \emph {et~al.}(2013)\citenamefont {Adami}, \citenamefont {Hu},\ and\ \citenamefont {Adams}}]{TVFAdami2013}%
  \BibitemOpen
  \bibfield  {author} {\bibinfo {author} {\bibfnamefont {S.}~\bibnamefont {Adami}}, \bibinfo {author} {\bibfnamefont {X.}~\bibnamefont {Hu}}, \ and\ \bibinfo {author} {\bibfnamefont {N.}~\bibnamefont {Adams}},\ }\bibfield  {title} {\enquote {\bibinfo {title} {{A transport-velocity formulation for smoothed particle hydrodynamics}},}\ }\href {\doibase 10.1016/j.jcp.2013.01.043} {\bibfield  {journal} {\bibinfo  {journal} {Journal of Computational Physics}\ }\textbf {\bibinfo {volume} {241}},\ \bibinfo {pages} {292--307} (\bibinfo {year} {2013})}\BibitemShut {NoStop}%
\bibitem [{\citenamefont {Negi}\ and\ \citenamefont {Ramachandran}(2022)}]{negiHowTrainYour2022a}%
  \BibitemOpen
  \bibfield  {author} {\bibinfo {author} {\bibfnamefont {P.}~\bibnamefont {Negi}}\ and\ \bibinfo {author} {\bibfnamefont {P.}~\bibnamefont {Ramachandran}},\ }\bibfield  {title} {\enquote {\bibinfo {title} {How to train your solver: {{Verification}} of boundary conditions for smoothed particle hydrodynamics},}\ }\href {\doibase 10.1063/5.0126234} {\bibfield  {journal} {\bibinfo  {journal} {Physics of Fluids}\ }\textbf {\bibinfo {volume} {34}},\ \bibinfo {pages} {117125} (\bibinfo {year} {2022})}\BibitemShut {NoStop}%
\bibitem [{\citenamefont {Lastiwka}\ \emph {et~al.}(2009)\citenamefont {Lastiwka}, \citenamefont {Basa},\ and\ \citenamefont {Quinlan}}]{lastiwkaPermeableNonreflectingBoundary2009a}%
  \BibitemOpen
  \bibfield  {author} {\bibinfo {author} {\bibfnamefont {M.}~\bibnamefont {Lastiwka}}, \bibinfo {author} {\bibfnamefont {M.}~\bibnamefont {Basa}}, \ and\ \bibinfo {author} {\bibfnamefont {N.~J.}\ \bibnamefont {Quinlan}},\ }\bibfield  {title} {\enquote {\bibinfo {title} {Permeable and non-reflecting boundary conditions in {{SPH}}},}\ }\href {\doibase 10.1002/fld.1971} {\bibfield  {journal} {\bibinfo  {journal} {International Journal for Numerical Methods in Fluids}\ }\textbf {\bibinfo {volume} {61}},\ \bibinfo {pages} {709--724} (\bibinfo {year} {2009})}\BibitemShut {NoStop}%
\bibitem [{\citenamefont {Federico}\ \emph {et~al.}(2012)\citenamefont {Federico}, \citenamefont {Marrone}, \citenamefont {Colagrossi}, \citenamefont {Aristodemo},\ and\ \citenamefont {Antuono}}]{federicoSimulating2DOpenchannel2012b}%
  \BibitemOpen
  \bibfield  {author} {\bibinfo {author} {\bibfnamefont {I.}~\bibnamefont {Federico}}, \bibinfo {author} {\bibfnamefont {S.}~\bibnamefont {Marrone}}, \bibinfo {author} {\bibfnamefont {A.}~\bibnamefont {Colagrossi}}, \bibinfo {author} {\bibfnamefont {F.}~\bibnamefont {Aristodemo}}, \ and\ \bibinfo {author} {\bibfnamefont {M.}~\bibnamefont {Antuono}},\ }\bibfield  {title} {\enquote {\bibinfo {title} {Simulating {{2D}} open-channel flows through an {{SPH}} model},}\ }\href {\doibase 10.1016/j.euromechflu.2012.02.002} {\bibfield  {journal} {\bibinfo  {journal} {European Journal of Mechanics - B/Fluids}\ }\textbf {\bibinfo {volume} {34}},\ \bibinfo {pages} {35--46} (\bibinfo {year} {2012})}\BibitemShut {NoStop}%
\bibitem [{\citenamefont {Tafuni}\ \emph {et~al.}(2018)\citenamefont {Tafuni}, \citenamefont {Dom{\'i}nguez}, \citenamefont {Vacondio},\ and\ \citenamefont {Crespo}}]{tafuniVersatileAlgorithmTreatment2018a}%
  \BibitemOpen
  \bibfield  {author} {\bibinfo {author} {\bibfnamefont {A.}~\bibnamefont {Tafuni}}, \bibinfo {author} {\bibfnamefont {J.~M.}\ \bibnamefont {Dom{\'i}nguez}}, \bibinfo {author} {\bibfnamefont {R.}~\bibnamefont {Vacondio}}, \ and\ \bibinfo {author} {\bibfnamefont {A.~J.~C.}\ \bibnamefont {Crespo}},\ }\bibfield  {title} {\enquote {\bibinfo {title} {A versatile algorithm for the treatment of open boundary conditions in {{Smoothed}} particle hydrodynamics {{GPU}} models},}\ }\href {\doibase 10.1016/j.cma.2018.08.004} {\bibfield  {journal} {\bibinfo  {journal} {Computer Methods in Applied Mechanics and Engineering}\ }\textbf {\bibinfo {volume} {342}},\ \bibinfo {pages} {604--624} (\bibinfo {year} {2018})}\BibitemShut {NoStop}%
\bibitem [{\citenamefont {Negi}\ \emph {et~al.}(2020)\citenamefont {Negi}, \citenamefont {Ramachandran},\ and\ \citenamefont {Haftu}}]{negiImprovedNonreflectingOutlet2020}%
  \BibitemOpen
  \bibfield  {author} {\bibinfo {author} {\bibfnamefont {P.}~\bibnamefont {Negi}}, \bibinfo {author} {\bibfnamefont {P.}~\bibnamefont {Ramachandran}}, \ and\ \bibinfo {author} {\bibfnamefont {A.}~\bibnamefont {Haftu}},\ }\bibfield  {title} {\enquote {\bibinfo {title} {An improved non-reflecting outlet boundary condition for weakly-compressible {{SPH}}},}\ }\href {\doibase 10.1016/j.cma.2020.113119} {\bibfield  {journal} {\bibinfo  {journal} {Computer Methods in Applied Mechanics and Engineering}\ }\textbf {\bibinfo {volume} {367}},\ \bibinfo {pages} {113119} (\bibinfo {year} {2020})}\BibitemShut {NoStop}%
\bibitem [{\citenamefont {Holmes}\ and\ \citenamefont {Pivonka}(2021)}]{holmesNovelPressureInlet2021}%
  \BibitemOpen
  \bibfield  {author} {\bibinfo {author} {\bibfnamefont {D.~W.}\ \bibnamefont {Holmes}}\ and\ \bibinfo {author} {\bibfnamefont {P.}~\bibnamefont {Pivonka}},\ }\bibfield  {title} {\enquote {\bibinfo {title} {Novel pressure inlet and outlet boundary conditions for {{Smoothed Particle Hydrodynamics}}, applied to real problems in porous media flow},}\ }\href {\doibase 10.1016/j.jcp.2020.110029} {\bibfield  {journal} {\bibinfo  {journal} {Journal of Computational Physics}\ }\textbf {\bibinfo {volume} {429}},\ \bibinfo {pages} {110029} (\bibinfo {year} {2021})}\BibitemShut {NoStop}%
\bibitem [{\citenamefont {Zhang}\ \emph {et~al.}(2023)\citenamefont {Zhang}, \citenamefont {Zhang}, \citenamefont {Zhang},\ and\ \citenamefont {Hu}}]{zhangLagrangianFreestreamBoundary2023}%
  \BibitemOpen
  \bibfield  {author} {\bibinfo {author} {\bibfnamefont {S.}~\bibnamefont {Zhang}}, \bibinfo {author} {\bibfnamefont {W.}~\bibnamefont {Zhang}}, \bibinfo {author} {\bibfnamefont {C.}~\bibnamefont {Zhang}}, \ and\ \bibinfo {author} {\bibfnamefont {X.}~\bibnamefont {Hu}},\ }\bibfield  {title} {\enquote {\bibinfo {title} {A {{Lagrangian}} free-stream boundary condition for weakly compressible smoothed particle hydrodynamics},}\ }\href {\doibase 10.1016/j.jcp.2023.112303} {\bibfield  {journal} {\bibinfo  {journal} {Journal of Computational Physics}\ }\textbf {\bibinfo {volume} {490}},\ \bibinfo {pages} {112303} (\bibinfo {year} {2023})}\BibitemShut {NoStop}%
\bibitem [{\citenamefont {Ferrand}\ \emph {et~al.}(2017)\citenamefont {Ferrand}, \citenamefont {Joly}, \citenamefont {Kassiotis}, \citenamefont {Violeau}, \citenamefont {Leroy}, \citenamefont {Morel},\ and\ \citenamefont {Rogers}}]{ferrandUnsteadyOpenBoundaries2017a}%
  \BibitemOpen
  \bibfield  {author} {\bibinfo {author} {\bibfnamefont {M.}~\bibnamefont {Ferrand}}, \bibinfo {author} {\bibfnamefont {A.}~\bibnamefont {Joly}}, \bibinfo {author} {\bibfnamefont {C.}~\bibnamefont {Kassiotis}}, \bibinfo {author} {\bibfnamefont {D.}~\bibnamefont {Violeau}}, \bibinfo {author} {\bibfnamefont {A.}~\bibnamefont {Leroy}}, \bibinfo {author} {\bibfnamefont {F.-X.}\ \bibnamefont {Morel}}, \ and\ \bibinfo {author} {\bibfnamefont {B.~D.}\ \bibnamefont {Rogers}},\ }\bibfield  {title} {\enquote {\bibinfo {title} {Unsteady open boundaries for {{SPH}} using semi-analytical conditions and {{Riemann}} solver in {{2D}}},}\ }\href {\doibase 10.1016/j.cpc.2016.09.009} {\bibfield  {journal} {\bibinfo  {journal} {Computer Physics Communications}\ }\textbf {\bibinfo {volume} {210}},\ \bibinfo {pages} {29--44} (\bibinfo {year} {2017})}\BibitemShut {NoStop}%
\bibitem [{\citenamefont {Werdelmann}\ \emph {et~al.}(2021)\citenamefont {Werdelmann}, \citenamefont {Koch}, \citenamefont {Krebs},\ and\ \citenamefont {Bauer}}]{werdelmannApproachPermeableBoundary2021}%
  \BibitemOpen
  \bibfield  {author} {\bibinfo {author} {\bibfnamefont {B.}~\bibnamefont {Werdelmann}}, \bibinfo {author} {\bibfnamefont {R.}~\bibnamefont {Koch}}, \bibinfo {author} {\bibfnamefont {W.}~\bibnamefont {Krebs}}, \ and\ \bibinfo {author} {\bibfnamefont {H.-J.}\ \bibnamefont {Bauer}},\ }\bibfield  {title} {\enquote {\bibinfo {title} {An approach for permeable boundary conditions in {{SPH}}},}\ }\href {\doibase 10.1016/j.jcp.2021.110562} {\bibfield  {journal} {\bibinfo  {journal} {Journal of Computational Physics}\ }\textbf {\bibinfo {volume} {444}},\ \bibinfo {pages} {110562} (\bibinfo {year} {2021})}\BibitemShut {NoStop}%
\bibitem [{\citenamefont {Monaghan}(1994)}]{monaghanSimulatingFreeSurface1994}%
  \BibitemOpen
  \bibfield  {author} {\bibinfo {author} {\bibfnamefont {J.}~\bibnamefont {Monaghan}},\ }\bibfield  {title} {\enquote {\bibinfo {title} {Simulating {{Free Surface Flows}} with {{SPH}}},}\ }\href {\doibase 10.1006/jcph.1994.1034} {\bibfield  {journal} {\bibinfo  {journal} {Journal of Computational Physics}\ }\textbf {\bibinfo {volume} {110}},\ \bibinfo {pages} {399--406} (\bibinfo {year} {1994})}\BibitemShut {NoStop}%
\bibitem [{\citenamefont {Monaghan}\ and\ \citenamefont {Kos}(1999)}]{monaghanSolitaryWavesCretan1999}%
  \BibitemOpen
  \bibfield  {author} {\bibinfo {author} {\bibfnamefont {J.~J.}\ \bibnamefont {Monaghan}}\ and\ \bibinfo {author} {\bibfnamefont {A.}~\bibnamefont {Kos}},\ }\bibfield  {title} {\enquote {\bibinfo {title} {Solitary {{Waves}} on a {{Cretan Beach}}},}\ }\href {\doibase 10.1061/(ASCE)0733-950X(1999)125:3(145)} {\bibfield  {journal} {\bibinfo  {journal} {Journal of Waterway, Port, Coastal, and Ocean Engineering}\ }\textbf {\bibinfo {volume} {125}},\ \bibinfo {pages} {145--155} (\bibinfo {year} {1999})}\BibitemShut {NoStop}%
\bibitem [{\citenamefont {Monaghan}\ and\ \citenamefont {Kajtar}(2009)}]{monaghanSPHParticleBoundary2009}%
  \BibitemOpen
  \bibfield  {author} {\bibinfo {author} {\bibfnamefont {J.}~\bibnamefont {Monaghan}}\ and\ \bibinfo {author} {\bibfnamefont {J.}~\bibnamefont {Kajtar}},\ }\bibfield  {title} {\enquote {\bibinfo {title} {{{SPH}} particle boundary forces for arbitrary boundaries},}\ }\href {\doibase 10.1016/j.cpc.2009.05.008} {\bibfield  {journal} {\bibinfo  {journal} {Computer Physics Communications}\ }\textbf {\bibinfo {volume} {180}},\ \bibinfo {pages} {1811--1820} (\bibinfo {year} {2009})}\BibitemShut {NoStop}%
\bibitem [{\citenamefont {Campbell}(1989)}]{campbellNewAlgorithmsBoundary1989}%
  \BibitemOpen
  \bibfield  {author} {\bibinfo {author} {\bibfnamefont {P.~M.}\ \bibnamefont {Campbell}},\ }\href {https://ui.adsabs.harvard.edu/abs/1989mrc..reptR....C} {\enquote {\bibinfo {title} {Some {{New Algorithms}} for {{Boundary Value Problems}} in {{Smooth Particle Hydrodynamics}}},}\ }\bibinfo {type} {Tech. Rep.}\ \bibinfo {number} {DNA-TR-88-286}\ (\bibinfo  {institution} {{Mission Research Corporation}},\ \bibinfo {address} {{1720 Randolph Road, SE Albuquerque, NM 87106-4245}},\ \bibinfo {year} {1989})\BibitemShut {NoStop}%
\bibitem [{\citenamefont {Marongiu}\ \emph {et~al.}(2007)\citenamefont {Marongiu}, \citenamefont {Leboeuf},\ and\ \citenamefont {Parkinson}}]{marongiuNumericalSimulationFlow2007}%
  \BibitemOpen
  \bibfield  {author} {\bibinfo {author} {\bibfnamefont {J.~C.}\ \bibnamefont {Marongiu}}, \bibinfo {author} {\bibfnamefont {F.}~\bibnamefont {Leboeuf}}, \ and\ \bibinfo {author} {\bibfnamefont {E.}~\bibnamefont {Parkinson}},\ }\bibfield  {title} {\enquote {\bibinfo {title} {Numerical simulation of the flow in a {{Pelton}} turbine using the meshless method smoothed particle hydrodynamics: {{A}} new simple solid boundary treatment},}\ }\href {\doibase 10.1243/09576509JPE465} {\bibfield  {journal} {\bibinfo  {journal} {Proceedings of the Institution of Mechanical Engineers, Part A: Journal of Power and Energy}\ }\textbf {\bibinfo {volume} {221}},\ \bibinfo {pages} {849--856} (\bibinfo {year} {2007})}\BibitemShut {NoStop}%
\bibitem [{\citenamefont {Hashemi}\ \emph {et~al.}(2012)\citenamefont {Hashemi}, \citenamefont {Fatehi},\ and\ \citenamefont {Manzari}}]{hashemiModifiedSPHMethod2012}%
  \BibitemOpen
  \bibfield  {author} {\bibinfo {author} {\bibfnamefont {M.~R.}\ \bibnamefont {Hashemi}}, \bibinfo {author} {\bibfnamefont {R.}~\bibnamefont {Fatehi}}, \ and\ \bibinfo {author} {\bibfnamefont {M.~T.}\ \bibnamefont {Manzari}},\ }\bibfield  {title} {\enquote {\bibinfo {title} {A modified {{SPH}} method for simulating motion of rigid bodies in {{Newtonian}} fluid flows},}\ }\href {\doibase 10.1016/j.ijnonlinmec.2011.10.007} {\bibfield  {journal} {\bibinfo  {journal} {International Journal of Non-Linear Mechanics}\ }\textbf {\bibinfo {volume} {47}},\ \bibinfo {pages} {626--638} (\bibinfo {year} {2012})}\BibitemShut {NoStop}%
\bibitem [{\citenamefont {Kulasegaram}\ \emph {et~al.}(2004)\citenamefont {Kulasegaram}, \citenamefont {Bonet}, \citenamefont {Lewis},\ and\ \citenamefont {Profit}}]{kulasegaramVariationalFormulationBased2004}%
  \BibitemOpen
  \bibfield  {author} {\bibinfo {author} {\bibfnamefont {S.}~\bibnamefont {Kulasegaram}}, \bibinfo {author} {\bibfnamefont {J.}~\bibnamefont {Bonet}}, \bibinfo {author} {\bibfnamefont {R.~W.}\ \bibnamefont {Lewis}}, \ and\ \bibinfo {author} {\bibfnamefont {M.}~\bibnamefont {Profit}},\ }\bibfield  {title} {\enquote {\bibinfo {title} {A variational formulation based contact algorithm for rigid boundaries in two-dimensional {{SPH}} applications},}\ }\href {\doibase 10.1007/s00466-003-0534-0} {\bibfield  {journal} {\bibinfo  {journal} {Computational Mechanics}\ }\textbf {\bibinfo {volume} {33}},\ \bibinfo {pages} {316--325} (\bibinfo {year} {2004})}\BibitemShut {NoStop}%
\bibitem [{\citenamefont {Feldman}\ and\ \citenamefont {Bonet}(2007)}]{feldmanDynamicRefinementBoundary2007}%
  \BibitemOpen
  \bibfield  {author} {\bibinfo {author} {\bibfnamefont {J.}~\bibnamefont {Feldman}}\ and\ \bibinfo {author} {\bibfnamefont {J.}~\bibnamefont {Bonet}},\ }\bibfield  {title} {\enquote {\bibinfo {title} {Dynamic refinement and boundary contact forces in {{SPH}} with applications in fluid flow problems},}\ }\href {\doibase 10.1002/nme.2010} {\bibfield  {journal} {\bibinfo  {journal} {International Journal for Numerical Methods in Engineering}\ }\textbf {\bibinfo {volume} {72}},\ \bibinfo {pages} {295--324} (\bibinfo {year} {2007})}\BibitemShut {NoStop}%
\bibitem [{\citenamefont {Ferrand}\ \emph {et~al.}(2013)\citenamefont {Ferrand}, \citenamefont {Laurence}, \citenamefont {Rogers}, \citenamefont {Violeau},\ and\ \citenamefont {Kassiotis}}]{ferrandUnifiedSemianalyticalWall2013}%
  \BibitemOpen
  \bibfield  {author} {\bibinfo {author} {\bibfnamefont {M.}~\bibnamefont {Ferrand}}, \bibinfo {author} {\bibfnamefont {D.~R.}\ \bibnamefont {Laurence}}, \bibinfo {author} {\bibfnamefont {B.~D.}\ \bibnamefont {Rogers}}, \bibinfo {author} {\bibfnamefont {D.}~\bibnamefont {Violeau}}, \ and\ \bibinfo {author} {\bibfnamefont {C.}~\bibnamefont {Kassiotis}},\ }\bibfield  {title} {\enquote {\bibinfo {title} {Unified semi-analytical wall boundary conditions for inviscid, laminar or turbulent flows in the meshless {{SPH}} method},}\ }\href {\doibase 10.1002/fld.3666} {\bibfield  {journal} {\bibinfo  {journal} {International Journal for Numerical Methods in Fluids}\ }\textbf {\bibinfo {volume} {71}},\ \bibinfo {pages} {446--472} (\bibinfo {year} {2013})}\BibitemShut {NoStop}%
\bibitem [{\citenamefont {Mayrhofer}\ \emph {et~al.}(2013)\citenamefont {Mayrhofer}, \citenamefont {Rogers}, \citenamefont {Violeau},\ and\ \citenamefont {Ferrand}}]{mayrhoferInvestigationWallBounded2013}%
  \BibitemOpen
  \bibfield  {author} {\bibinfo {author} {\bibfnamefont {A.}~\bibnamefont {Mayrhofer}}, \bibinfo {author} {\bibfnamefont {B.~D.}\ \bibnamefont {Rogers}}, \bibinfo {author} {\bibfnamefont {D.}~\bibnamefont {Violeau}}, \ and\ \bibinfo {author} {\bibfnamefont {M.}~\bibnamefont {Ferrand}},\ }\bibfield  {title} {\enquote {\bibinfo {title} {Investigation of wall bounded flows using {{SPH}} and the unified semi-analytical wall boundary conditions},}\ }\href {\doibase 10.1016/j.cpc.2013.07.004} {\bibfield  {journal} {\bibinfo  {journal} {Computer Physics Communications}\ }\textbf {\bibinfo {volume} {184}},\ \bibinfo {pages} {2515--2527} (\bibinfo {year} {2013})}\BibitemShut {NoStop}%
\bibitem [{\citenamefont {Leroy}\ \emph {et~al.}(2014)\citenamefont {Leroy}, \citenamefont {Violeau}, \citenamefont {Ferrand},\ and\ \citenamefont {Kassiotis}}]{leroyUnifiedSemianalyticalWall2014}%
  \BibitemOpen
  \bibfield  {author} {\bibinfo {author} {\bibfnamefont {A.}~\bibnamefont {Leroy}}, \bibinfo {author} {\bibfnamefont {D.}~\bibnamefont {Violeau}}, \bibinfo {author} {\bibfnamefont {M.}~\bibnamefont {Ferrand}}, \ and\ \bibinfo {author} {\bibfnamefont {C.}~\bibnamefont {Kassiotis}},\ }\bibfield  {title} {\enquote {\bibinfo {title} {Unified semi-analytical wall boundary conditions applied to 2-{{D}} incompressible {{SPH}}},}\ }\href {\doibase 10.1016/j.jcp.2013.12.035} {\bibfield  {journal} {\bibinfo  {journal} {Journal of Computational Physics}\ }\textbf {\bibinfo {volume} {261}},\ \bibinfo {pages} {106--129} (\bibinfo {year} {2014})}\BibitemShut {NoStop}%
\bibitem [{\citenamefont {Mayrhofer}\ \emph {et~al.}(2015)\citenamefont {Mayrhofer}, \citenamefont {Ferrand}, \citenamefont {Kassiotis}, \citenamefont {Violeau},\ and\ \citenamefont {Morel}}]{mayrhoferUnifiedSemianalyticalWall2015}%
  \BibitemOpen
  \bibfield  {author} {\bibinfo {author} {\bibfnamefont {A.}~\bibnamefont {Mayrhofer}}, \bibinfo {author} {\bibfnamefont {M.}~\bibnamefont {Ferrand}}, \bibinfo {author} {\bibfnamefont {C.}~\bibnamefont {Kassiotis}}, \bibinfo {author} {\bibfnamefont {D.}~\bibnamefont {Violeau}}, \ and\ \bibinfo {author} {\bibfnamefont {F.-X.}\ \bibnamefont {Morel}},\ }\bibfield  {title} {\enquote {\bibinfo {title} {Unified semi-analytical wall boundary conditions in {{SPH}}: Analytical extension to 3-{{D}}},}\ }\href {\doibase 10.1007/s11075-014-9835-y} {\bibfield  {journal} {\bibinfo  {journal} {Numerical Algorithms}\ }\textbf {\bibinfo {volume} {68}},\ \bibinfo {pages} {15--34} (\bibinfo {year} {2015})}\BibitemShut {NoStop}%
\bibitem [{\citenamefont {Chiron}\ \emph {et~al.}(2019)\citenamefont {Chiron}, \citenamefont {De~Leffe}, \citenamefont {Oger},\ and\ \citenamefont {Le~Touz{\'e}}}]{chironFastAccurateSPH2019}%
  \BibitemOpen
  \bibfield  {author} {\bibinfo {author} {\bibfnamefont {L.}~\bibnamefont {Chiron}}, \bibinfo {author} {\bibfnamefont {M.}~\bibnamefont {De~Leffe}}, \bibinfo {author} {\bibfnamefont {G.}~\bibnamefont {Oger}}, \ and\ \bibinfo {author} {\bibfnamefont {D.}~\bibnamefont {Le~Touz{\'e}}},\ }\bibfield  {title} {\enquote {\bibinfo {title} {Fast and accurate {{SPH}} modelling of {{3D}} complex wall boundaries in viscous and non viscous flows},}\ }\href {\doibase 10.1016/j.cpc.2018.08.001} {\bibfield  {journal} {\bibinfo  {journal} {Computer Physics Communications}\ }\textbf {\bibinfo {volume} {234}},\ \bibinfo {pages} {93--111} (\bibinfo {year} {2019})}\BibitemShut {NoStop}%
\bibitem [{\citenamefont {Boregowda}\ and\ \citenamefont {Liu}(2023)}]{boregowdaInsightsUsingBoundary2023a}%
  \BibitemOpen
  \bibfield  {author} {\bibinfo {author} {\bibfnamefont {P.}~\bibnamefont {Boregowda}}\ and\ \bibinfo {author} {\bibfnamefont {G.-R.}\ \bibnamefont {Liu}},\ }\bibfield  {title} {\enquote {\bibinfo {title} {Insights on using the boundary integral {{SPH}} formulations to calculate {{Laplacians}} with {{Dirichlet}} boundaries},}\ }\href {\doibase 10.1016/j.enganabound.2023.07.011} {\bibfield  {journal} {\bibinfo  {journal} {Engineering Analysis with Boundary Elements}\ }\textbf {\bibinfo {volume} {155}},\ \bibinfo {pages} {652--667} (\bibinfo {year} {2023})}\BibitemShut {NoStop}%
\bibitem [{\citenamefont {Leroy}\ \emph {et~al.}(2016)\citenamefont {Leroy}, \citenamefont {Violeau}, \citenamefont {Ferrand}, \citenamefont {Fratter},\ and\ \citenamefont {Joly}}]{leroyNewOpenBoundary2016}%
  \BibitemOpen
  \bibfield  {author} {\bibinfo {author} {\bibfnamefont {A.}~\bibnamefont {Leroy}}, \bibinfo {author} {\bibfnamefont {D.}~\bibnamefont {Violeau}}, \bibinfo {author} {\bibfnamefont {M.}~\bibnamefont {Ferrand}}, \bibinfo {author} {\bibfnamefont {L.}~\bibnamefont {Fratter}}, \ and\ \bibinfo {author} {\bibfnamefont {A.}~\bibnamefont {Joly}},\ }\bibfield  {title} {\enquote {\bibinfo {title} {A new open boundary formulation for incompressible {{SPH}}},}\ }\href {\doibase 10.1016/j.camwa.2016.09.008} {\bibfield  {journal} {\bibinfo  {journal} {Computers \& Mathematics with Applications}\ }\textbf {\bibinfo {volume} {72}},\ \bibinfo {pages} {2417--2432} (\bibinfo {year} {2016})}\BibitemShut {NoStop}%
\bibitem [{\citenamefont {Ferrari}\ \emph {et~al.}(2009)\citenamefont {Ferrari}, \citenamefont {Dumbser}, \citenamefont {Toro},\ and\ \citenamefont {Armanini}}]{ferrariNew3DParallel2009}%
  \BibitemOpen
  \bibfield  {author} {\bibinfo {author} {\bibfnamefont {A.}~\bibnamefont {Ferrari}}, \bibinfo {author} {\bibfnamefont {M.}~\bibnamefont {Dumbser}}, \bibinfo {author} {\bibfnamefont {E.~F.}\ \bibnamefont {Toro}}, \ and\ \bibinfo {author} {\bibfnamefont {A.}~\bibnamefont {Armanini}},\ }\bibfield  {title} {\enquote {\bibinfo {title} {A new {{3D}} parallel {{SPH}} scheme for free surface flows},}\ }\href {\doibase 10.1016/j.compfluid.2008.11.012} {\bibfield  {journal} {\bibinfo  {journal} {Computers \& Fluids}\ }\textbf {\bibinfo {volume} {38}},\ \bibinfo {pages} {1203--1217} (\bibinfo {year} {2009})}\BibitemShut {NoStop}%
\bibitem [{\citenamefont {Vacondio}\ \emph {et~al.}(2012)\citenamefont {Vacondio}, \citenamefont {Rogers},\ and\ \citenamefont {Stansby}}]{vacondioSmoothedParticleHydrodynamics2012}%
  \BibitemOpen
  \bibfield  {author} {\bibinfo {author} {\bibfnamefont {R.}~\bibnamefont {Vacondio}}, \bibinfo {author} {\bibfnamefont {B.~D.}\ \bibnamefont {Rogers}}, \ and\ \bibinfo {author} {\bibfnamefont {P.~K.}\ \bibnamefont {Stansby}},\ }\bibfield  {title} {\enquote {\bibinfo {title} {{Smoothed Particle Hydrodynamics: Approximate zero-consistent 2-D boundary conditions and still shallow-water tests}},}\ }\href {\doibase 10.1002/fld.2559} {\bibfield  {journal} {\bibinfo  {journal} {International Journal for Numerical Methods in Fluids}\ }\textbf {\bibinfo {volume} {69}},\ \bibinfo {pages} {226--253} (\bibinfo {year} {2012})}\BibitemShut {NoStop}%
\bibitem [{\citenamefont {Fourtakas}\ \emph {et~al.}(2015)\citenamefont {Fourtakas}, \citenamefont {Vacondio},\ and\ \citenamefont {Rogers}}]{fourtakasApproximateZerothFirstorder2015}%
  \BibitemOpen
  \bibfield  {author} {\bibinfo {author} {\bibfnamefont {G.}~\bibnamefont {Fourtakas}}, \bibinfo {author} {\bibfnamefont {R.}~\bibnamefont {Vacondio}}, \ and\ \bibinfo {author} {\bibfnamefont {B.~D.}\ \bibnamefont {Rogers}},\ }\bibfield  {title} {\enquote {\bibinfo {title} {On the approximate zeroth and first-order consistency in the presence of 2-{{D}} irregular boundaries in {{SPH}} obtained by the virtual boundary particle methods},}\ }\href {\doibase 10.1002/fld.4026} {\bibfield  {journal} {\bibinfo  {journal} {International Journal for Numerical Methods in Fluids}\ }\textbf {\bibinfo {volume} {78}},\ \bibinfo {pages} {475--501} (\bibinfo {year} {2015})}\BibitemShut {NoStop}%
\bibitem [{\citenamefont {Fourtakas}\ \emph {et~al.}(2019)\citenamefont {Fourtakas}, \citenamefont {Dominguez}, \citenamefont {Vacondio},\ and\ \citenamefont {Rogers}}]{fourtakasLocalUniformStencil2019}%
  \BibitemOpen
  \bibfield  {author} {\bibinfo {author} {\bibfnamefont {G.}~\bibnamefont {Fourtakas}}, \bibinfo {author} {\bibfnamefont {J.~M.}\ \bibnamefont {Dominguez}}, \bibinfo {author} {\bibfnamefont {R.}~\bibnamefont {Vacondio}}, \ and\ \bibinfo {author} {\bibfnamefont {B.~D.}\ \bibnamefont {Rogers}},\ }\bibfield  {title} {\enquote {\bibinfo {title} {Local uniform stencil ({{LUST}}) boundary condition for arbitrary 3-{{D}} boundaries in parallel smoothed particle hydrodynamics ({{SPH}}) models},}\ }\href {\doibase 10.1016/j.compfluid.2019.06.009} {\bibfield  {journal} {\bibinfo  {journal} {Computers \& Fluids}\ }\textbf {\bibinfo {volume} {190}},\ \bibinfo {pages} {346--361} (\bibinfo {year} {2019})}\BibitemShut {NoStop}%
\bibitem [{\citenamefont {Crespo}\ \emph {et~al.}(2007)\citenamefont {Crespo}, \citenamefont {{G{\'o}mez-Gesteira}},\ and\ \citenamefont {Dalrymple}}]{crespoBoundaryConditionsGenerated2007b}%
  \BibitemOpen
  \bibfield  {author} {\bibinfo {author} {\bibfnamefont {A.~J.~C.}\ \bibnamefont {Crespo}}, \bibinfo {author} {\bibfnamefont {M.}~\bibnamefont {{G{\'o}mez-Gesteira}}}, \ and\ \bibinfo {author} {\bibfnamefont {R.~A.}\ \bibnamefont {Dalrymple}},\ }\bibfield  {title} {\enquote {\bibinfo {title} {Boundary {{Conditions Generated}} by {{Dynamic Particles}} in {{SPH Methods}}},}\ }\href {\doibase 10.3970/cmc.2007.005.173} {\bibfield  {journal} {\bibinfo  {journal} {Computers, Materials \& Continua}\ }\textbf {\bibinfo {volume} {5}},\ \bibinfo {pages} {173--184} (\bibinfo {year} {2007})}\BibitemShut {NoStop}%
\bibitem [{\citenamefont {Ren}\ \emph {et~al.}(2015)\citenamefont {Ren}, \citenamefont {He}, \citenamefont {Dong},\ and\ \citenamefont {Wen}}]{renNonlinearSimulationsWaveinduced2015}%
  \BibitemOpen
  \bibfield  {author} {\bibinfo {author} {\bibfnamefont {B.}~\bibnamefont {Ren}}, \bibinfo {author} {\bibfnamefont {M.}~\bibnamefont {He}}, \bibinfo {author} {\bibfnamefont {P.}~\bibnamefont {Dong}}, \ and\ \bibinfo {author} {\bibfnamefont {H.}~\bibnamefont {Wen}},\ }\bibfield  {title} {\enquote {\bibinfo {title} {Nonlinear simulations of wave-induced motions of a freely floating body using {{WCSPH}} method},}\ }\href {\doibase 10.1016/j.apor.2014.12.003} {\bibfield  {journal} {\bibinfo  {journal} {Applied Ocean Research}\ }\textbf {\bibinfo {volume} {50}},\ \bibinfo {pages} {1--12} (\bibinfo {year} {2015})}\BibitemShut {NoStop}%
\bibitem [{\citenamefont {Akinci}\ \emph {et~al.}(2012)\citenamefont {Akinci}, \citenamefont {Ihmsen}, \citenamefont {Akinci}, \citenamefont {Solenthaler},\ and\ \citenamefont {Teschner}}]{akinciVersatileRigidfluidCoupling2012}%
  \BibitemOpen
  \bibfield  {author} {\bibinfo {author} {\bibfnamefont {N.}~\bibnamefont {Akinci}}, \bibinfo {author} {\bibfnamefont {M.}~\bibnamefont {Ihmsen}}, \bibinfo {author} {\bibfnamefont {G.}~\bibnamefont {Akinci}}, \bibinfo {author} {\bibfnamefont {B.}~\bibnamefont {Solenthaler}}, \ and\ \bibinfo {author} {\bibfnamefont {M.}~\bibnamefont {Teschner}},\ }\bibfield  {title} {\enquote {\bibinfo {title} {Versatile rigid-fluid coupling for incompressible {{SPH}}},}\ }\href {\doibase 10.1145/2185520.2185558} {\bibfield  {journal} {\bibinfo  {journal} {ACM Transactions on Graphics}\ }\textbf {\bibinfo {volume} {31}},\ \bibinfo {pages} {1--8} (\bibinfo {year} {2012})}\BibitemShut {NoStop}%
\bibitem [{\citenamefont {Liu}\ \emph {et~al.}(2014)\citenamefont {Liu}, \citenamefont {Lin},\ and\ \citenamefont {Shao}}]{liuISPHSimulationCoupled2014}%
  \BibitemOpen
  \bibfield  {author} {\bibinfo {author} {\bibfnamefont {X.}~\bibnamefont {Liu}}, \bibinfo {author} {\bibfnamefont {P.}~\bibnamefont {Lin}}, \ and\ \bibinfo {author} {\bibfnamefont {S.}~\bibnamefont {Shao}},\ }\bibfield  {title} {\enquote {\bibinfo {title} {An {{ISPH}} simulation of coupled structure interaction with free surface flows},}\ }\href {\doibase 10.1016/j.jfluidstructs.2014.02.002} {\bibfield  {journal} {\bibinfo  {journal} {Journal of Fluids and Structures}\ }\textbf {\bibinfo {volume} {48}},\ \bibinfo {pages} {46--61} (\bibinfo {year} {2014})}\BibitemShut {NoStop}%
\bibitem [{\citenamefont {Li}\ \emph {et~al.}(2021)\citenamefont {Li}, \citenamefont {Zhang},\ and\ \citenamefont {Yuan}}]{liImprovedDynamicBoundary2021}%
  \BibitemOpen
  \bibfield  {author} {\bibinfo {author} {\bibfnamefont {X.}~\bibnamefont {Li}}, \bibinfo {author} {\bibfnamefont {H.}~\bibnamefont {Zhang}}, \ and\ \bibinfo {author} {\bibfnamefont {D.}~\bibnamefont {Yuan}},\ }\bibfield  {title} {\enquote {\bibinfo {title} {An {{Improved Dynamic Boundary Condition}} in {{SPH Method}}},}\ }\href {\doibase 10.5755/j02.mech.28674} {\bibfield  {journal} {\bibinfo  {journal} {Mechanics}\ }\textbf {\bibinfo {volume} {27}},\ \bibinfo {pages} {465--474} (\bibinfo {year} {2021})}\BibitemShut {NoStop}%
\bibitem [{\citenamefont {English}\ \emph {et~al.}(2022)\citenamefont {English}, \citenamefont {Dom{\'i}nguez}, \citenamefont {Vacondio}, \citenamefont {Crespo}, \citenamefont {Stansby}, \citenamefont {Lind}, \citenamefont {Chiapponi},\ and\ \citenamefont {{G{\'o}mez-Gesteira}}}]{englishModifiedDynamicBoundary2022}%
  \BibitemOpen
  \bibfield  {author} {\bibinfo {author} {\bibfnamefont {A.}~\bibnamefont {English}}, \bibinfo {author} {\bibfnamefont {J.~M.}\ \bibnamefont {Dom{\'i}nguez}}, \bibinfo {author} {\bibfnamefont {R.}~\bibnamefont {Vacondio}}, \bibinfo {author} {\bibfnamefont {A.~J.~C.}\ \bibnamefont {Crespo}}, \bibinfo {author} {\bibfnamefont {P.~K.}\ \bibnamefont {Stansby}}, \bibinfo {author} {\bibfnamefont {S.~J.}\ \bibnamefont {Lind}}, \bibinfo {author} {\bibfnamefont {L.}~\bibnamefont {Chiapponi}}, \ and\ \bibinfo {author} {\bibfnamefont {M.}~\bibnamefont {{G{\'o}mez-Gesteira}}},\ }\bibfield  {title} {\enquote {\bibinfo {title} {Modified dynamic boundary conditions ({{mDBC}}) for general-purpose smoothed particle hydrodynamics ({{SPH}}): Application to tank sloshing, dam break and fish pass problems},}\ }\href {\doibase 10.1007/s40571-021-00403-3} {\bibfield  {journal} {\bibinfo  {journal} {Computational Particle Mechanics}\ }\textbf {\bibinfo {volume} {9}},\ \bibinfo {pages} {1--15} (\bibinfo {year} {2022})}\BibitemShut {NoStop}%
\bibitem [{\citenamefont {Sikarudi}\ and\ \citenamefont {Nikseresht}(2016)}]{sikarudiNeumannRobinBoundary2016}%
  \BibitemOpen
  \bibfield  {author} {\bibinfo {author} {\bibfnamefont {M.~E.}\ \bibnamefont {Sikarudi}}\ and\ \bibinfo {author} {\bibfnamefont {A.}~\bibnamefont {Nikseresht}},\ }\bibfield  {title} {\enquote {\bibinfo {title} {Neumann and {{Robin}} boundary conditions for heat conduction modeling using smoothed particle hydrodynamics},}\ }\href {\doibase 10.1016/j.cpc.2015.07.004} {\bibfield  {journal} {\bibinfo  {journal} {Computer Physics Communications}\ }\textbf {\bibinfo {volume} {198}},\ \bibinfo {pages} {1--11} (\bibinfo {year} {2016})}\BibitemShut {NoStop}%
\bibitem [{\citenamefont {Wang}\ \emph {et~al.}(2019)\citenamefont {Wang}, \citenamefont {Hu}, \citenamefont {Zhang},\ and\ \citenamefont {Pan}}]{wangModelingHeatTransfer2019}%
  \BibitemOpen
  \bibfield  {author} {\bibinfo {author} {\bibfnamefont {J.}~\bibnamefont {Wang}}, \bibinfo {author} {\bibfnamefont {W.}~\bibnamefont {Hu}}, \bibinfo {author} {\bibfnamefont {X.}~\bibnamefont {Zhang}}, \ and\ \bibinfo {author} {\bibfnamefont {W.}~\bibnamefont {Pan}},\ }\bibfield  {title} {\enquote {\bibinfo {title} {Modeling heat transfer subject to inhomogeneous {{Neumann}} boundary conditions by smoothed particle hydrodynamics and peridynamics},}\ }\href {\doibase 10.1016/j.ijheatmasstransfer.2019.05.054} {\bibfield  {journal} {\bibinfo  {journal} {International Journal of Heat and Mass Transfer}\ }\textbf {\bibinfo {volume} {139}},\ \bibinfo {pages} {948--962} (\bibinfo {year} {2019})}\BibitemShut {NoStop}%
\bibitem [{\citenamefont {Valizadeh}\ and\ \citenamefont {Monaghan}(2015)}]{valizadehStudySolidWall2015}%
  \BibitemOpen
  \bibfield  {author} {\bibinfo {author} {\bibfnamefont {A.}~\bibnamefont {Valizadeh}}\ and\ \bibinfo {author} {\bibfnamefont {J.~J.}\ \bibnamefont {Monaghan}},\ }\bibfield  {title} {\enquote {\bibinfo {title} {A study of solid wall models for weakly compressible {{SPH}}},}\ }\href {\doibase 10.1016/j.jcp.2015.07.033} {\bibfield  {journal} {\bibinfo  {journal} {Journal of Computational Physics}\ }\textbf {\bibinfo {volume} {300}},\ \bibinfo {pages} {5--19} (\bibinfo {year} {2015})}\BibitemShut {NoStop}%
\bibitem [{\citenamefont {Randles}\ and\ \citenamefont {Libersky}(1996)}]{randlesSmoothedParticleHydrodynamics1996}%
  \BibitemOpen
  \bibfield  {author} {\bibinfo {author} {\bibfnamefont {P.~W.}\ \bibnamefont {Randles}}\ and\ \bibinfo {author} {\bibfnamefont {L.~D.}\ \bibnamefont {Libersky}},\ }\bibfield  {title} {\enquote {\bibinfo {title} {Smoothed {{Particle Hydrodynamics}}: {{Some}} recent improvements and applications},}\ }\href {\doibase 10.1016/S0045-7825(96)01090-0} {\bibfield  {journal} {\bibinfo  {journal} {Computer Methods in Applied Mechanics and Engineering}\ }\textbf {\bibinfo {volume} {139}},\ \bibinfo {pages} {375--408} (\bibinfo {year} {1996})}\BibitemShut {NoStop}%
\bibitem [{\citenamefont {Antuono}\ \emph {et~al.}(2010)\citenamefont {Antuono}, \citenamefont {Colagrossi}, \citenamefont {Marrone},\ and\ \citenamefont {Molteni}}]{antuonoFreesurfaceFlowsSolved2010}%
  \BibitemOpen
  \bibfield  {author} {\bibinfo {author} {\bibfnamefont {M.}~\bibnamefont {Antuono}}, \bibinfo {author} {\bibfnamefont {A.}~\bibnamefont {Colagrossi}}, \bibinfo {author} {\bibfnamefont {S.}~\bibnamefont {Marrone}}, \ and\ \bibinfo {author} {\bibfnamefont {D.}~\bibnamefont {Molteni}},\ }\bibfield  {title} {\enquote {\bibinfo {title} {Free-surface flows solved by means of {{SPH}} schemes with numerical diffusive terms},}\ }\href {\doibase 10.1016/j.cpc.2009.11.002} {\bibfield  {journal} {\bibinfo  {journal} {Computer Physics Communications}\ }\textbf {\bibinfo {volume} {181}},\ \bibinfo {pages} {532--549} (\bibinfo {year} {2010})}\BibitemShut {NoStop}%
\bibitem [{\citenamefont {Price}(2012)}]{priceSmoothedParticleHydrodynamics2012}%
  \BibitemOpen
  \bibfield  {author} {\bibinfo {author} {\bibfnamefont {D.~J.}\ \bibnamefont {Price}},\ }\bibfield  {title} {\enquote {\bibinfo {title} {Smoothed particle hydrodynamics and magnetohydrodynamics},}\ }\href {\doibase 10.1016/j.jcp.2010.12.011} {\bibfield  {journal} {\bibinfo  {journal} {Journal of Computational Physics}\ }\textbf {\bibinfo {volume} {231}},\ \bibinfo {pages} {759--794} (\bibinfo {year} {2012})}\BibitemShut {NoStop}%
\bibitem [{\citenamefont {Puri}\ and\ \citenamefont {Ramachandran}(2014)}]{puriComparisonSPHSchemes2014}%
  \BibitemOpen
  \bibfield  {author} {\bibinfo {author} {\bibfnamefont {K.}~\bibnamefont {Puri}}\ and\ \bibinfo {author} {\bibfnamefont {P.}~\bibnamefont {Ramachandran}},\ }\bibfield  {title} {\enquote {\bibinfo {title} {A comparison of {{SPH}} schemes for the compressible {{Euler}} equations},}\ }\href {\doibase 10.1016/j.jcp.2013.08.060} {\bibfield  {journal} {\bibinfo  {journal} {Journal of Computational Physics}\ }\textbf {\bibinfo {volume} {256}},\ \bibinfo {pages} {308--333} (\bibinfo {year} {2014})}\BibitemShut {NoStop}%
\bibitem [{\citenamefont {Sun}\ \emph {et~al.}(2019)\citenamefont {Sun}, \citenamefont {Colagrossi}, \citenamefont {Marrone}, \citenamefont {Antuono},\ and\ \citenamefont {Zhang}}]{sunConsistentApproachParticle2019}%
  \BibitemOpen
  \bibfield  {author} {\bibinfo {author} {\bibfnamefont {P.}~\bibnamefont {Sun}}, \bibinfo {author} {\bibfnamefont {A.}~\bibnamefont {Colagrossi}}, \bibinfo {author} {\bibfnamefont {S.}~\bibnamefont {Marrone}}, \bibinfo {author} {\bibfnamefont {M.}~\bibnamefont {Antuono}}, \ and\ \bibinfo {author} {\bibfnamefont {A.-M.}\ \bibnamefont {Zhang}},\ }\bibfield  {title} {\enquote {\bibinfo {title} {A consistent approach to particle shifting in the {$\delta$} - {{Plus}} -{{SPH}} model},}\ }\href {\doibase 10.1016/j.cma.2019.01.045} {\bibfield  {journal} {\bibinfo  {journal} {Computer Methods in Applied Mechanics and Engineering}\ }\textbf {\bibinfo {volume} {348}},\ \bibinfo {pages} {912--934} (\bibinfo {year} {2019})}\BibitemShut {NoStop}%
\bibitem [{\citenamefont {Adepu}\ and\ \citenamefont {Ramachandran}(2023)}]{adepuCorrectedTransportvelocityFormulation2023}%
  \BibitemOpen
  \bibfield  {author} {\bibinfo {author} {\bibfnamefont {D.}~\bibnamefont {Adepu}}\ and\ \bibinfo {author} {\bibfnamefont {P.}~\bibnamefont {Ramachandran}},\ }\bibfield  {title} {\enquote {\bibinfo {title} {A corrected transport-velocity formulation for fluid and structural mechanics with {{SPH}}},}\ }\href {\doibase 10.1007/s40571-023-00631-9} {\bibfield  {journal} {\bibinfo  {journal} {Computational Particle Mechanics}\ } (\bibinfo {year} {2023}),\ 10.1007/s40571-023-00631-9}\BibitemShut {NoStop}%
\bibitem [{\citenamefont {Morris}\ and\ \citenamefont {Monaghan}(1997)}]{morrisSwitchReduceSPH1997}%
  \BibitemOpen
  \bibfield  {author} {\bibinfo {author} {\bibfnamefont {J.~P.}\ \bibnamefont {Morris}}\ and\ \bibinfo {author} {\bibfnamefont {J.~J.}\ \bibnamefont {Monaghan}},\ }\bibfield  {title} {\enquote {\bibinfo {title} {A {{Switch}} to {{Reduce SPH Viscosity}}},}\ }\href {\doibase 10.1006/jcph.1997.5690} {\bibfield  {journal} {\bibinfo  {journal} {Journal of Computational Physics}\ }\textbf {\bibinfo {volume} {136}},\ \bibinfo {pages} {41--50} (\bibinfo {year} {1997})}\BibitemShut {NoStop}%
\bibitem [{\citenamefont {Cullen}\ and\ \citenamefont {Dehnen}(2010)}]{cullenInviscidSmoothedParticle2010}%
  \BibitemOpen
  \bibfield  {author} {\bibinfo {author} {\bibfnamefont {L.}~\bibnamefont {Cullen}}\ and\ \bibinfo {author} {\bibfnamefont {W.}~\bibnamefont {Dehnen}},\ }\bibfield  {title} {\enquote {\bibinfo {title} {Inviscid smoothed particle hydrodynamics: {{Inviscid}} smoothed particle hydrodynamics},}\ }\href {\doibase 10.1111/j.1365-2966.2010.17158.x} {\bibfield  {journal} {\bibinfo  {journal} {Monthly Notices of the Royal Astronomical Society}\ }\textbf {\bibinfo {volume} {408}},\ \bibinfo {pages} {669--683} (\bibinfo {year} {2010})}\BibitemShut {NoStop}%
\bibitem [{\citenamefont {Read}\ and\ \citenamefont {Hayfield}(2012)}]{readSPHSSmoothedParticle2012}%
  \BibitemOpen
  \bibfield  {author} {\bibinfo {author} {\bibfnamefont {J.~I.}\ \bibnamefont {Read}}\ and\ \bibinfo {author} {\bibfnamefont {T.}~\bibnamefont {Hayfield}},\ }\bibfield  {title} {\enquote {\bibinfo {title} {{{SPHS}}: Smoothed particle hydrodynamics with a higher order dissipation switch: {{SPH}} with a higher order dissipation switch},}\ }\href {\doibase 10.1111/j.1365-2966.2012.20819.x} {\bibfield  {journal} {\bibinfo  {journal} {Monthly Notices of the Royal Astronomical Society}\ }\textbf {\bibinfo {volume} {422}},\ \bibinfo {pages} {3037--3055} (\bibinfo {year} {2012})}\BibitemShut {NoStop}%
\bibitem [{\citenamefont {Rosswog}(2020)}]{rosswogSimpleEntropybasedDissipation2020}%
  \BibitemOpen
  \bibfield  {author} {\bibinfo {author} {\bibfnamefont {S.}~\bibnamefont {Rosswog}},\ }\bibfield  {title} {\enquote {\bibinfo {title} {A {{Simple}}, {{Entropy-based Dissipation Trigger}} for {{SPH}}},}\ }\href {\doibase 10.3847/1538-4357/ab9a2e} {\bibfield  {journal} {\bibinfo  {journal} {The Astrophysical Journal}\ }\textbf {\bibinfo {volume} {898}},\ \bibinfo {pages} {60} (\bibinfo {year} {2020})}\BibitemShut {NoStop}%
\bibitem [{\citenamefont {Khayyer}\ \emph {et~al.}(2019)\citenamefont {Khayyer}, \citenamefont {Gotoh},\ and\ \citenamefont {Shimizu}}]{khayyerProjectionbasedParticleMethod2019}%
  \BibitemOpen
  \bibfield  {author} {\bibinfo {author} {\bibfnamefont {A.}~\bibnamefont {Khayyer}}, \bibinfo {author} {\bibfnamefont {H.}~\bibnamefont {Gotoh}}, \ and\ \bibinfo {author} {\bibfnamefont {Y.}~\bibnamefont {Shimizu}},\ }\bibfield  {title} {\enquote {\bibinfo {title} {A projection-based particle method with optimized particle shifting for multiphase flows with large density ratios and discontinuous density fields},}\ }\href {\doibase 10.1016/j.compfluid.2018.10.018} {\bibfield  {journal} {\bibinfo  {journal} {Computers \& Fluids}\ }\textbf {\bibinfo {volume} {179}},\ \bibinfo {pages} {356--371} (\bibinfo {year} {2019})}\BibitemShut {NoStop}%
\bibitem [{\citenamefont {Rastelli}\ \emph {et~al.}(2022)\citenamefont {Rastelli}, \citenamefont {Vacondio}, \citenamefont {Marongiu}, \citenamefont {Fourtakas},\ and\ \citenamefont {Rogers}}]{rastelliImplicitIterativeParticle2022}%
  \BibitemOpen
  \bibfield  {author} {\bibinfo {author} {\bibfnamefont {P.}~\bibnamefont {Rastelli}}, \bibinfo {author} {\bibfnamefont {R.}~\bibnamefont {Vacondio}}, \bibinfo {author} {\bibfnamefont {J.}~\bibnamefont {Marongiu}}, \bibinfo {author} {\bibfnamefont {G.}~\bibnamefont {Fourtakas}}, \ and\ \bibinfo {author} {\bibfnamefont {B.~D.}\ \bibnamefont {Rogers}},\ }\bibfield  {title} {\enquote {\bibinfo {title} {Implicit iterative particle shifting for meshless numerical schemes using kernel basis functions},}\ }\href {\doibase 10.1016/j.cma.2022.114716} {\bibfield  {journal} {\bibinfo  {journal} {Computer Methods in Applied Mechanics and Engineering}\ }\textbf {\bibinfo {volume} {393}},\ \bibinfo {pages} {114716} (\bibinfo {year} {2022})}\BibitemShut {NoStop}%
\bibitem [{\citenamefont {Ramachandran}\ and\ \citenamefont {Puri}(2019)}]{ramachandranEntropicallyDampedArtificial2019}%
  \BibitemOpen
  \bibfield  {author} {\bibinfo {author} {\bibfnamefont {P.}~\bibnamefont {Ramachandran}}\ and\ \bibinfo {author} {\bibfnamefont {K.}~\bibnamefont {Puri}},\ }\bibfield  {title} {\enquote {\bibinfo {title} {Entropically damped artificial compressibility for {{SPH}}},}\ }\href {\doibase 10.1016/j.compfluid.2018.11.023} {\bibfield  {journal} {\bibinfo  {journal} {Computers \& Fluids}\ }\textbf {\bibinfo {volume} {179}},\ \bibinfo {pages} {579--594} (\bibinfo {year} {2019})}\BibitemShut {NoStop}%
\bibitem [{\citenamefont {Muta}\ and\ \citenamefont {Ramachandran}(2022)}]{mutaEfficientAccurateAdaptive2022}%
  \BibitemOpen
  \bibfield  {author} {\bibinfo {author} {\bibfnamefont {A.}~\bibnamefont {Muta}}\ and\ \bibinfo {author} {\bibfnamefont {P.}~\bibnamefont {Ramachandran}},\ }\bibfield  {title} {\enquote {\bibinfo {title} {Efficient and accurate adaptive resolution for weakly-compressible {{SPH}}},}\ }\href {\doibase 10.1016/j.cma.2022.115019} {\bibfield  {journal} {\bibinfo  {journal} {Computer Methods in Applied Mechanics and Engineering}\ }\textbf {\bibinfo {volume} {395}},\ \bibinfo {pages} {115019} (\bibinfo {year} {2022})}\BibitemShut {NoStop}%
\bibitem [{\citenamefont {Haftu}\ \emph {et~al.}(2022)\citenamefont {Haftu}, \citenamefont {Muta},\ and\ \citenamefont {Ramachandran}}]{haftuParallelAdaptiveWeaklycompressible2022}%
  \BibitemOpen
  \bibfield  {author} {\bibinfo {author} {\bibfnamefont {A.}~\bibnamefont {Haftu}}, \bibinfo {author} {\bibfnamefont {A.}~\bibnamefont {Muta}}, \ and\ \bibinfo {author} {\bibfnamefont {P.}~\bibnamefont {Ramachandran}},\ }\bibfield  {title} {\enquote {\bibinfo {title} {Parallel adaptive weakly-compressible {{SPH}} for complex moving geometries},}\ }\href {\doibase 10.1016/j.cpc.2022.108377} {\bibfield  {journal} {\bibinfo  {journal} {Computer Physics Communications}\ }\textbf {\bibinfo {volume} {277}},\ \bibinfo {pages} {108377} (\bibinfo {year} {2022})}\BibitemShut {NoStop}%
\bibitem [{\citenamefont {Liu}\ and\ \citenamefont {Liu}(2006)}]{liuRestoringParticleConsistency2006}%
  \BibitemOpen
  \bibfield  {author} {\bibinfo {author} {\bibfnamefont {M.}~\bibnamefont {Liu}}\ and\ \bibinfo {author} {\bibfnamefont {G.}~\bibnamefont {Liu}},\ }\bibfield  {title} {\enquote {\bibinfo {title} {Restoring particle consistency in smoothed particle hydrodynamics},}\ }\href {\doibase 10.1016/j.apnum.2005.02.012} {\bibfield  {journal} {\bibinfo  {journal} {Applied Numerical Mathematics}\ }\textbf {\bibinfo {volume} {56}},\ \bibinfo {pages} {19--36} (\bibinfo {year} {2006})}\BibitemShut {NoStop}%
\bibitem [{\citenamefont {Vacondio}\ \emph {et~al.}(2013)\citenamefont {Vacondio}, \citenamefont {Rogers}, \citenamefont {Stansby}, \citenamefont {Mignosa},\ and\ \citenamefont {Feldman}}]{vacondioVariableResolutionSPH2013}%
  \BibitemOpen
  \bibfield  {author} {\bibinfo {author} {\bibfnamefont {R.}~\bibnamefont {Vacondio}}, \bibinfo {author} {\bibfnamefont {B.}~\bibnamefont {Rogers}}, \bibinfo {author} {\bibfnamefont {P.}~\bibnamefont {Stansby}}, \bibinfo {author} {\bibfnamefont {P.}~\bibnamefont {Mignosa}}, \ and\ \bibinfo {author} {\bibfnamefont {J.}~\bibnamefont {Feldman}},\ }\bibfield  {title} {\enquote {\bibinfo {title} {Variable resolution for {{SPH}}: {{A}} dynamic particle coalescing and splitting scheme},}\ }\href {\doibase 10.1016/j.cma.2012.12.014} {\bibfield  {journal} {\bibinfo  {journal} {Computer Methods in Applied Mechanics and Engineering}\ }\textbf {\bibinfo {volume} {256}},\ \bibinfo {pages} {132--148} (\bibinfo {year} {2013})}\BibitemShut {NoStop}%
\bibitem [{\citenamefont {Yang}\ and\ \citenamefont {Kong}(2019)}]{yangAdaptiveResolutionMultiphase2019}%
  \BibitemOpen
  \bibfield  {author} {\bibinfo {author} {\bibfnamefont {X.}~\bibnamefont {Yang}}\ and\ \bibinfo {author} {\bibfnamefont {S.-C.}\ \bibnamefont {Kong}},\ }\bibfield  {title} {\enquote {\bibinfo {title} {Adaptive resolution for multiphase smoothed particle hydrodynamics},}\ }\href {\doibase 10.1016/j.cpc.2019.01.002} {\bibfield  {journal} {\bibinfo  {journal} {Computer Physics Communications}\ }\textbf {\bibinfo {volume} {239}},\ \bibinfo {pages} {112--125} (\bibinfo {year} {2019})}\BibitemShut {NoStop}%
\bibitem [{\citenamefont {Muta}\ \emph {et~al.}(2020)\citenamefont {Muta}, \citenamefont {Ramachandran},\ and\ \citenamefont {Negi}}]{mutaEfficientOpenSource2020}%
  \BibitemOpen
  \bibfield  {author} {\bibinfo {author} {\bibfnamefont {A.}~\bibnamefont {Muta}}, \bibinfo {author} {\bibfnamefont {P.}~\bibnamefont {Ramachandran}}, \ and\ \bibinfo {author} {\bibfnamefont {P.}~\bibnamefont {Negi}},\ }\bibfield  {title} {\enquote {\bibinfo {title} {An efficient, open source, iterative {{ISPH}} scheme},}\ }\href {\doibase 10.1016/j.cpc.2020.107283} {\bibfield  {journal} {\bibinfo  {journal} {Computer Physics Communications}\ }\textbf {\bibinfo {volume} {255}},\ \bibinfo {pages} {107283} (\bibinfo {year} {2020})},\ \Eprint {http://arxiv.org/abs/1908.01762} {arxiv:1908.01762} \BibitemShut {NoStop}%
\bibitem [{\citenamefont {Ramachandran}\ \emph {et~al.}(2021{\natexlab{a}})\citenamefont {Ramachandran}, \citenamefont {Muta},\ and\ \citenamefont {Ramakrishna}}]{ramachandranDualtimeSmoothedParticle2021}%
  \BibitemOpen
  \bibfield  {author} {\bibinfo {author} {\bibfnamefont {P.}~\bibnamefont {Ramachandran}}, \bibinfo {author} {\bibfnamefont {A.}~\bibnamefont {Muta}}, \ and\ \bibinfo {author} {\bibfnamefont {M.}~\bibnamefont {Ramakrishna}},\ }\bibfield  {title} {\enquote {\bibinfo {title} {Dual-time smoothed particle hydrodynamics for incompressible fluid simulation},}\ }\href {\doibase 10.1016/j.compfluid.2021.105031} {\bibfield  {journal} {\bibinfo  {journal} {Computers \& Fluids}\ }\textbf {\bibinfo {volume} {227}},\ \bibinfo {pages} {105031} (\bibinfo {year} {2021}{\natexlab{a}})}\BibitemShut {NoStop}%
\bibitem [{\citenamefont {Ramachandran}\ \emph {et~al.}(2021{\natexlab{b}})\citenamefont {Ramachandran}, \citenamefont {Bhosale}, \citenamefont {Puri}, \citenamefont {Negi}, \citenamefont {Muta}, \citenamefont {Dinesh}, \citenamefont {Menon}, \citenamefont {Govind}, \citenamefont {Sanka}, \citenamefont {Sebastian}, \citenamefont {Sen}, \citenamefont {Kaushik}, \citenamefont {Kumar}, \citenamefont {Kurapati}, \citenamefont {Patil}, \citenamefont {Tavker}, \citenamefont {Pandey}, \citenamefont {Kaushik}, \citenamefont {Dutt},\ and\ \citenamefont {Agarwal}}]{ramachandranPySPHPythonbasedFramework2021a}%
  \BibitemOpen
  \bibfield  {author} {\bibinfo {author} {\bibfnamefont {P.}~\bibnamefont {Ramachandran}}, \bibinfo {author} {\bibfnamefont {A.}~\bibnamefont {Bhosale}}, \bibinfo {author} {\bibfnamefont {K.}~\bibnamefont {Puri}}, \bibinfo {author} {\bibfnamefont {P.}~\bibnamefont {Negi}}, \bibinfo {author} {\bibfnamefont {A.}~\bibnamefont {Muta}}, \bibinfo {author} {\bibfnamefont {A.}~\bibnamefont {Dinesh}}, \bibinfo {author} {\bibfnamefont {D.}~\bibnamefont {Menon}}, \bibinfo {author} {\bibfnamefont {R.}~\bibnamefont {Govind}}, \bibinfo {author} {\bibfnamefont {S.}~\bibnamefont {Sanka}}, \bibinfo {author} {\bibfnamefont {A.~S.}\ \bibnamefont {Sebastian}}, \bibinfo {author} {\bibfnamefont {A.}~\bibnamefont {Sen}}, \bibinfo {author} {\bibfnamefont {R.}~\bibnamefont {Kaushik}}, \bibinfo {author} {\bibfnamefont {A.}~\bibnamefont {Kumar}}, \bibinfo {author} {\bibfnamefont {V.}~\bibnamefont {Kurapati}}, \bibinfo {author} {\bibfnamefont {M.}~\bibnamefont {Patil}}, \bibinfo {author} {\bibfnamefont {D.}~\bibnamefont {Tavker}}, \bibinfo {author} {\bibfnamefont {P.}~\bibnamefont {Pandey}}, \bibinfo {author} {\bibfnamefont {C.}~\bibnamefont {Kaushik}}, \bibinfo {author} {\bibfnamefont {A.}~\bibnamefont {Dutt}}, \ and\ \bibinfo {author} {\bibfnamefont {A.}~\bibnamefont {Agarwal}},\ }\bibfield  {title} {\enquote {\bibinfo {title} {{{PySPH}}: {{A Python-based Framework}} for {{Smoothed Particle Hydrodynamics}}},}\ }\href {\doibase 10.1145/3460773} {\bibfield  {journal} {\bibinfo  {journal} {ACM Transactions on Mathematical Software}\ }\textbf {\bibinfo {volume} {47}},\ \bibinfo {pages} {34:1--34:38} (\bibinfo {year} {2021}{\natexlab{b}})}\BibitemShut {NoStop}%
\bibitem [{\citenamefont {Negi}\ and\ \citenamefont {Ramachandran}(2021)}]{negiAlgorithmsUniformParticle2021}%
  \BibitemOpen
  \bibfield  {author} {\bibinfo {author} {\bibfnamefont {P.}~\bibnamefont {Negi}}\ and\ \bibinfo {author} {\bibfnamefont {P.}~\bibnamefont {Ramachandran}},\ }\bibfield  {title} {\enquote {\bibinfo {title} {Algorithms for uniform particle initialization in domains with complex boundaries},}\ }\href {\doibase 10.1016/j.cpc.2021.108008} {\bibfield  {journal} {\bibinfo  {journal} {Computer Physics Communications}\ }\textbf {\bibinfo {volume} {265}},\ \bibinfo {pages} {108008} (\bibinfo {year} {2021})}\BibitemShut {NoStop}%
\bibitem [{\citenamefont {Ramachandran}(2018)}]{ramachandranAutomanPythonBasedAutomation2018}%
  \BibitemOpen
  \bibfield  {author} {\bibinfo {author} {\bibfnamefont {P.}~\bibnamefont {Ramachandran}},\ }\bibfield  {title} {\enquote {\bibinfo {title} {Automan: {{A Python-Based Automation Framework}} for {{Numerical Computing}}},}\ }\href {\doibase 10.1109/MCSE.2018.05329818} {\bibfield  {journal} {\bibinfo  {journal} {Computing in Science \& Engineering}\ }\textbf {\bibinfo {volume} {20}},\ \bibinfo {pages} {81--97} (\bibinfo {year} {2018})}\BibitemShut {NoStop}%
\bibitem [{\citenamefont {Roe}(1981)}]{roeApproximateRiemannSolvers1981}%
  \BibitemOpen
  \bibfield  {author} {\bibinfo {author} {\bibfnamefont {P.~L.}\ \bibnamefont {Roe}},\ }\bibfield  {title} {\enquote {\bibinfo {title} {Approximate {{Riemann}} solvers, parameter vectors, and difference schemes},}\ }\href {\doibase 10.1016/0021-9991(81)90128-5} {\bibfield  {journal} {\bibinfo  {journal} {Journal of Computational Physics}\ }\textbf {\bibinfo {volume} {43}},\ \bibinfo {pages} {357--372} (\bibinfo {year} {1981})}\BibitemShut {NoStop}%
\bibitem [{\citenamefont {Gibbons}\ \emph {et~al.}(2023)\citenamefont {Gibbons}, \citenamefont {Damm}, \citenamefont {Jacobs},\ and\ \citenamefont {Gollan}}]{gibbonsEilmerOpensourceMultiphysics2023}%
  \BibitemOpen
  \bibfield  {author} {\bibinfo {author} {\bibfnamefont {N.~N.}\ \bibnamefont {Gibbons}}, \bibinfo {author} {\bibfnamefont {K.~A.}\ \bibnamefont {Damm}}, \bibinfo {author} {\bibfnamefont {P.~A.}\ \bibnamefont {Jacobs}}, \ and\ \bibinfo {author} {\bibfnamefont {R.~J.}\ \bibnamefont {Gollan}},\ }\bibfield  {title} {\enquote {\bibinfo {title} {Eilmer: {{An}} open-source multi-physics hypersonic flow solver},}\ }\href {\doibase 10.1016/j.cpc.2022.108551} {\bibfield  {journal} {\bibinfo  {journal} {Computer Physics Communications}\ }\textbf {\bibinfo {volume} {282}},\ \bibinfo {pages} {108551} (\bibinfo {year} {2023})}\BibitemShut {NoStop}%
\bibitem [{\citenamefont {Economon}\ \emph {et~al.}(2016)\citenamefont {Economon}, \citenamefont {Palacios}, \citenamefont {Copeland}, \citenamefont {Lukaczyk},\ and\ \citenamefont {Alonso}}]{economonSU2OpenSourceSuite2016}%
  \BibitemOpen
  \bibfield  {author} {\bibinfo {author} {\bibfnamefont {T.~D.}\ \bibnamefont {Economon}}, \bibinfo {author} {\bibfnamefont {F.}~\bibnamefont {Palacios}}, \bibinfo {author} {\bibfnamefont {S.~R.}\ \bibnamefont {Copeland}}, \bibinfo {author} {\bibfnamefont {T.~W.}\ \bibnamefont {Lukaczyk}}, \ and\ \bibinfo {author} {\bibfnamefont {J.~J.}\ \bibnamefont {Alonso}},\ }\bibfield  {title} {\enquote {\bibinfo {title} {{{SU2}}: {{An Open-Source Suite}} for {{Multiphysics Simulation}} and {{Design}}},}\ }\href {\doibase 10.2514/1.J053813} {\bibfield  {journal} {\bibinfo  {journal} {AIAA Journal}\ }\textbf {\bibinfo {volume} {54}},\ \bibinfo {pages} {828--846} (\bibinfo {year} {2016})}\BibitemShut {NoStop}%
\bibitem [{\citenamefont {Geuzaine}\ and\ \citenamefont {Remacle}(2009)}]{geuzaineGmsh3DFinite2009}%
  \BibitemOpen
  \bibfield  {author} {\bibinfo {author} {\bibfnamefont {C.}~\bibnamefont {Geuzaine}}\ and\ \bibinfo {author} {\bibfnamefont {J.-F.}\ \bibnamefont {Remacle}},\ }\bibfield  {title} {\enquote {\bibinfo {title} {Gmsh: {{A}} 3-{{D}} finite element mesh generator with built-in pre- and post-processing facilities},}\ }\href {\doibase 10.1002/nme.2579} {\bibfield  {journal} {\bibinfo  {journal} {International Journal for Numerical Methods in Engineering}\ }\textbf {\bibinfo {volume} {79}},\ \bibinfo {pages} {1309--1331} (\bibinfo {year} {2009})}\BibitemShut {NoStop}%
\bibitem [{\citenamefont {Anderson}(2021)}]{andersonModernCompressibleFlow2021}%
  \BibitemOpen
  \bibfield  {author} {\bibinfo {author} {\bibfnamefont {J.~D.}\ \bibnamefont {Anderson}},\ }\href@noop {} {\emph {\bibinfo {title} {Modern Compressible Flow: With Historical Perspective}}},\ \bibinfo {edition} {fourth edition}\ ed.\ (\bibinfo  {publisher} {{McGraw Hill}},\ \bibinfo {address} {{New York, NY}},\ \bibinfo {year} {2021})\BibitemShut {NoStop}%
\bibitem [{\citenamefont {Sod}(1978)}]{sodSurveySeveralFinite1978}%
  \BibitemOpen
  \bibfield  {author} {\bibinfo {author} {\bibfnamefont {G.~A.}\ \bibnamefont {Sod}},\ }\bibfield  {title} {\enquote {\bibinfo {title} {A {{Survey}} of {{Several Finite Difference Methods}} for {{Systems}} of {{Nonlinear Hyperbolic Conservation Laws}}},}\ }\href {\doibase 10.1016/0021-9991(78)90023-2} {\bibfield  {journal} {\bibinfo  {journal} {Journal of Computational Physics}\ }\textbf {\bibinfo {volume} {27}},\ \bibinfo {pages} {1--31} (\bibinfo {year} {1978})}\BibitemShut {NoStop}%
\bibitem [{\citenamefont {Billig}(1967)}]{billigShockwaveShapesSphericaland1967}%
  \BibitemOpen
  \bibfield  {author} {\bibinfo {author} {\bibfnamefont {F.~S.}\ \bibnamefont {Billig}},\ }\bibfield  {title} {\enquote {\bibinfo {title} {Shock-wave shapes around spherical-and cylindrical-nosed bodies.}}\ }\href {\doibase 10.2514/3.28969} {\bibfield  {journal} {\bibinfo  {journal} {Journal of Spacecraft and Rockets}\ }\textbf {\bibinfo {volume} {4}},\ \bibinfo {pages} {822--823} (\bibinfo {year} {1967})}\BibitemShut {NoStop}%
\bibitem [{\citenamefont {Anderson}(2019)}]{andersonHypersonicHighTemperatureGas2019}%
  \BibitemOpen
  \bibfield  {author} {\bibinfo {author} {\bibfnamefont {J.~D.}\ \bibnamefont {Anderson}},\ }\href {\doibase 10.2514/4.105142} {\emph {\bibinfo {title} {Hypersonic and {{High-Temperature Gas Dynamics}}, {{Third Edition}}}}}\ (\bibinfo  {publisher} {{American Institute of Aeronautics and Astronautics, Inc.}},\ \bibinfo {address} {{Reston, VA}},\ \bibinfo {year} {2019})\BibitemShut {NoStop}%
\bibitem [{\citenamefont {Back}\ \emph {et~al.}(1965)\citenamefont {Back}, \citenamefont {Gier},\ and\ \citenamefont {Massier}}]{backComparisonMeasuredPredicted1965b}%
  \BibitemOpen
  \bibfield  {author} {\bibinfo {author} {\bibfnamefont {L.~H.}\ \bibnamefont {Back}}, \bibinfo {author} {\bibfnamefont {H.~L.}\ \bibnamefont {Gier}}, \ and\ \bibinfo {author} {\bibfnamefont {P.~F.}\ \bibnamefont {Massier}},\ }\bibfield  {title} {\enquote {\bibinfo {title} {Comparison of measured and predicted flows through conical supersonic nozzles, with emphasis on the transonic region},}\ }\href {\doibase 10.2514/3.3216} {\bibfield  {journal} {\bibinfo  {journal} {AIAA Journal}\ }\textbf {\bibinfo {volume} {3}},\ \bibinfo {pages} {1606--1614} (\bibinfo {year} {1965})}\BibitemShut {NoStop}%
\bibitem [{\citenamefont {Emery}(1968)}]{emeryEvaluationSeveralDifferencing1968}%
  \BibitemOpen
  \bibfield  {author} {\bibinfo {author} {\bibfnamefont {A.~F.}\ \bibnamefont {Emery}},\ }\bibfield  {title} {\enquote {\bibinfo {title} {An evaluation of several differencing methods for inviscid fluid flow problems},}\ }\href {\doibase 10.1016/0021-9991(68)90060-0} {\bibfield  {journal} {\bibinfo  {journal} {Journal of Computational Physics}\ }\textbf {\bibinfo {volume} {2}},\ \bibinfo {pages} {306--331} (\bibinfo {year} {1968})}\BibitemShut {NoStop}%
\bibitem [{\citenamefont {Woodward}\ and\ \citenamefont {Colella}(1984)}]{woodwardNumericalSimulationTwodimensional1984a}%
  \BibitemOpen
  \bibfield  {author} {\bibinfo {author} {\bibfnamefont {P.}~\bibnamefont {Woodward}}\ and\ \bibinfo {author} {\bibfnamefont {P.}~\bibnamefont {Colella}},\ }\bibfield  {title} {\enquote {\bibinfo {title} {The numerical simulation of two-dimensional fluid flow with strong shocks},}\ }\href {\doibase 10.1016/0021-9991(84)90142-6} {\bibfield  {journal} {\bibinfo  {journal} {Journal of Computational Physics}\ }\textbf {\bibinfo {volume} {54}},\ \bibinfo {pages} {115--173} (\bibinfo {year} {1984})}\BibitemShut {NoStop}%
\bibitem [{\citenamefont {Gao}\ \emph {et~al.}(2023)\citenamefont {Gao}, \citenamefont {Liang},\ and\ \citenamefont {Fu}}]{gaoNewSmoothedParticle2023a}%
  \BibitemOpen
  \bibfield  {author} {\bibinfo {author} {\bibfnamefont {T.}~\bibnamefont {Gao}}, \bibinfo {author} {\bibfnamefont {T.}~\bibnamefont {Liang}}, \ and\ \bibinfo {author} {\bibfnamefont {L.}~\bibnamefont {Fu}},\ }\bibfield  {title} {\enquote {\bibinfo {title} {A new smoothed particle hydrodynamics method based on high-order moving-least-square targeted essentially non-oscillatory scheme for compressible flows},}\ }\href {\doibase 10.1016/j.jcp.2023.112270} {\bibfield  {journal} {\bibinfo  {journal} {Journal of Computational Physics}\ }\textbf {\bibinfo {volume} {489}},\ \bibinfo {pages} {112270} (\bibinfo {year} {2023})}\BibitemShut {NoStop}%
\bibitem [{\citenamefont {Tan}\ and\ \citenamefont {Shu}(2010)}]{tanInverseLaxWendroffProcedure2010}%
  \BibitemOpen
  \bibfield  {author} {\bibinfo {author} {\bibfnamefont {S.}~\bibnamefont {Tan}}\ and\ \bibinfo {author} {\bibfnamefont {C.-W.}\ \bibnamefont {Shu}},\ }\bibfield  {title} {\enquote {\bibinfo {title} {Inverse {{Lax-Wendroff}} procedure for numerical boundary conditions of conservation laws},}\ }\href {\doibase 10.1016/j.jcp.2010.07.014} {\bibfield  {journal} {\bibinfo  {journal} {Journal of Computational Physics}\ }\textbf {\bibinfo {volume} {229}},\ \bibinfo {pages} {8144--8166} (\bibinfo {year} {2010})}\BibitemShut {NoStop}%
\bibitem [{\citenamefont {Vevek}\ \emph {et~al.}(2019)\citenamefont {Vevek}, \citenamefont {Zang},\ and\ \citenamefont {New}}]{vevekAlternativeSetupsDouble2019b}%
  \BibitemOpen
  \bibfield  {author} {\bibinfo {author} {\bibfnamefont {U.~S.}\ \bibnamefont {Vevek}}, \bibinfo {author} {\bibfnamefont {B.}~\bibnamefont {Zang}}, \ and\ \bibinfo {author} {\bibfnamefont {T.~H.}\ \bibnamefont {New}},\ }\bibfield  {title} {\enquote {\bibinfo {title} {On {{Alternative Setups}} of the {{Double Mach Reflection Problem}}},}\ }\href {\doibase 10.1007/s10915-018-0803-x} {\bibfield  {journal} {\bibinfo  {journal} {Journal of Scientific Computing}\ }\textbf {\bibinfo {volume} {78}},\ \bibinfo {pages} {1291--1303} (\bibinfo {year} {2019})}\BibitemShut {NoStop}%
\bibitem [{\citenamefont {Prasanna~Kumar}\ and\ \citenamefont {Patnaik}(2018)}]{prasannakumarMultimassCorrectionMulticomponent2018}%
  \BibitemOpen
  \bibfield  {author} {\bibinfo {author} {\bibfnamefont {S.~S.}\ \bibnamefont {Prasanna~Kumar}}\ and\ \bibinfo {author} {\bibfnamefont {B.~S.~V.}\ \bibnamefont {Patnaik}},\ }\bibfield  {title} {\enquote {\bibinfo {title} {A multimass correction for multicomponent fluid flow simulation using {{Smoothed Particle Hydrodynamics}}},}\ }\href {\doibase 10.1002/nme.5727} {\bibfield  {journal} {\bibinfo  {journal} {International Journal for Numerical Methods in Engineering}\ }\textbf {\bibinfo {volume} {113}},\ \bibinfo {pages} {1929--1949} (\bibinfo {year} {2018})}\BibitemShut {NoStop}%
\bibitem [{\citenamefont {Moss}\ \emph {et~al.}(2006)\citenamefont {Moss}, \citenamefont {Glass},\ and\ \citenamefont {Greene}}]{mossDSMCSimulationsApollo2006}%
  \BibitemOpen
  \bibfield  {author} {\bibinfo {author} {\bibfnamefont {J.}~\bibnamefont {Moss}}, \bibinfo {author} {\bibfnamefont {C.}~\bibnamefont {Glass}}, \ and\ \bibinfo {author} {\bibfnamefont {F.}~\bibnamefont {Greene}},\ }\bibfield  {title} {\enquote {\bibinfo {title} {{{DSMC Simulations}} of {{Apollo Capsule Aerodynamics}} for {{Hypersonic Rarefied Conditions}}},}\ }in\ \href {\doibase 10.2514/6.2006-3577} {\emph {\bibinfo {booktitle} {9th {{AIAA}}/{{ASME Joint Thermophysics}} and {{Heat Transfer Conference}}}}}\ (\bibinfo  {publisher} {{American Institute of Aeronautics and Astronautics}},\ \bibinfo {address} {{San Francisco, California}},\ \bibinfo {year} {2006})\BibitemShut {NoStop}%
\bibitem [{\citenamefont {Price}(2007)}]{priceSPLASHInteractiveVisualisation2007}%
  \BibitemOpen
  \bibfield  {author} {\bibinfo {author} {\bibfnamefont {D.~J.}\ \bibnamefont {Price}},\ }\bibfield  {title} {\enquote {\bibinfo {title} {{{SPLASH}}: {{An Interactive Visualisation Tool}} for {{Smoothed Particle Hydrodynamics Simulations}}},}\ }\href {\doibase 10.1071/AS07022} {\bibfield  {journal} {\bibinfo  {journal} {Publications of the Astronomical Society of Australia}\ }\textbf {\bibinfo {volume} {24}},\ \bibinfo {pages} {159--173} (\bibinfo {year} {2007})}\BibitemShut {NoStop}%
\bibitem [{\citenamefont {Hopkins}(2015)}]{hopkinsNewClassAccurate2015}%
  \BibitemOpen
  \bibfield  {author} {\bibinfo {author} {\bibfnamefont {P.~F.}\ \bibnamefont {Hopkins}},\ }\bibfield  {title} {\enquote {\bibinfo {title} {A new class of accurate, mesh-free hydrodynamic simulation methods},}\ }\href {\doibase 10.1093/mnras/stv195} {\bibfield  {journal} {\bibinfo  {journal} {Monthly Notices of the Royal Astronomical Society}\ }\textbf {\bibinfo {volume} {450}},\ \bibinfo {pages} {53--110} (\bibinfo {year} {2015})}\BibitemShut {NoStop}%
\end{thebibliography}%

\end{document}
%